\documentclass{aastex61}
\usepackage{graphicx}
\usepackage{amsmath}
\usepackage{epsfig}
\usepackage{bm}

\shorttitle{Galactic Center Radio bubbles}
\shortauthors{M.F.Zhang et al.}
\begin{document}
\title{A Supernova-driven, Magnetically-collimated Outflow as the Origin of the Galactic Center Radio Bubbles}

\author[0000-0001-8261-3254]{Mengfei Zhang}
\affil{School of Astronomy and Space Science, Nanjing University, Nanjing 210023, China}
\affil{Key Laboratory of Modern Astronomy and Astrophysics (Nanjing University), Ministry of Education, Nanjing 210023, China}
\email{zmf@nju.edu.cn}

\author[
0000-0003-0355-6437]{Zhiyuan Li}
\affil{School of Astronomy and Space Science, Nanjing University, Nanjing 210023, China}
\affil{Key Laboratory of Modern Astronomy and Astrophysics (Nanjing University), Ministry of Education, Nanjing 210023, China}
\email{lizy@nju.edu.cn}

\author[0000-0002-6753-2066]{Mark R. Morris}
\affil{Department of Physics and Astronomy, University of California, Los Angeles, CA 90095, USA}

%\correspondingauthor{M.F.Zhang}
%\email{zmf@nju.edu.cn}

% \author[0000-0001-8261-3254]{
% M.F.Zhang\altaffilmark{1,2},
% W.W.Tian\altaffilmark{1},
% D.A.Leahy\altaffilmark{3},
% H.Zhu\altaffilmark{1,4},
% X.H.Cui\altaffilmark{1}
% }
% \affil{
% \sp{1} Key Laboratory of Optical Astronomy, National Astronomical Observatories, Chinese Academy of Sciences,
% Beijing 100012, China \\
% \sp{2} University of Chinese Academy of Sciences, 19A Yuquan Road, Shijingshan District, Beijing 100049, China\\
% \sp{3} Department of Physics $\&$ Astronomy, University of Calgary, Calgary, Alberta T2N 1N4, Canada\\
% \sp{4} Harvard-Smithsonian Center for Astrophysics, 60 Garden Street, Cambridge, MA 02138, USA\\
% }

% \email{tww@bao.ac.cn}

% \maketitle
\label{firstpage}
\begin{abstract}
A pair of non-thermal radio bubbles recently discovered in the inner few hundred parsecs of the Galactic center bears a close spatial association with elongated, thermal X-ray features called the X-ray chimneys.
While their morphology, position, and orientation vividly point to an outflow from the Galactic center, the physical processes responsible for the outflow remain to be understood.
%The origin and evolution of these features are essential to understanding the violent activities in the Galactic center.
We use three-dimensional magnetohydrodynamic simulations to test the hypothesis that the radio bubbles/X-ray chimneys are the manifestation of an energetic outflow driven by multiple core-collapsed supernovae in the nuclear stellar disk, where numerous massive stars are known to be present.
%The supernovae (SNe) are set to explode at 1 kyr intervals in a cylindrical region with a diameter of 50 pc and a height of 10 pc in the central region.
Our simulations are run with different combinations of two main parameters, the supernova birth rate and the strength of a global magnetic field being vertically oriented with respect to the disk.
The simulation results show that a hot gas outflow can naturally form and acquire a vertically elongated shape due to collimation by the magnetic pressure.
In particular, the simulation with an initial magnetic field strength of 80 $\mu$G and a supernova rate of 1 kyr$^{-1}$
can well reproduce the observed morphology, internal energy and X-ray luminosity of the bubbles after an evolutionary time of 330 kyr.
On the other hand, a magnetic field strength of 200 $\mu$G gives rise to an overly elongated outflow that is inconsistent with the observed bubbles.
The simulations also reveal that, inside the bubbles, mutual collisions between the shock waves of individual supernovae produce dense filaments of locally amplified magnetic field. Such filaments may account for a fraction of the synchrotron-emitting radio filaments known to exist in the Galactic center.
\end{abstract}


\section{Introduction} \label{sec:intro}
Galactic outflows driven by energy and momentum of an active galactic nucleus (AGN) and/or supernovae (SNe) are now understood to be an indispensable component of the galactic ecosystem \citep{2012ARA&A..50..455F, 2014ARA&A..52..589H, 2017hsn..book.2431H, 2018Galax...6..114Z}.
%The AGN can be divided into two distinct populations, the radiative mode and the jet mode, the outflows of which are driven by radiation and jets, respectively \citep{2012ARA&A..50..455F, 2014ARA&A..52..589H}.
%The active star formation can produce many massive stars, leading to the subsequent supernovae explosions, which will also produce outflows by hot winds, radiation pressure of starlight and cosmic rays \citep{2017hsn..book.2431H, 2018Galax...6..114Z}.
%The large-scale outflows driven by the active star formation are multi-phase and difficult to be directly linked with the small-scale stellar activity.
%To understand these outflows, we need to figure out more details of different outflows.
Multi-wavelength observations over the past decades have established an ever-growing inventory of galactic outflows, leading to the recognition that these outflows typically involve multi-scales and multi-phases.
However, our physical understanding of galactic outflows, in particular their mass budget, energetics and life cycle, is still far from complete.

The Galactic center, loosely defined here as the innermost few hundred parsec region of our Galaxy, provides the closest and perhaps the best laboratory for studying the formation and early evolution of a galactic outflow.
%Sgr A$^*$ is presently in a quiescent state, but it was likely much more active in the recent past \citep{2010ApJ...714..732P, Ponti2013, 2018ApJ...856..180C}.
%Also, there was possibly a short-duration burst of star formation a few millions of years ago \citep{2010RvMP...82.3121G}, implying multi-supernovae (SNe) explosions after many years.
%In addition, \citet{NoguerasLara2019} suggested an increased star formation rate (SFR) of $0.2-0.8\rm~M_{\odot}~yr^{-1}$ during the past 30 Myr in the Galactic center, but this is a long-duration mean value and possibly much larger or smaller in a short duration.
%These activities would influence the surrounding Galactic center environment and the large scale of the Milky Way, then form outflows and various diffuse structures.
Observational evidence has accumulated over recent years for a multi-phase outflow from the Galactic center \citep{2003ApJ...582..246B, 2010ApJ...708..474L, 2019ApJ...875...32N}, collectively known as the Galactic Center Lobe (GCL; \citealp{1984Natur.310..568S}), a loop-like feature extending vertically out to $\gtrsim$ 1 degree (at a presumed distance of 8 kpc, $1^\circ$ corresponds to 140 pc) north of the disk mid-plane.
Compelling evidence also exists for outflows at still larger (kiloparsec-) scales \citep{2010ApJ...724.1044S, 2013Natur.493...66C, 2018ApJ...855...33D, 2020Natur.584..364D, Predehl2020}, but the physical relation between the outflows on different scales, e.g., whether they were produced by the same mechanism, remains an open question.

More recently, our view of the Galactic center outflow is further sharpened.
Based on high-resolution radio continuum observations afforded by the MeerKAT radio telescope, \citet{2019Natur.573..235H} found evidence for a pair of radio bubbles in the Galactic center,
%likely associated with the radio lobe and NTFs, which enriches these structures.
which are roughly symmetric about the disk mid-plane with a width of 140 pc and a full length of 430 pc.
The northern bubble is spatially coincident with the GCL, but it is more clearly limb-brightened. In particular, the eastern side of the radio bubbles is delineated by the famous Radio Arc \citep{1984Natur.310..557Y} and its northern and southern extension toward higher latitudes; the western side is also bounded by prominent non-thermal filaments (NTFs; \citealp{1984Natur.310..557Y}).
Non-thermal emission is predominant in the radio bubbles at the observed frequency of 1284 MHz, although the GCL is known to show substantial thermal emission at different wavebands  \citep{2003ApJ...582..246B, 2010ApJ...708..474L,2019PASJ...71...80N}.
%Moreover, X-ray observations show us some different patterns \citep{2003ANS...324..167M}.
Strikingly, the shells of the radio bubbles delineate the so-called ``X-ray chimneys'' recently discovered by X-ray observations \citep{2019Natur.567..347P}, which is a pair of diffuse, thermal X-ray features extending above and below the mid-plane.
This strongly suggests a physical relation between the two features, reminiscent of a collimated hot gas outflow with an expanding shell \citep{2021A&A...646A..66P}.
%At much larger scale, the Fermi bubbles and eROSITA bubbles were discovered by \citet{2010ApJ...724.1044S} and \citet{Predehl2020}, whose relations with the small-scale structures deserve more discussion.
%In addition, \citet{1984Natur.310..568S} and \citet{1984Natur.310..557Y} discovered a radio lobe and the non-thermal filaments (NTFs), respectively.
%They both have a size of tens of parsecs, but some NTFs are longer.
%And the radio lobe is possibly related to an X-ray lobe detected inside the X-ray chimney \citep{Ponti2015, 2019Natur.567..347P}.
%Recent observations show more details of these radio structures \citep{2004ApJS..155..421Y, 2013Natur.493...66C}, but their origin is still in debate.
% These multi-wavelength results can help us better understand these structures, their origin and evolution.

Proposed origins for the Galactic center outflow as well as for the outflows on larger scales (i.e., the Fermi bubbles and the recently discovered eROSITA bubbles; \citealp{2010ApJ...724.1044S, Predehl2020}) fall in two categories \citep{2019Natur.573..235H}: (i) past activity from the central
super-massive black hole (SMBH), commonly known as Sgr A*, which is currently in a quiescent state \citep{Cheng2011, Zubovas2011, 2012MNRAS.424..666Z, Zhang2020, 2020ApJ...904...46K}; or (ii) episodic or continuous nuclear star formation  \citep{2010RvMP...82.3121G, 2014MNRAS.444L..39L, Crocker2015}.
%be related with the unique environment and high-energy events only happening in the Galactic center.
%For the radio bubbles, one possible origin is a jet or a burst caused by the accretion around the super massive black hole (SMBH), and the other is a multi-SNe burst
In principle, both processes can drive an energetic outflow and produce the bubble-like structures observed at multi-wavelengths and multi-scales.
Therefore a quantitative modeling and close comparison with the observations are crucial to distinguish between the two scenarios.
In the literature, there have been a number of numerical simulations of a large-scale outflow from the Galactic center, which focuses on the formation of the Fermi bubbles by AGN jets or AGN winds \citep{Guo2012, Mou2014, Mou2015, Cheng2015, Zhang2020}.
In addition, \cite{Sarkar2015} and \cite{2017MNRAS.467.3544S} investigate the formation of the Fermi bubbles by simulating a nuclear starburst-driven wind.
In this work, we investigate the specific scenario that the radio bubbles/X-ray chimneys are the manifestation of an outflow driven by sequential SN explosions concentrated in the Galactic center, using three-dimensional magnetohydrodynamic (MHD) simulations, which is the first attempt of this kind to our knowledge.
Recently, \citet{2017ApJ...841..101L} and \citet{Li2020} have performed advanced numerical simulations to study SNe-driven outflows on a similar physical scale, but these simulations were run with physical conditions typical of galactic disks.
The Galactic center, on the other hand, is a unique environment characterized by a strong gravity, a concentration of massive stars, and a strong and ordered magnetic field.
In particular, the presence of the NTFs, which have a strong tendency to be vertically oriented with respect to the disk, points to a vertical magnetic field in the Galactic center (see review by \citealp{Ferriere2009}).
Theoretical studies have demonstrated that a strong external magnetic field can significantly affect the evolution of a supernova remnant (SNR; \citealp{1991MNRAS.252...82I, 1995MNRAS.274.1157R, Wu2019}), as
the magnetic pressure confines the expansion of the SN ejecta in such a way that they preferentially propagate along the direction of the magnetic field.

%The environments surrounding these structures are pretty complicated, therefore, an analytical solution is not convincing to investigate and verify different models.
%The numerical simulations can provide more quantitative information, but choosing initial conditions is difficult.
%\citet{Fielding2017} and \citet{Hu2019} simulated the supernova-driven winds from centers in galaxies, while \citet{2017ApJ...841..101L} and \citet{Li2020} also numerically studied the outflows of SNe in disks.
%However, they pay more attention to kpc structures and the initial conditions are not suitable for the Galactic center.
%A multi-SNe explosion simulation in the Galactic center is needed to study the influence of SNe and those small-scale structures, such as radio bubbles, the X-ray chimney and the NTFs.

We are thus motivated to perform numerical simulations to test the scenario of an SN-driven, magnetically-collimated outflow for the radio bubbles/X-ray chimneys.
In Section \ref{sec:sim}, we describe our basic model and settings of the simulation.
In Section \ref{sec:res}, we present the simulation results and confront them with the observations.
In Section \ref{sec:dis}, we discuss the implications as well as limitations of our results.
A summary is given in Section \ref{sec:sum}.

\section{Simulation} \label{sec:sim}
We use the publicly available MHD code \textit{PLUTO}\footnote{http://plutocode.ph.unito.it/} \citep{Mignone2007, Mignone2012} to simulate sequential SNe explosions in the Galactic center and the formation of an SN-driven bubble.
The global dynamical evolution and fine structures of the bubble necessarily depend on many physical processes and physical quantities of the Galactic center, some of which are not well constrained.
Rather than pursuing a full degree of realism or a thorough exploration of the parameter space, our main aim here is to test a simplified but well-motivated model for the bubble formation.

\subsection{Basic MHD Equations and Magnetic Field Configuration}
\label{subsec:Bfield}
The simulation is based on a three-dimensional (3D) MHD cartesian frame with a grid of $512^3$, equivalent to a physical volume of 200$^3$~pc$^3$ and a linear resolution of 0.39 pc.
We set the $z$-axis to be perpendicular to the Galactic disk (north as positive), the $y$-axis to be parallel to the line-of-sight (the observer at the negative side), and the $x$-axis to run along decreasing Galactic longitude.
Because the radio bubbles are roughly symmetric about the Galactic plane, we only simulate the $z>0$ volume, sufficient to enclose the northern bubble, which exhibits a size of $\sim$120 pc (width) $\times$ 190 pc (height).
We adopt an outflow boundary condition.

The simulation is governed by the ideal MHD conservation equations,
\begin{equation}
    \begin{cases}
      \dfrac{\partial \rho}{\partial t} + \nabla \cdot (\rho \bm{v}) = 0 ,\\
      \\
      \dfrac{\partial (\rho\bm{v})}{\partial t}+\nabla \cdot\left[\rho\bm{vv}-\dfrac{\bm{B B}}{4\pi}+\bm{1}\left(p+\dfrac{\bm{B}^{2}}{8\pi}\right)\right]^{T}=-\rho \nabla \Phi, \\
      \\
      \dfrac{\partial E_{t}}{\partial t}+\nabla \cdot\left[\left(\dfrac{\rho \bm{v}^{2}}{2}+\rho e+p+\rho \Phi\right) \dfrac{\bm{v}-\bm{v} \times \bm{B} \times \bm{B}}{4\pi}\right] \\
      = -\dfrac{\partial\left( \rho \Phi\right)}{\partial t}, \\
      \\
      \dfrac{\partial \bm{B}}{\partial t} - \nabla \times (\bm{v} \times \bm{B}) = 0,
    \end{cases}
\end{equation}
where $\rho$ is the mass density, $p$ the thermal pressure, $\bm{v}$ the velocity, $\bm{B}$ the magnetic field, $\bm{1}$ the dyadic tensor, $\Phi$ the gravitational potential, and $E_t$ the total energy density, defined as:
%A factor of $1/\sqrt{4\pi}$ has been absorbed in the definition of magnetic field, i.e., $\bm{B} = \bm{B_{\rm G}}/\sqrt{4\pi}$, where $\bm{B_{\rm G}}$ is the magnetic field in unit of Gauss.
%This is a non-relativistic equation set, which is appropriate for studying the morphology and radiation of the radio bubbles.
%The total energy density $E_t$ is defined as:
\begin{equation}
  E_t = \rho \epsilon + \dfrac{(\rho\bm{v})^2}{2\rho} + \dfrac{\bm{B}^2}{8\pi},
\end{equation}
where $\epsilon$ is the internal energy.
We use an ideal equation of state, i.e., $\epsilon = p/ (\Gamma -1)$, in which the ratio of specific heats $\Gamma$ = 5/3.

As mentioned in Section~\ref{sec:intro}, the orientation of the NTFs indicates that a vertical magnetic field is prevalent in the Galactic center.
We adopt a dipole magnetic field structure generated by a current loop with a diameter of 300 pc,  which can be expressed analytically \citep{simpson2001simple}.
With this large diameter, the magnetic field lines remain approximately vertical to the disk within our simulation volume.
%so we can simplify the analytical expression as $B_x,B_y=C_1xyz$ and $B_z=C_2z$, in which, x, y, z indicates the coordinates, $B_x, B_y, B_z$ are the magnetic field intensity along different axis, $C_1, C_2$ are constants.
There are ample evidence that the Galactic center has an average magnetic field strength substantially higher than in the disk \citep{Ferriere2009}.
\citet{2010Natur.463...65C} derived a lower limit of 50 $\mu$G for the central 400 pc, based on an upper limit in the detected diffuse $\gamma$-ray flux.
Given the observed radio spectral energy distribution of the Galactic center, a weaker magnetic field would lead to more relativistic electrons and consequently a higher $\gamma$-ray flux due to inverse Compton emission.
In fact, energy equipartition between the magnetic field, X-ray-emitting hot plasma and turbulent gas implies a magnetic field strength of $\sim$100 $\mu$G \citep{2010Natur.463...65C}.
On the other hand, \citet{Thomas2020} suggested a stronger magnetic strength of 200 $\mu$G in the NTFs.
In our fiducial run of simulation, the initial magnetic field strength at the origin ($x=y=z=0$) is set as $B_{\rm 0}=80~\mu$G.
Values of $50~\mu$G and $200~\mu$G are also tested to examine the effect of a weaker/stronger magnetic field (see Section~\ref{subsec:synthetic}).

The simulation neglects viscosity and thermal conduction, but takes into account radiative cooling. We adopt the TABULATED cooling function implemented in {\it PLUTO}, which is generated with \textit{Cloudy} for an optically thin plasma and solar abundances \citep{2017RMxAA..53..385F}.
% Although a more realistic cooling function may be implemented in a tabulated form, our preliminary test shows that this is rather time-consuming while having little effect on the dynamical evolution of the bubble, since the involved radiative cooling is only moderate.
% Therefore, we adopt the simple power-law cooling function.
We neglect the synchrotron cooling of relativistic electrons, which are presumably produced by the SN shocks (see Section~\ref{subsec:synthetic}).

\begin{table*}
  \caption{Simulation Parameters for the Radio Bubbles}
  \label{table:parameters}
  \centering
  \begin{tabular}{l l l l l}
      \hline\hline
      Fiducial Parameters                      & Value          & &                \\
      \hline
      SN Ejecta Mass                     & 10 M$_{\odot}$ & &\\
      SN Kinetic Energy        & 1$\times$ 10$^{51}$ erg & & \\
      Injection Radius        & 4 pc           & & \\
      Ambient Temperature                     & 1$\times$ 10$^{6}$ K    & &       \\
      Diameter of Explosion Region    & 50 pc & & \\
      Height of Explosion Region      & 10 pc & & \\
      \hline\hline
      Simulation Runs                 & B80I1      & B80I2     &  B50I1     & B200I1  \\
      \hline
      Magnetic Field Strength         & 80 $\mu$G  & 80 $\mu$G & 50 $\mu$G  & 200 $\mu$G    \\
      Explosion Interval              & 1 kyr      & 2 kyr     & 1 kyr      & 1 kyr  \\
      \hline
  \end{tabular}\\
\end{table*}

\subsection{Gravitational Potential and Initial ISM Conditions}
\label{subsec:gravity}
%The gravitation of the inner galaxy
The gravitation in the Galactic center mainly originates from two components, namely, the nuclear star cluster (NSC), which dominates the innermost $\sim$20 pc, and the nuclear stellar disk (ND) that occupies the inner few hundred parsecs.
We neglect larger-scale structures such as the bar and the Galactic disk.
The SMBH, which has four million solar masses and a sphere of influence of a few parsecs in radius, can also be ignored given the scales of interest here.
The NSC/ND will not evolve significantly on the timescale involved in our simulations, hence we adopt a fixed gravitational potential, which,
following \citet{2008ApJ...675.1278S}, can be approximated by a logarithmic form,
\begin{equation}
      \Phi = 0.5v_0^2\log(R_c^2+\dfrac{x^2}{a^2} + \dfrac{y^2}{b^2} +\dfrac{z^2}{c^2}),
\end{equation}
% the stable velocity on the flattened portion of the rotation curve
where $v_0$ is the asymptotic velocity of a flat rotation curve, $R_c$ is the core radius, and $a$, $b$ and $c$ are stretching parameters.
We adopt $v_0 = 98.6\rm~km~s^{-1}$, $R_c = 2\rm~pc$, $a=b=c=1$ for the NSC, and $v_0 = 190\rm~km~s^{-1}$, $R_c = 90\rm~pc$, $a=b=1$, $c=0.71$ for the ND, from Table 1 of \citet{2008ApJ...675.1278S}.
The combined NSC+ND potential has been found to provide a good match to the observed stellar mass distribution in the Galactic center \citep{2002A&A...384..112L}.

%The adopted gravitational potential is shown in Figure~\ref{fig:grav}. { [maybe omit this figure]}

% \begin{figure}
%     \centering
%     \includegraphics[width=0.472\textwidth]{phi_yz.eps}
%     \caption{The adopted Galactic center gravitational potential, which includes contributions from the NSC and the ND. }
% \label{fig:grav}
% \end{figure}

\begin{figure*}
    \centering
    \includegraphics[width=0.323\textwidth]{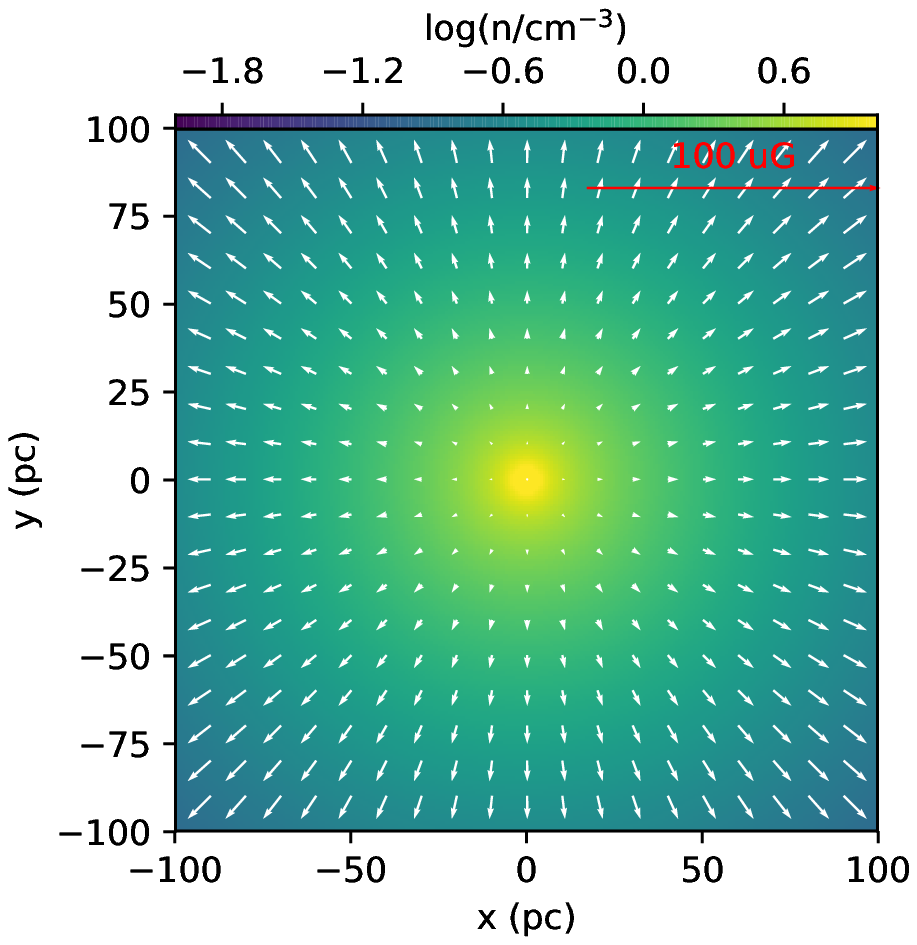}
    \includegraphics[width=0.323\textwidth]{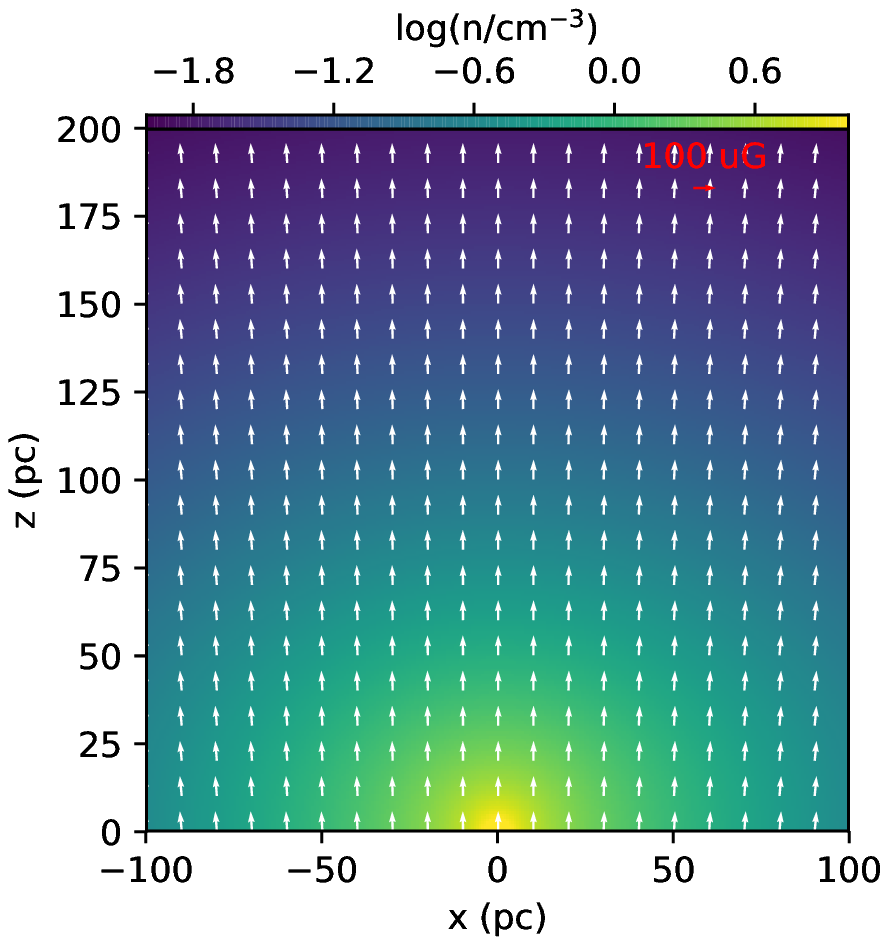}
    \includegraphics[width=0.323\textwidth]{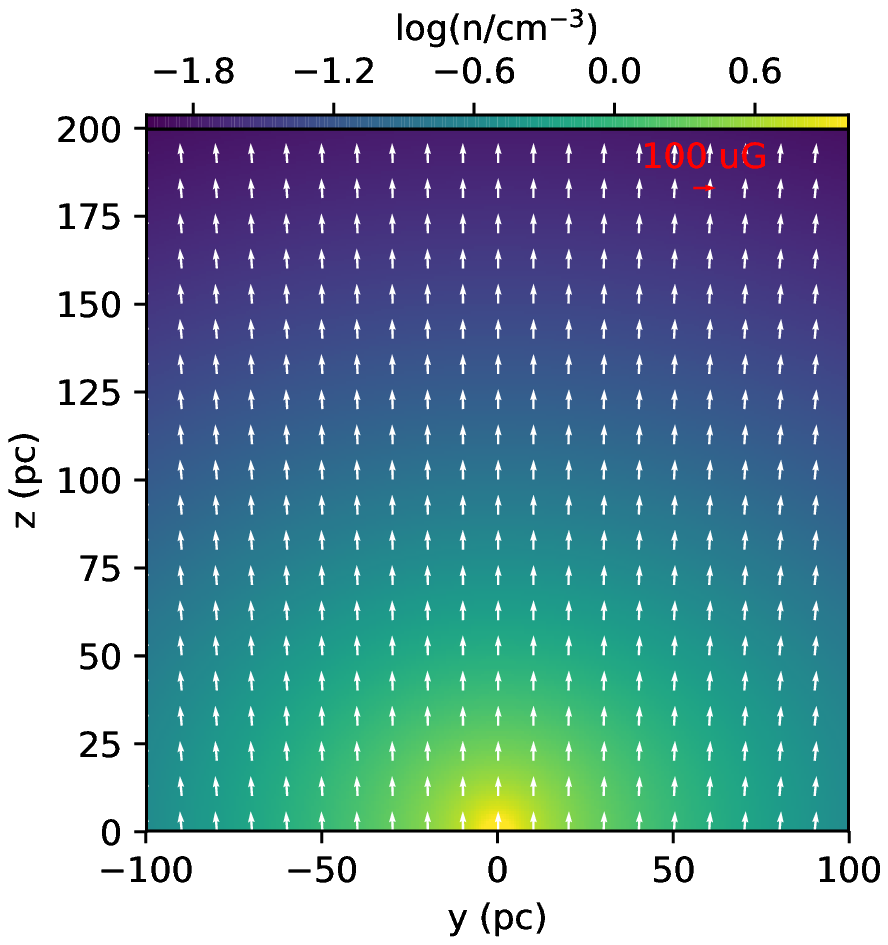}
    \caption{Initial distribution of gas density, plotted in logarithmic scale and in units of cm$^{-3}$. The white arrows indicate the initial magnetic field distribution.
    The three panels are slices through the $z=0$, $y=0$ and $x=0$ planes, respectively.
    %The x-z and y-z panels are slices through the center of the box along each axis, while the x-y panel show the slice at z = 0. The background is the density distribution with a unit of log(cm$^{-3}$), and the white arrows indicate the magnetic field.
    }
\label{fig:density}
\end{figure*}

At the beginning of the simulation, the interstellar medium (ISM) is assumed to be isothermal and in hydrostatic equilibrium with the gravitational potential,
\begin{equation}
      \dfrac{\nabla P}{\rho} = - \nabla \Phi,
\label{eqn:balance}
\end{equation}
where $P=n_tkT$ is the thermal pressure, and $n_t$ is the total number density of gas particles including protons, electrons and heavy elements.
As usual we define $\rho = \mu m_p n_t$, where $m_p$ is the proton mass and $\mu \approx 0.6$ is the mean molecular weight for solar abundance.
The initial temperature is set to be $10^6$~K, which is roughly the virial temperature given the enclosed gravitational mass of $1 \times10^9\rm~M_\odot$ within 100 pc.
 The prevalence of hot gas (with temperatures $\gtrsim10^6$ K) in the Galactic center has been established observationally (e.g., \citealp{2003ApJ...591..891B, Ponti2015}).
While cooler gas (with temperatures $\lesssim 10^4$ K) is also known to exist in the Galactic center, it tends to concentrate in dense filaments and clouds near the midplane and is not expected to play a significant role in the bubble formation. We discuss possible effects of a multi-phase ISM on the observed properties of the bubble in Section~\ref{subsec:caveat}.
The initial density distribution can then be derived by solving Eqn.~\ref{eqn:balance}, as shown in Figure~\ref{fig:density} along with the initial magnetic field distribution.
From the adopted initial conditions, it can be shown that the thermal pressure of the ISM ($n_tkT \sim 10^{-12}-10^{-10}\rm~dyn~cm^{-2}$) is everywhere significantly lower than the magnetic pressure ($B_0^2/8\pi \sim 2.5\times10^{-10}\rm~dyn~cm^{-2}$), perhaps except in the innermost few parsecs. In the meantime, the Alfv{\'e}n speed, $V_{\rm A}=(B_0^2/4\pi\rho)^{\frac{1}{2}}  \lesssim 10^{3}\rm~km~s^{-1}$, is much lower than the typical expansion velocity of the SN. Therefore, the present case of the Galactic center satisfies the {\it moderately strong field} condition defined by \citet{1991MNRAS.252...82I}.

%By solving Equation~\ref{eqn:balance}, we can obtain the density distribution:
%\begin{equation}
%  \begin{split}
%      n(x,y,z) = n_0(R_{c,c}^2+x^2+y^2+z^2)^{-\tfrac{v_{0,c}^2\mu m_p}{2kT}} \times \\
%      (R_{c,d}^2+x^2+y^2+\dfrac{z^2}{c^2})^{-\tfrac{v_{0,d}^2\mu m_p}{2kT}},
%  \end{split}
%\label{eqn:distribution}
%\end{equation}

%in which $n_0$ is a constant, and the subscripts `c' and `d' represent the NSC and the ND, respectively.
%$R_{cc}$ the core radius of the NSC, $R_{cd}$ the core radius of the disk, $v_{0c}$ the stable velocity of the cluster, $v_{0d}$ the stable velocity of the disk.
%In such a distribution, the magnitude of the temperature is assumed to be $1\times10^6$ K, or the density gradient will become unphysical.
% By substituting the values of Table 1 of \citet{2008ApJ...675.1278S} into Equation~\ref{eqn:distribution} and using the unit setting in the simulation, we set $r^2=x^2+y^2+z^2$ and simplify the equation as
% \begin{equation}
%   \begin{split}
%       n(x,y,z) = 10^7(4+r^2)^{-0.39} \times \\ (8100+x^2+y^2+2z^2)^{-1.43}.
%   \end{split}
% \label{eqn:distribution}
% \end{equation}

\subsection{Supernova Input}
\label{subsec:SN}
In the simulations, SNe are set to explode within a predefined cylindrical volume.
The cylinder has a diameter of 50 pc in the $x-y$ plane and a thickness of 10 pc along the $z$-axis, to mimic the concentration of massive stars near the Galactic plane \citep{2015MNRAS.447.1059K}.
We have tested a wider explosion area in the $x-y$ plane (e.g., 100 pc in diameter, closer to that of the CMZ), finding that the
resultant bubble would become significantly fatter, inconsistent with the observed morphology.
In reality, the CMZ may provide a horizontal confinement to the bubble. However, a self-consistent implementation of the CMZ would necessarily introduce more free parameters, and is beyond the scope of the present work.
The base of the radio bubbles shows a small but appreciable offset to the west of Sgr A* \citep{2019Natur.573..235H}.
Thus we place the center of the cylinder at $x = 5$ pc to mimic this behavior.
Due to the otherwise axisymmetry in the simulation, this appears to be the most viable way to reproduce the observed offset.

%In Milky Way, there is at least one supernova explosion per century \citep{Tammann1994}, while \citet{Adams2013} suggest a total Galactic supernova rate of 4.6$^{+7.4}_{-2.7}$ per century.
The fiducial SN birth rate is set to be $1\rm~kyr^{-1}$ \citep{2018ApJ...855...33D}, which is estimated by assuming an SFR of 0.1 M$_{\odot}$ yr$^{-1}$, a \citet{Kroupa2001} initial mass function (IMF) and a minimum mass of 8 M$_{\odot}$ for the progenitor star of a core-collapse SN.
\citet{Barnes2017} and \citet{2020MNRAS.497.5024S} estimated a current SFR of 0.1 M$_{\odot}$ yr$^{-1}$ inside the CMZ, while \citet{NoguerasLara2019} found that star formation in the ND (which has a similar radial extent as the CMZ) has been relatively active in the past 30 Myr, with an SFR of $0.2-0.8\rm~M_{\odot}~yr^{-1}$.
Our assumed SFR of 0.1 M$_{\odot}$ yr$^{-1}$ is compatible with the smaller radial extent of our adopted exploding region, which may be the case if SN events have been episodic and clustering on a $\lesssim$ Myr timescale.
We also test the effect of a lower SN birth rate of $0.5\rm~kyr^{-1}$ (see below).
We have neglected Type Ia SNe, which have a birth rate of $\lesssim0.05\rm~kyr^{-1}$ according to the enclosed stellar mass in the ND/NSC \citep{2005A&A...433..807M}, though a recent study by \citet{2021ApJ...908...31Z} found evidence that Sgr A East, one of the few currently known SNRs in the Galactic center, was created by a Type Iax SN.

Individual SNe are thus injected at random positions inside the cylindrical volume, one after another with a fixed interval according to the assumed birth rate.
Each SN has an ejecta mass of $M_{\rm ej}=10 \rm~M_{\odot}$ and a kinetic energy of $E_{\rm ej}=1\times$ 10$^{51}$ erg \citep{Poznanski2013}.
This energy is deposited into a sphere with a radius of $R_{\rm SN}=4$ pc, ignoring any intrinsic anisotropy.
The analytic solution within $R_{\rm SN}$ is derived from \citet{Truelove1999}, in which the newly born SN is divided into two parts, the inner uniform density core region and the outer power-law density envelope region.
The radius of the former is 10 times that of the latter, and the power-law index is set as zero.

% We decide to choose a lower energy input $4.8\times10^{52} erg$ and a dynamical age of 1.5 Myr.
% In the code, we need random numbers to simulate the random explosions, but we also want to make the results repeatable.
% Therefore, the random seed is chosen based on the simulation time instead of the real system time.
% The system time, i.e., the universal time, is always lapsing, so the obtained random numbers will be different in different simulations.
% However, the simulation time will not change with the universal time goes by, as long as the time step does not change.
% As a result, the random numbers are kept the same in different simulations.

\subsection{Simulation Runs and Synthetic Emission Maps}
\label{subsec:synthetic}
In this work, we perform four runs of simulation, each with a unique combination of magnetic field strength and SN explosion interval.
Our fiducial simulation is represented by run \textit{B80I1}, where \textit{B} and \textit{I} indicate the magnetic field and explosion interval, respectively.
The fiducial run has $B_0=80\rm~{\mu}G$ and $I=1$ kyr.
The other three runs have either one of the two parameters varied.
\textit{B50I1} has $B_0=50\rm~{\mu}G$ and \textit{B200I1} has $B_0=200\rm~{\mu}G$, covering the empirical lower and upper limits inferred for the Galactic center (Section~\ref{subsec:Bfield}).
Finally, \textit{B80I2} has an explosion interval of 2 kyr.
The total elapsed time is set to be 330 kyr for all four runs.
In the fiducial simulation, this is about the time when the top of the bubble approaches the edge of the simulation box.
The time step is adaptive and ranges between $1-40$ yr.
The simulation parameters are summarized in Table~\ref{table:parameters}.

To facilitate comparison with the observations, we generate synthetic radio and X-ray maps for the final snapshot (i.e., $t$ = 330 kyr) of the simulation.
We include synchrotron radiation and free-free emission in the radio band (default at 1284 MHz, to be consistent with the MeerKAT observation), while for the X-ray band only thermal emission from a collisionally-ionized, optically-thin plasma is considered.
%However, the simulation cannot provide all parameters used to calculate the radiation, so there are some assumptions.

First we need to distinguish regions inside and outside the evolving bubble.
This is realized by adding a tracer parameter, $Q$, evaluated at every pixel in the simulation, which obeys a simple conservation equation:
\begin{equation}
      \dfrac{\partial (\rho Q)}{\partial t} + \nabla \cdot (\rho Q \bm{v}) = 0.
\label{eqn:tracer}
\end{equation}
$Q$ has a value of 1 for pure SN ejecta and 0 for the unpolluted ISM, and a value between 0--1 for pixels with mixed ejecta and ISM.
%In one pixel, this parameter will increase largely with the injection of ejecta.
We further calculate the Mach number for every pixel.
%The synchrotron component is  restricted to pixels with a Mach number greater than 2, i.e., a shocked region.
The synthetic maps only take into account pixels with a non-zero tracer parameter or a Mach number greater than 2.
The latter condition is employed to ensure that pixels with a high Mach number but a zero tracer parameter, such as those at or immediately behind the shock, are included.
% In fact, some high-energy cosmic rays in ejecta will escape beyond the shock front and mix with the ISM, which will improve the tracer value outside the bubbles, so we actually choose the regions with a tracer larger than $10^{-3}$ to calculate emission.

Synchrotron emissivity depends on the magnetic field strength and the density of relativistic electrons. However, the latter cannot be directly obtained from our simulation and thus requires some working assumption.
Here, we assume that the relativistic electron density at a given pixel of interest is proportional to the local gas density \citep{Orlando2007, Zhang2017}, normalized to have a mean energy density of $0.1 \rm~eV~cm^{-3}$ across the bubble volume.
This is compatible with the estimated mean cosmic-ray energy density of $10 \rm~eV~cm^{-3}$ in the bubble \citep{2019Natur.573..235H} and the empirical fact that relativistic electrons account for $\sim 1\%$ of the total cosmic-ray energy density in the GeV band \citep{Blasi2013}.
 %which is about ten times higher than that in the solar neighborhood.
We calculate the synchrotron emissivity in each pixel and integrate along the light-of-sight (i.e., the $y$-axis) to derive the synchrotron intensity map. In this calculation the $y$-component of the magnetic field is neglected due to the nature of synchrotron radiation.
%Synchrotron self-absorption is taken into account, but turns out to be negligible.
%in the estimation, but its optical depth is about $10^{-4}$ in this work, a negligible value.

Radio free-free emission is calculated following the standard formula of \citet{Longair2011}, which, at a give pixel, scales with density squared and is a function of temperature.
%In addition, we calculate the X-ray emission radiated by the $i$th pixel in the simulated box as $L_i=1.2 n_i^2 \Lambda(T_i, Z) V_i$, where $n_i$ is number density, $T_i$ is the temperature, $\Lambda$ is the volume emissivity as a function of $T_i$, abundance $Z$, and $V_i$ the physical volume of the pixel (same for all pixels).
 A temperature threshold of $10^4$ K is adopted when calculating the free-free emission.
We find that only a tiny fraction of all pixels in any of our simulations has a temperature below $10^5$ K.
The X-ray emissivity of an optically-thin thermal plasma in collisional ionization equilibrium \citep{2001ApJ...556L..91S}, also scaling with density squared, is extracted from \textit{ATOMDB}\footnote{http://www.atomdb.org}, version 3.0.9, for which we adopt a solar abundance.
%The regions with temperatures lower than 10$^4$ K are ignored when we calculate the X-ray emission.
The free-free and X-ray intensity maps are again derived by integrating along the $y$-axis.
We find that self-absorption is negligible in both the radio and X-ray bands, thanks to the relatively low column density involved.

\section{Results} \label{sec:res}
In this section we present the simulation results. We first describe the formation and subsequent evolution of the bubble in the fiducial run, showing that a good agreement on the overall morphology of the bubble is achieved between the simulation and observation  (Section \ref{subsec:formation}).
We then present the other three runs of simulations and examine the effect of varying magnetic field strength or SN birth rate on the bubble formation (Section~\ref{subsec:compset}).
Lastly, we confront the synthetic emission maps with the radio and X-ray observations (Section~\ref{subsec:compobs}).

\subsection{Bubble Formation and Evolution in the Fiducial Simulation}
\label{subsec:formation}

\begin{figure*}
    \centering
    \includegraphics[width=0.323\textwidth]{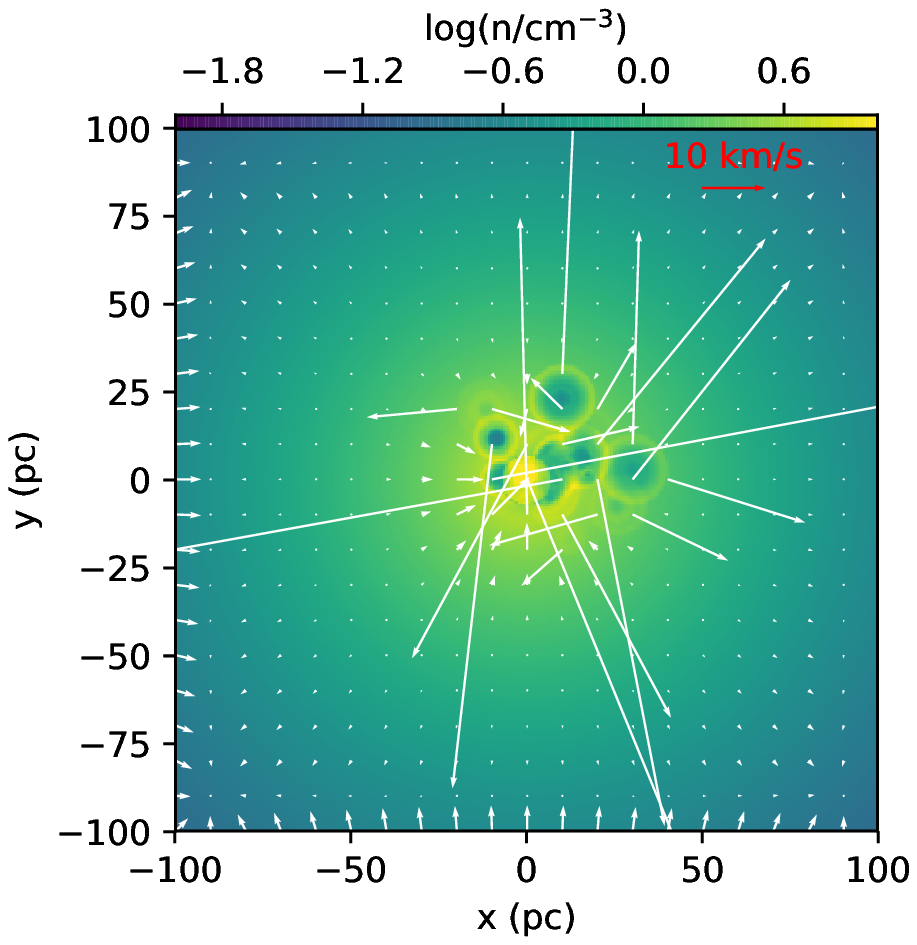}
    \includegraphics[width=0.323\textwidth]{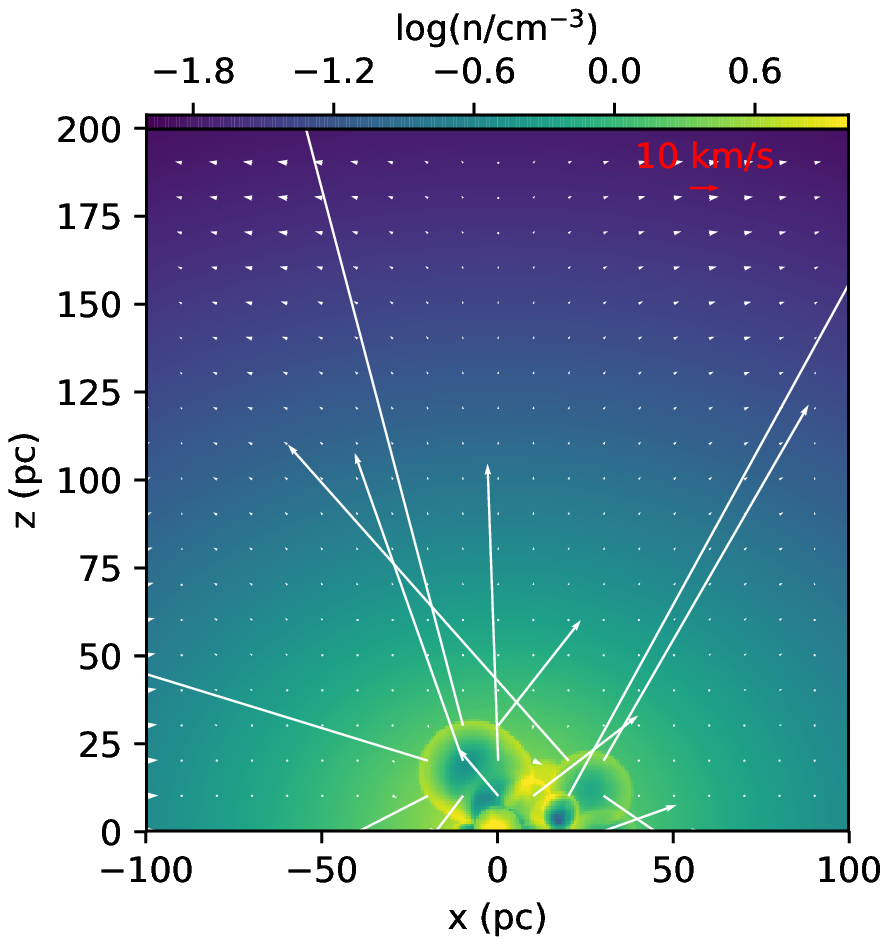}
    \includegraphics[width=0.323\textwidth]{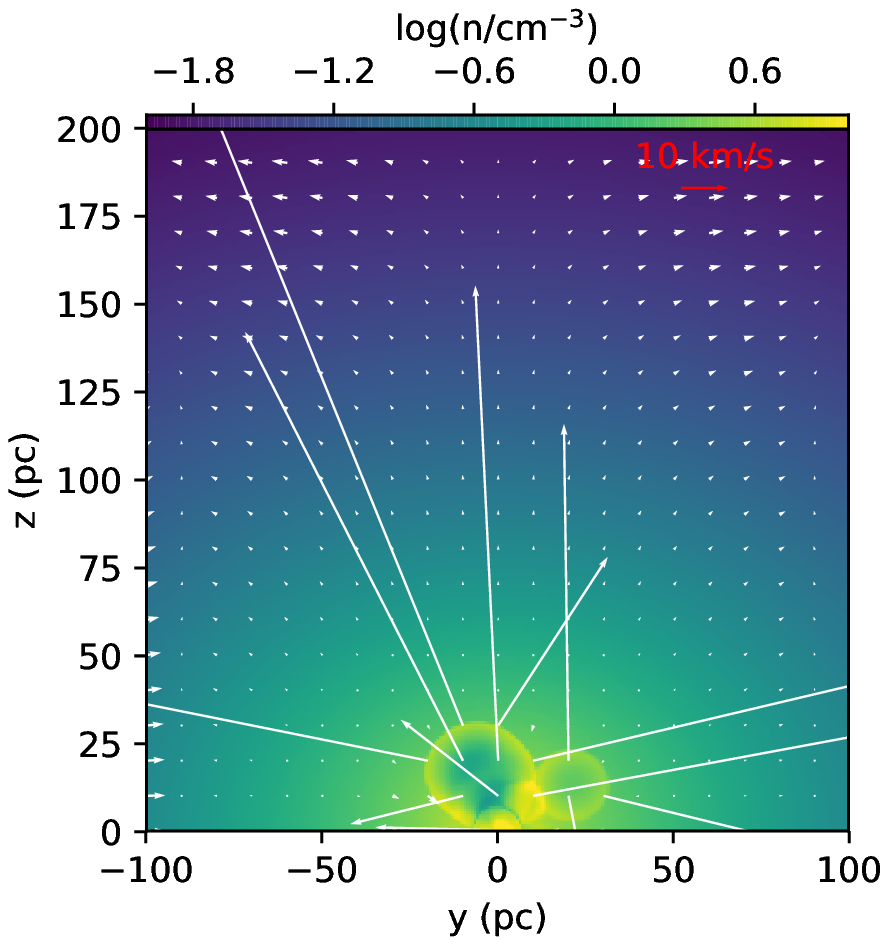}\newline
    \includegraphics[width=0.323\textwidth]{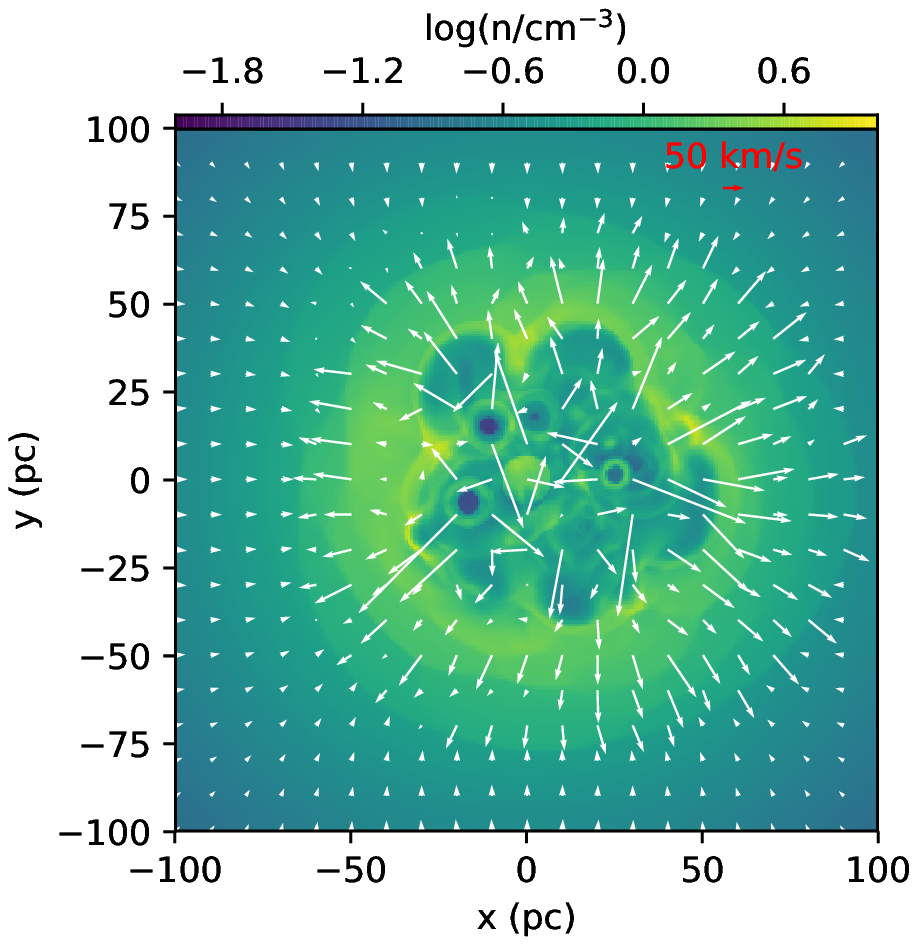}
    \includegraphics[width=0.323\textwidth]{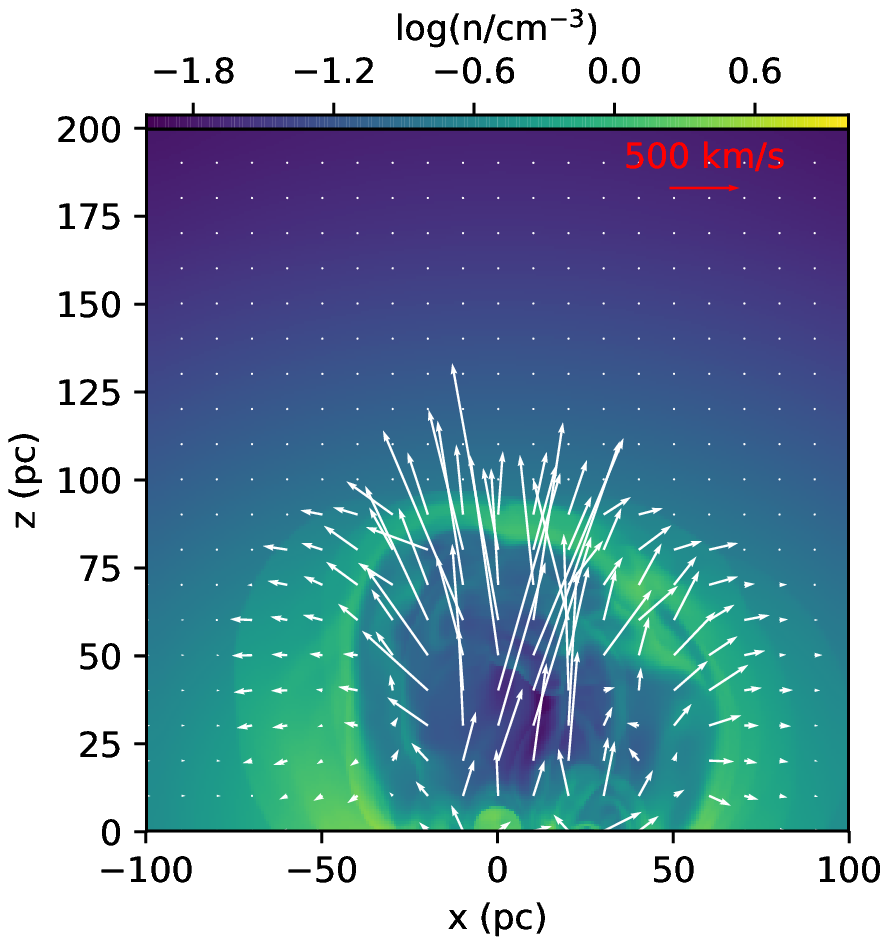}
    \includegraphics[width=0.323\textwidth]{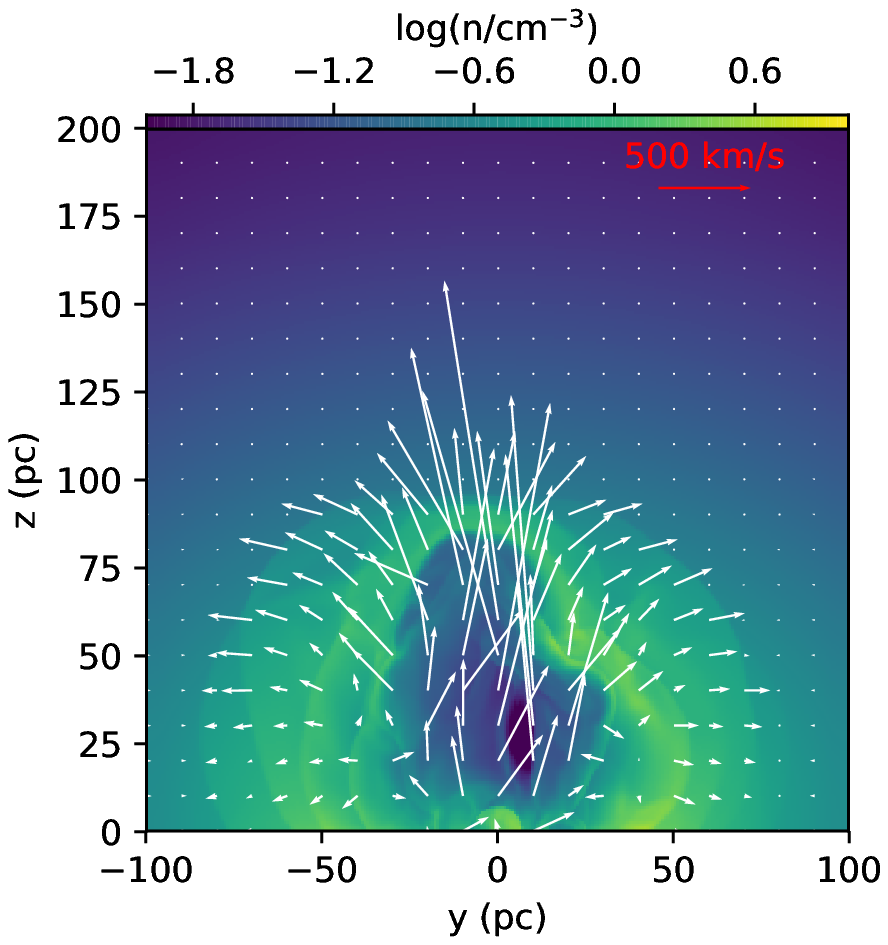}\newline
    \includegraphics[width=0.323\textwidth]{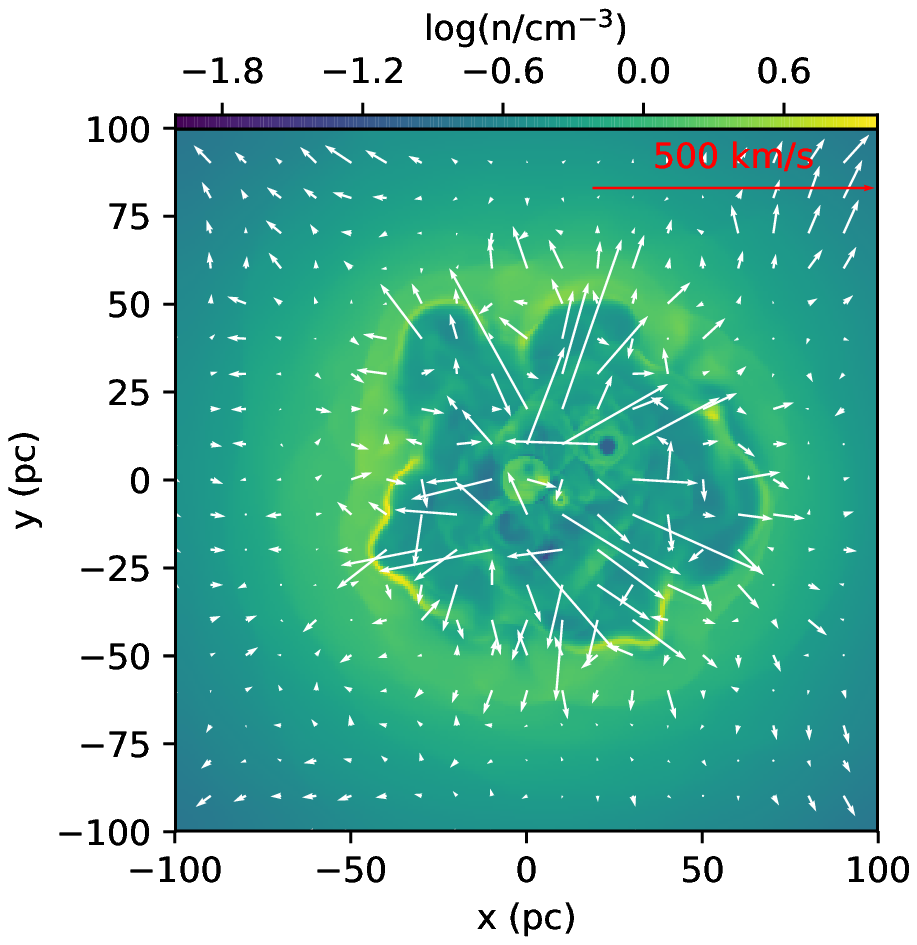}
    \includegraphics[width=0.323\textwidth]{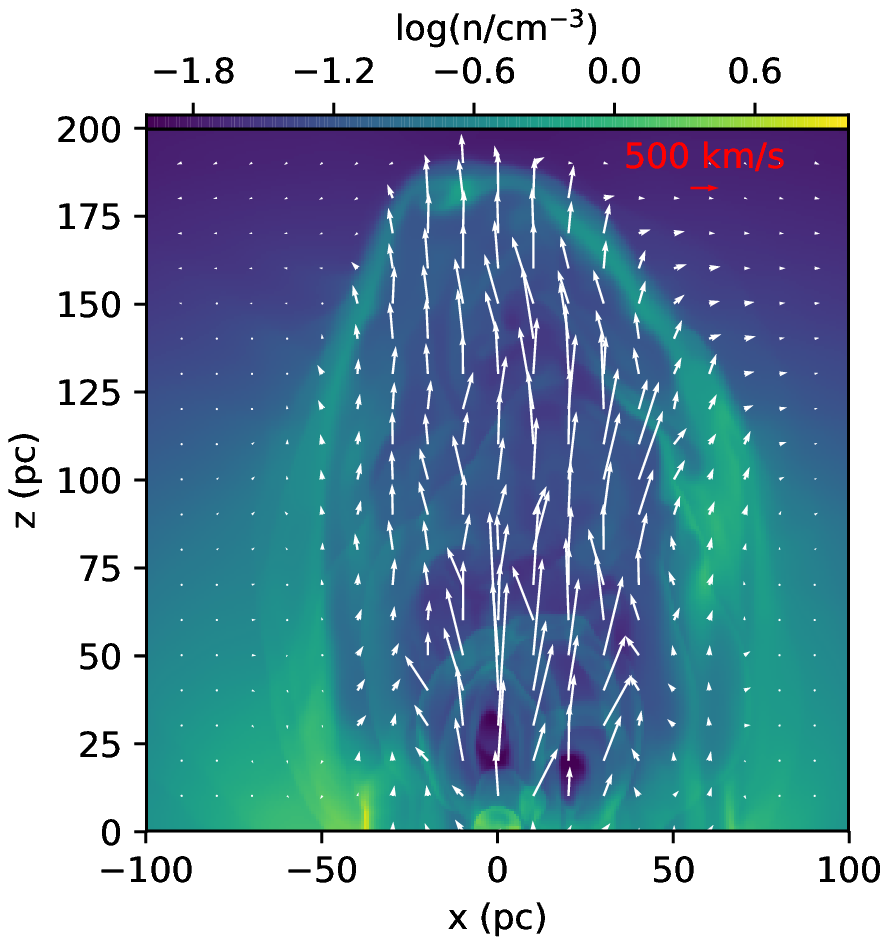}
    \includegraphics[width=0.323\textwidth]{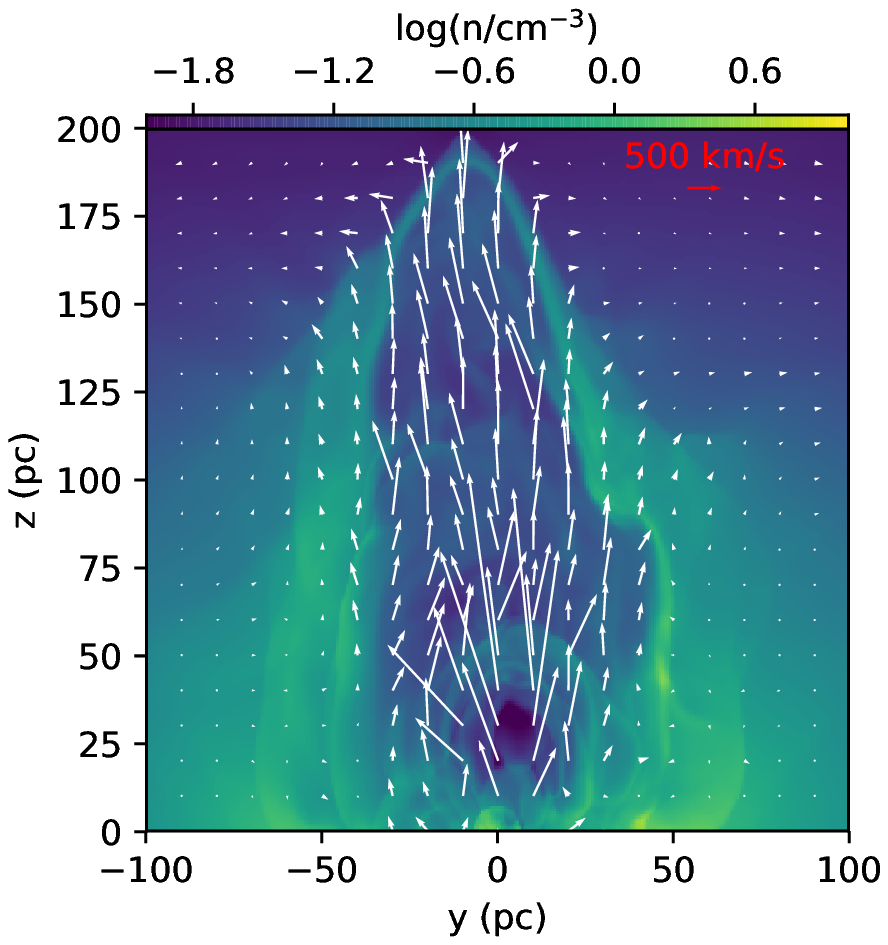}
    \caption{Density-velocity distributions after 30 (top row), 180 (middle row) and 330 (bottom row) kyr, for simulation \textit{B80I1}, i.e., with initial magnetic field strength of 80 $\mu$G and an explosion interval of 1 kyr.
    The gas density is plotted in logarithmic scale and in units of cm$^{-3}$.
    The white arrows indicate the velocity vector.
    The left, middle and right columns are slices through the $z=0$, $y=0$ and $x=0$ planes, respectively.
    %The x-z and y-z panels are slices through the center of the box along each axis, while the x-y panel show the slice at z = 0.
    %The background is the density distribution with a unit of log(cm$^{-3}$), and the white arrows indicate the velocity.
    }
\label{fig:rho}
\end{figure*}

\begin{figure*}
    \centering
    \includegraphics[width=0.323\textwidth]{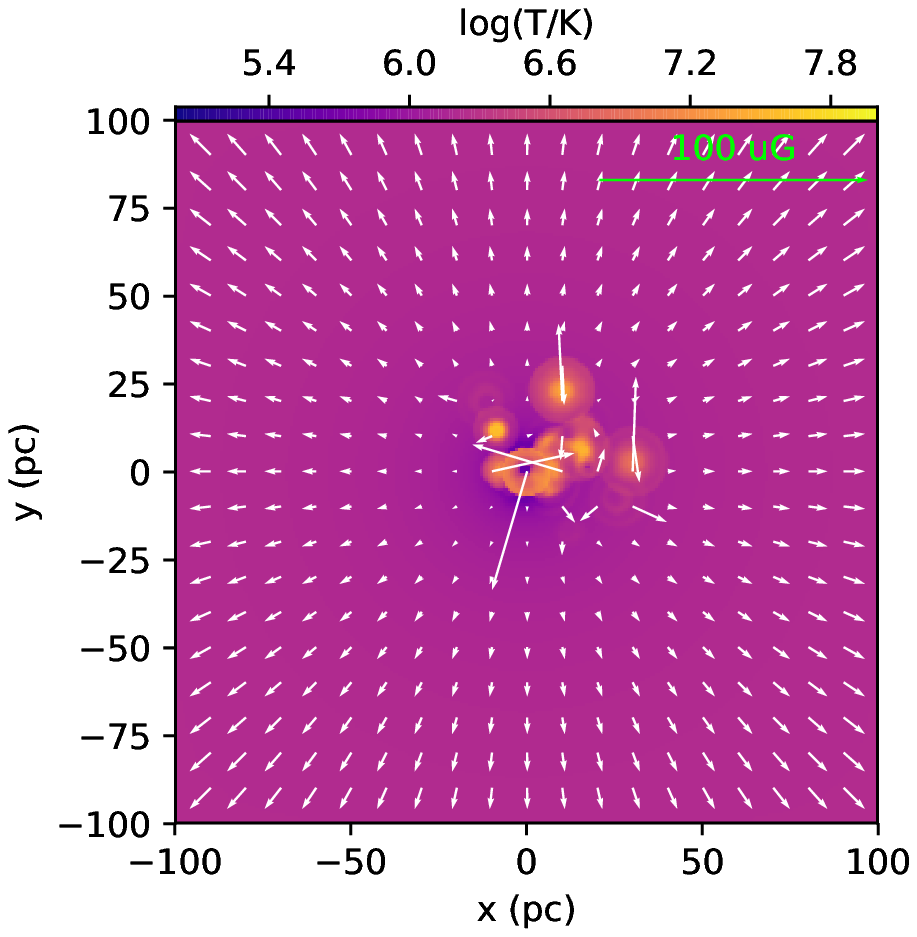}
    \includegraphics[width=0.323\textwidth]{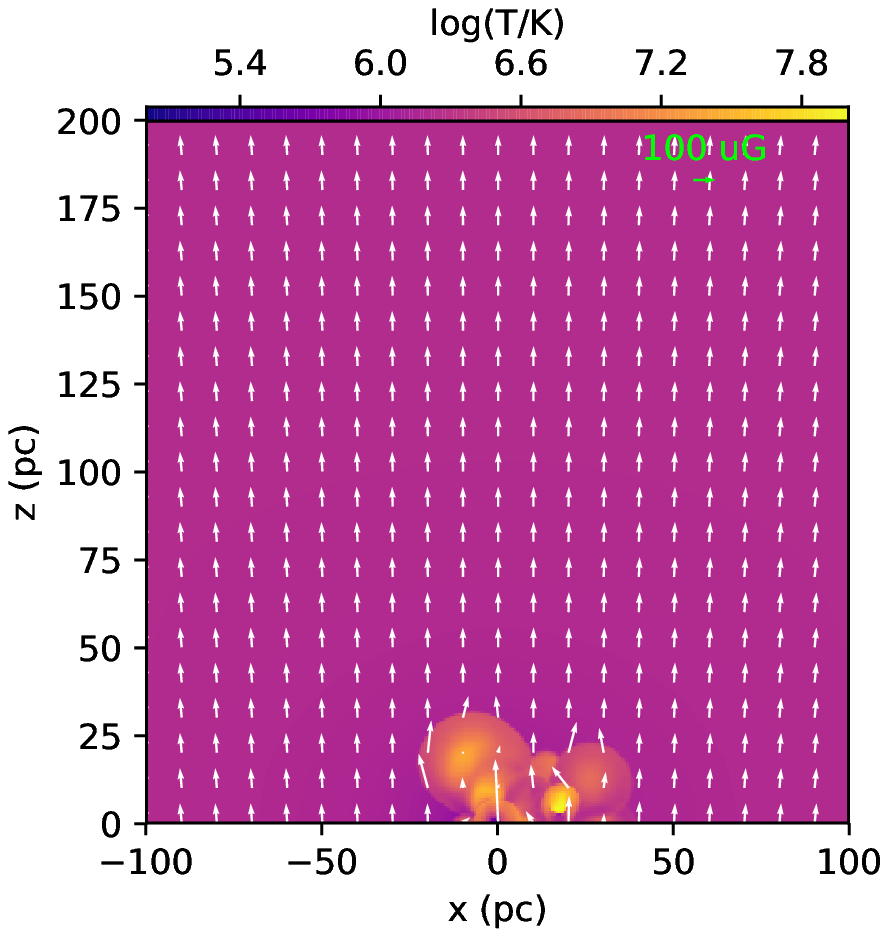}
    \includegraphics[width=0.323\textwidth]{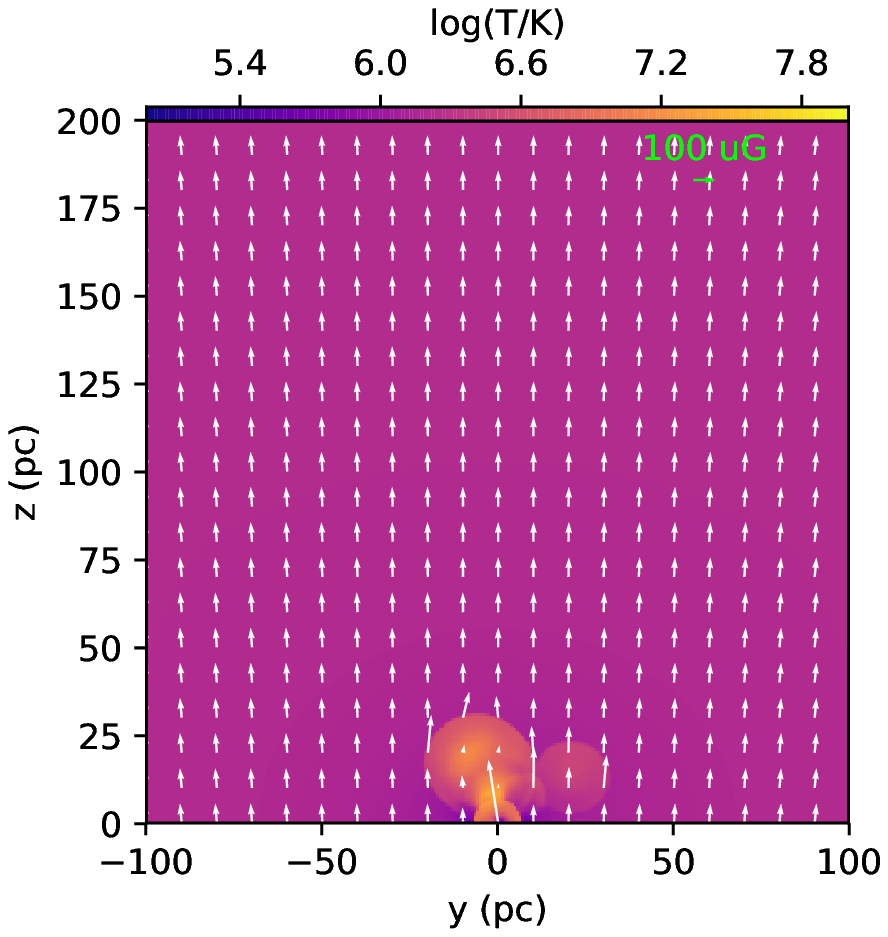}\newline
    \includegraphics[width=0.323\textwidth]{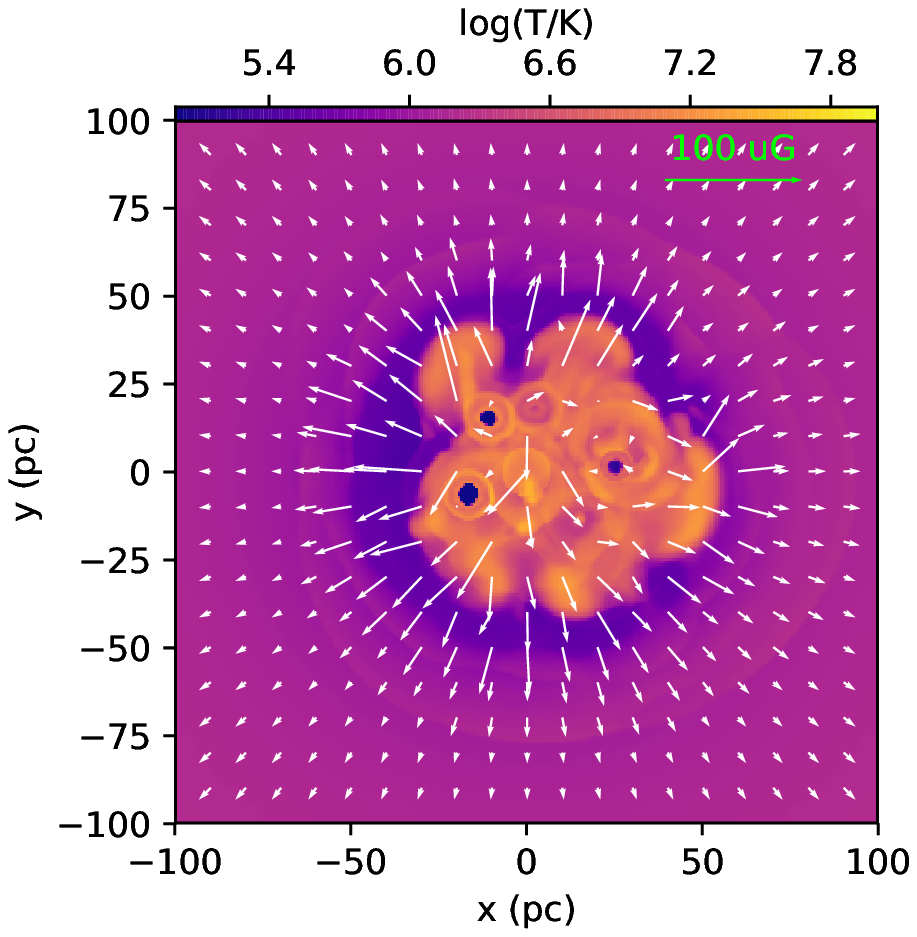}
    \includegraphics[width=0.323\textwidth]{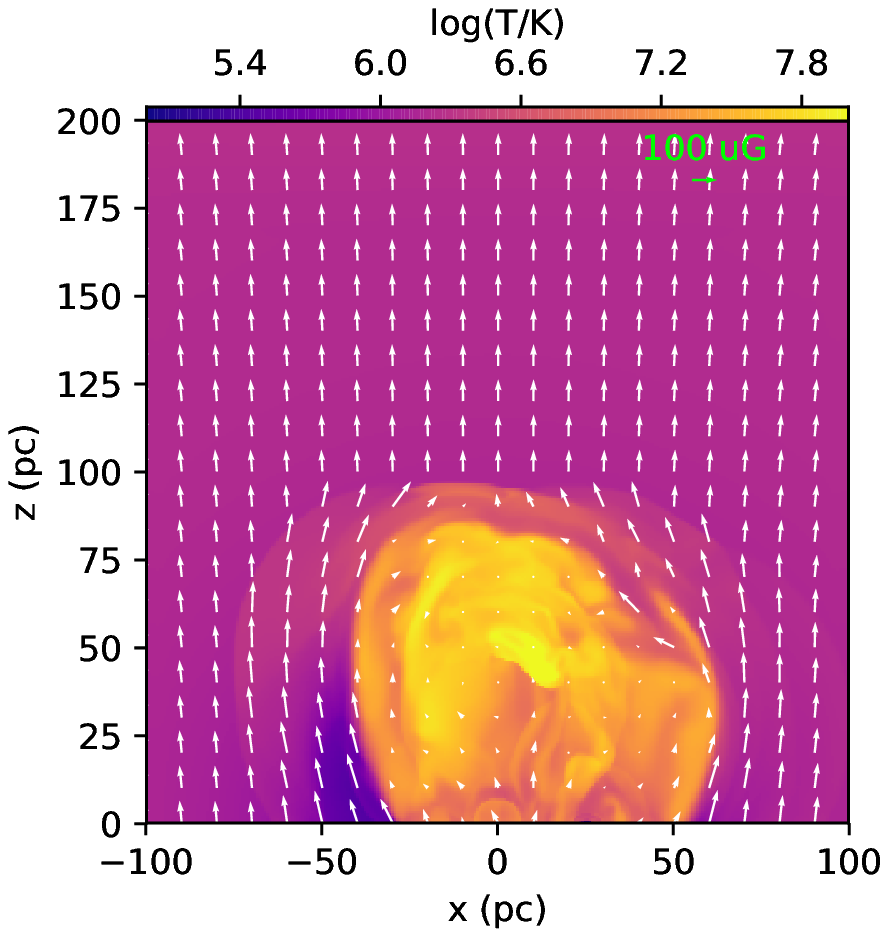}
    \includegraphics[width=0.323\textwidth]{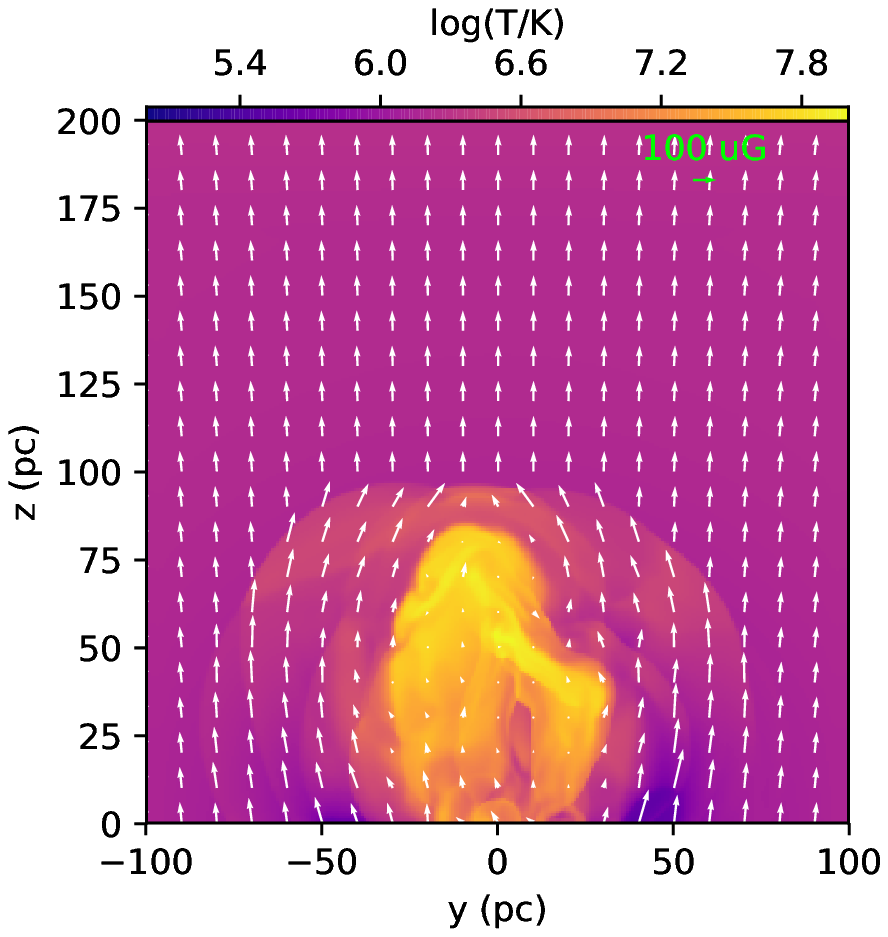}\newline
    \includegraphics[width=0.323\textwidth]{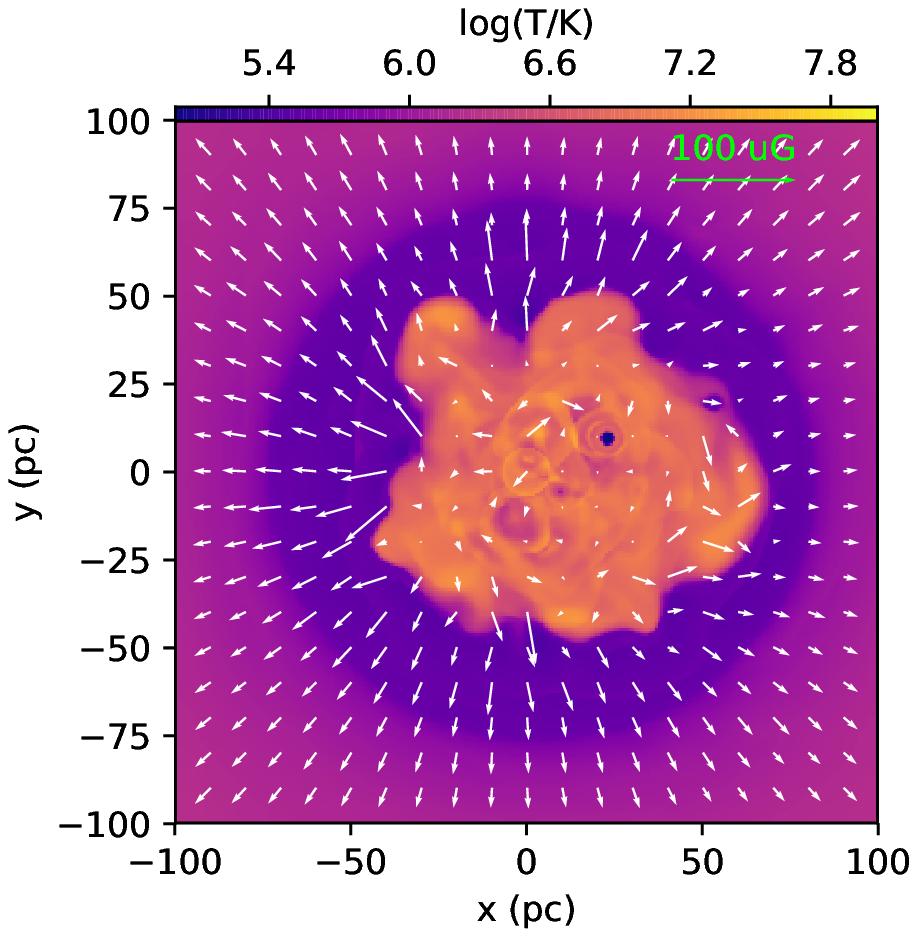}
    \includegraphics[width=0.323\textwidth]{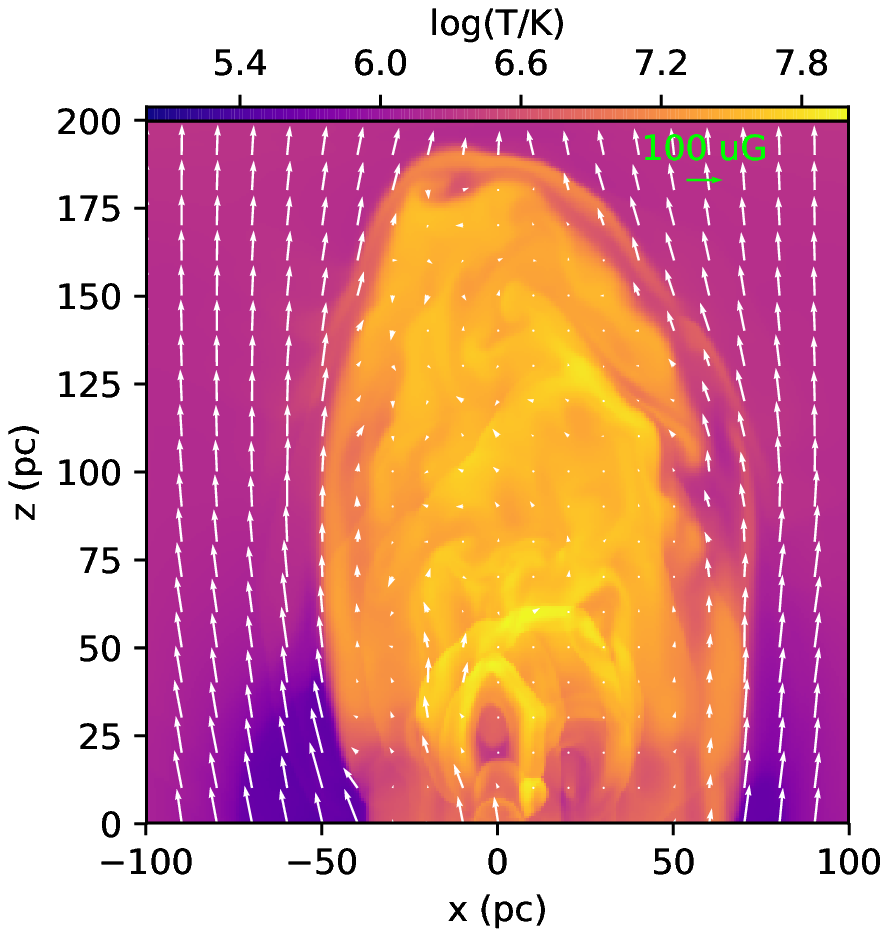}
    \includegraphics[width=0.323\textwidth]{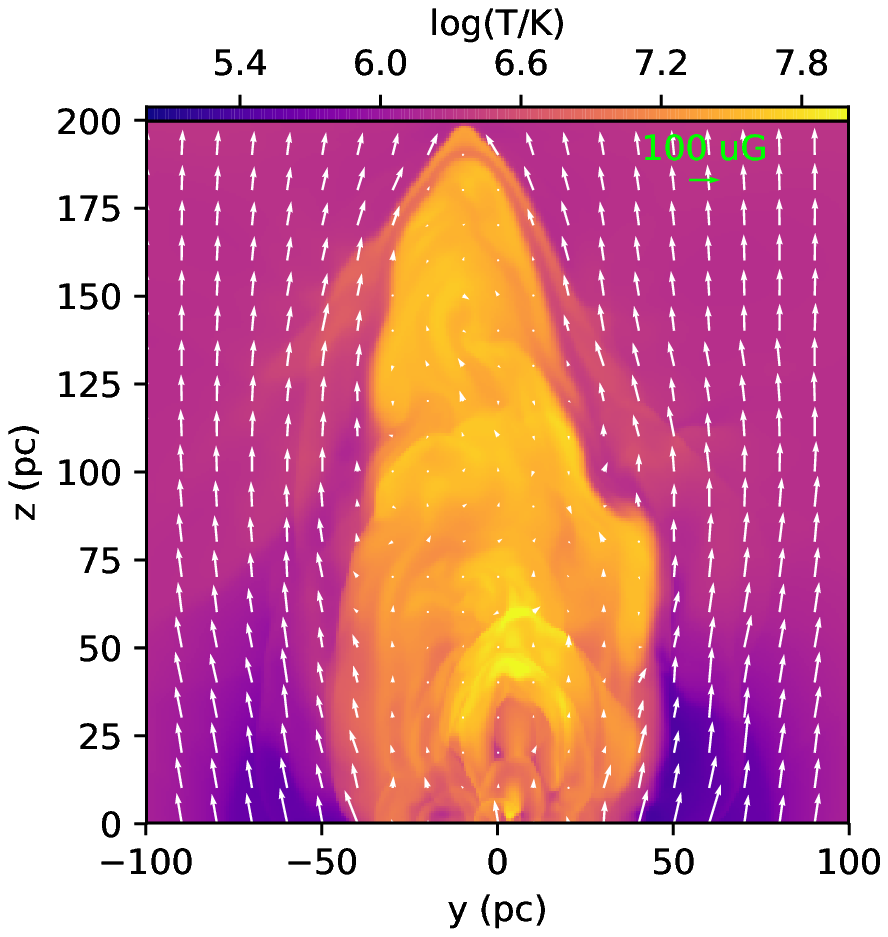}
    \caption{Temperature-magnetic field distributions after 30 (top row), 180 (middle row) and 330 (bottom row) kyr, for simulation \textit{B80I1},
    The gas temperature is plotted in logarithmic scale and in units of Kelvin.
    The white arrows indicate the magnetic field vector.
    The left, middle and right columns are slices through the $z=0$, $y=0$ and $x=0$ planes, respectively.
    }
\label{fig:T}
\end{figure*}

\begin{figure*}
    \centering
    \includegraphics[width=0.323\textwidth]{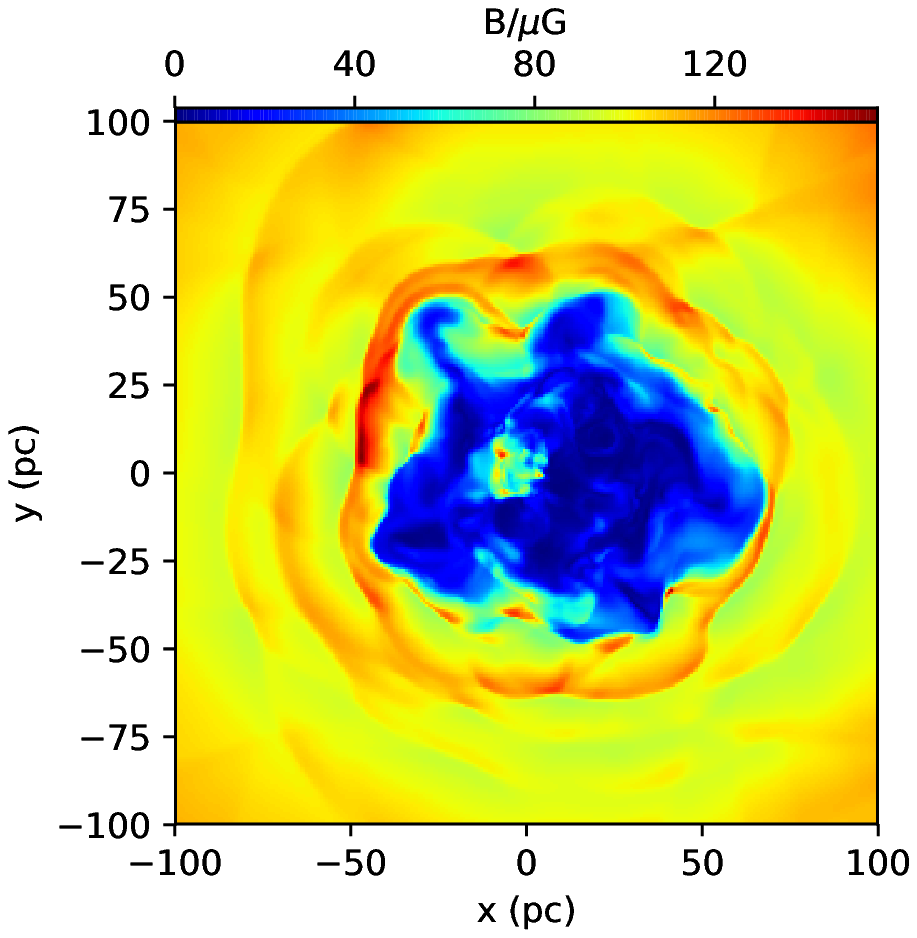}
    \includegraphics[width=0.323\textwidth]{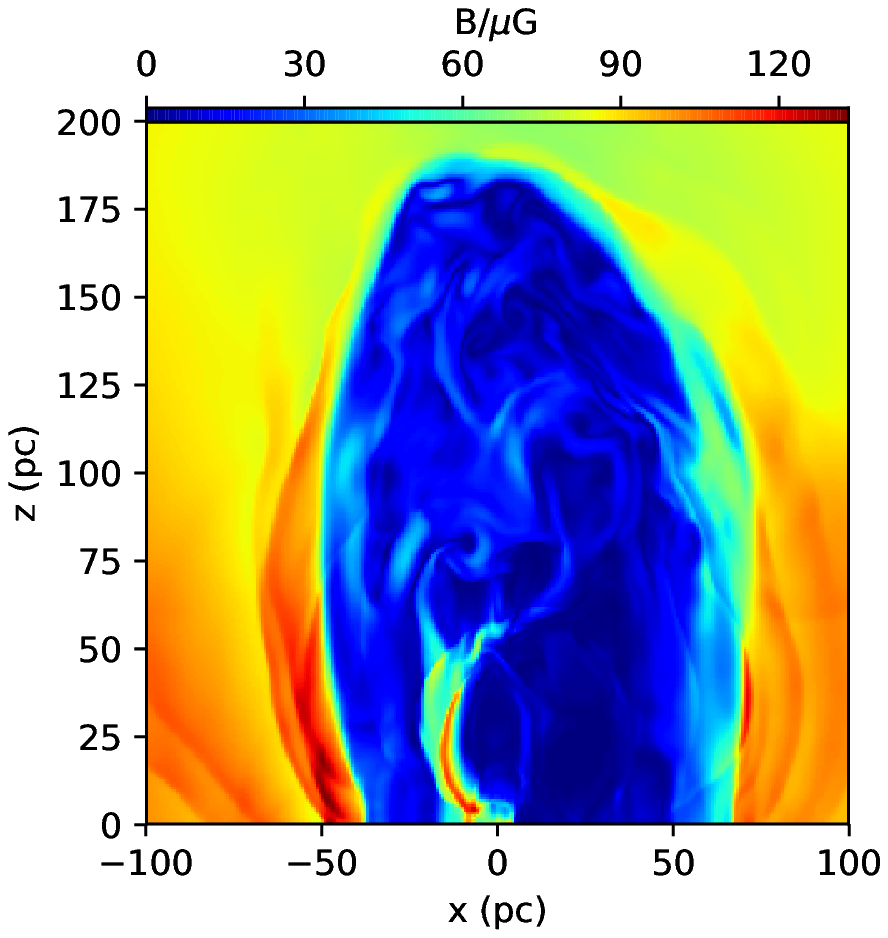}
    \includegraphics[width=0.323\textwidth]{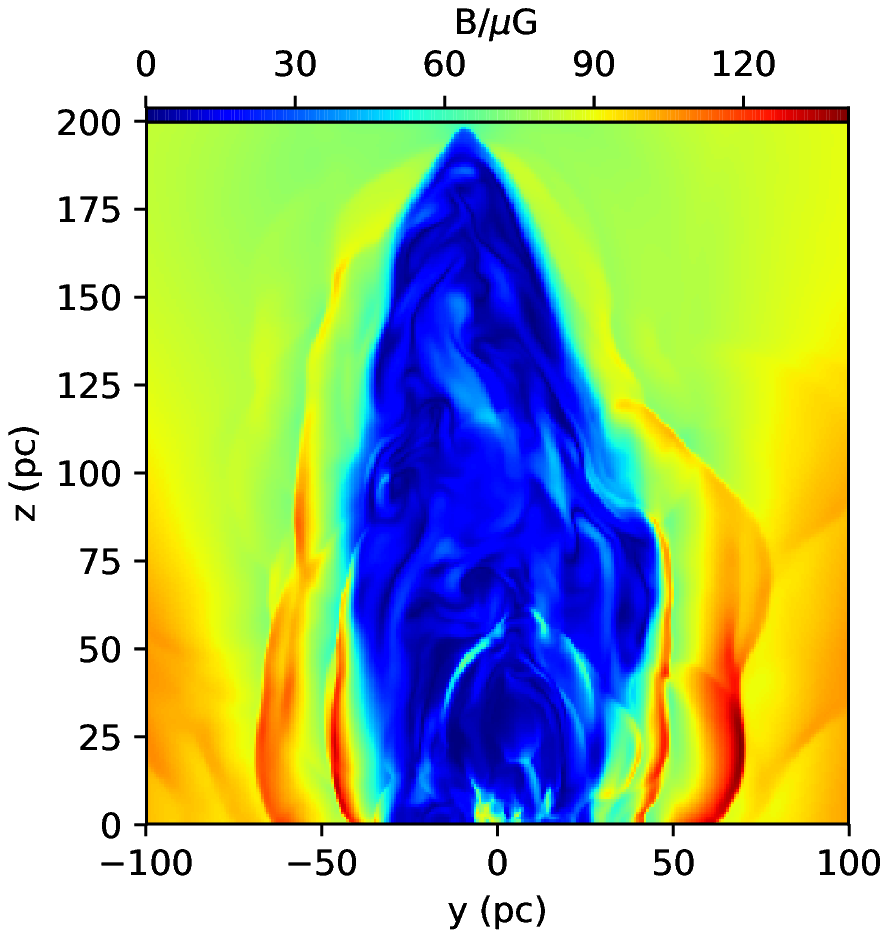}
    \caption{Magnetic strength distributions after 330 kyr for simulation \textit{B80I1},
    The left, middle and right columns are slices through the $z=0$, $y=0$ and $x=0$ planes, respectively.
    }
\label{fig:B}
\end{figure*}

%
%The model \textit{B80I1} will spend 330 kyr to reach the scale of the radio bubbles, and for simplicity, we show the results after 330 kyr for all models.
%\citet{2019Natur.573..235H} estimate a dynamical age of 7 Myr for the radio bubbles based on the observed dispersion velocity of $30 km\ s^{-1}$, but they assume that the  velocity of the bubble shells has been constant during the past 7 Myr since the progenitor events happened, which is impossible.
%The initial velocity of the progenitor events will decrease quickly in the Galactic center due to the dense ISM.
%Moreover, the real expansion velocity is usually larger than the dispersion velocity in this region.
%In short, the dynamical age of the bubbles should be much smaller than 7 Myr and 330 kyr is reasonable.

In Figures~\ref{fig:rho} and \ref{fig:T},
we show the gas density and temperature maps of run \textit{B80I1}.
In each figure, the density or temperature distribution is shown for a slice through the $z=0$ (left columns), $y=0$ (middle columns) and $x=0$ (right columns) plane,
after a simulation time of $t$ = 30 (top rows), 180 (middle rows) and 330 (bottom rows) kyr.

By design, 30 SNe have exploded by the time of 30 kyr.
The forward shock front of several youngest SNe are clearly revealed in the density map, as well as by
the overlaid projected velocity vectors.
A high-density region forms and persists around the origin ($x=y=z=0$), because of the steep gravitational potential even in the presumed absence of an SMBH.
%Moreover, the gravity also makes this region stable during the simulation, even if there is a supernova exploded in it.
As the shocks propagate, they compress and heat the ambient gas and also frequently collide with each other, eventually forming an expanding complex of post-shock gas with temperatures of $10^{7-8}$ K.
%This leads to dramatic increase of the gas temperature behind the shock fronts, as illustrated in \ref{fig:T}.

By the time of 180 kyr, this hot gas complex has developed into a bubble structure with a common dense shell, most clearly seen in the $x-z$ and $y-z$ planes.
Inside the bubble, the density is low as a result of expansion, while the temperature remains high due to repeated shock heating.
Numerous arc-like features are evident in the temperature map, especially in the $x-z$ and $y-z$ planes, which are the relic of individual SN shocks.
At this stage, the bubble looks fat, with a similar extent ($\sim100$ pc) along the three dimensions.
However, the overall expansion starts to show a preference along the vertical (positive $z$) direction, with the vertical expansion velocity of the shell now being $\sim690\rm~km~s^{-1}$, substantially larger than the average expansion velocity of $\sim120\rm~km~s^{-1}$ in the $x-y$ plane.
%It can also be seen that the velocity vectors of the bubble interior are preferentially pointed along the vertical (positive $z$) direction.
This is primarily due to the collimation effect by an ordered magnetic field \citep{1991MNRAS.252...82I,1992ApJ...389..297S,1995MNRAS.274.1157R,Wu2019}.
Specifically, the SN shocks tend to push the semi-vertical magnetic field to the sides, greatly suppressing the magnetic field inside the bubble and in the meantime amplifying the magnetic field near the bubble shell. In turn, the latter decelerates and even halts the horizontal expansion of the bubble. The vertical expansion, on the other hand, feels no such magnetic confinement, thus a high velocity along this direction remains. The relatively strong gravitational potential in the $x-y$ plane also contributes to retarding the horizontal expansion and facilitates the bubble collimation along the $z$-axis.

As a result, by the time of 330 kyr, the bubble becomes much more elongated. The top of the bubble almost reaches the edge of the simulation box ($z=200$ pc), with a vertical expansion velocity still as high as $\sim600\rm~km~s^{-1}$, whereas its horizontal extent has not grown significantly since $t$ = 180 yr.
The width of the bubble at its base is about 120 pc, with a small but appreciable offset towards the positive $x$-axis, both in agreement with the observed bubble.
Arc-like features tracing the sequential SN shocks remain prominent throughout the bubble interior.
Near some of these arcs, locally enhanced magnetic fields are evident, which is the result of shock compression, as illustrated in Figure~\ref{fig:B}.
%we can see the C-type shock driven by a hydromagnetic wave resulting from the expansion of the bubble, some magnetized filaments and the magnetic field dipping below the ambient field strength just outside the shock which is because the gravity beats the thermal pressure and is pulling materials into Galactic center.}
The magnetic field strength takes a highest value of 175 $\mu$G across the bubble.
Our simulation ends at this point.

\subsection{Comparison with Other Simulation Runs}
\label{subsec:compset}

\begin{figure*}
    \centering
    \includegraphics[width=0.323\textwidth]{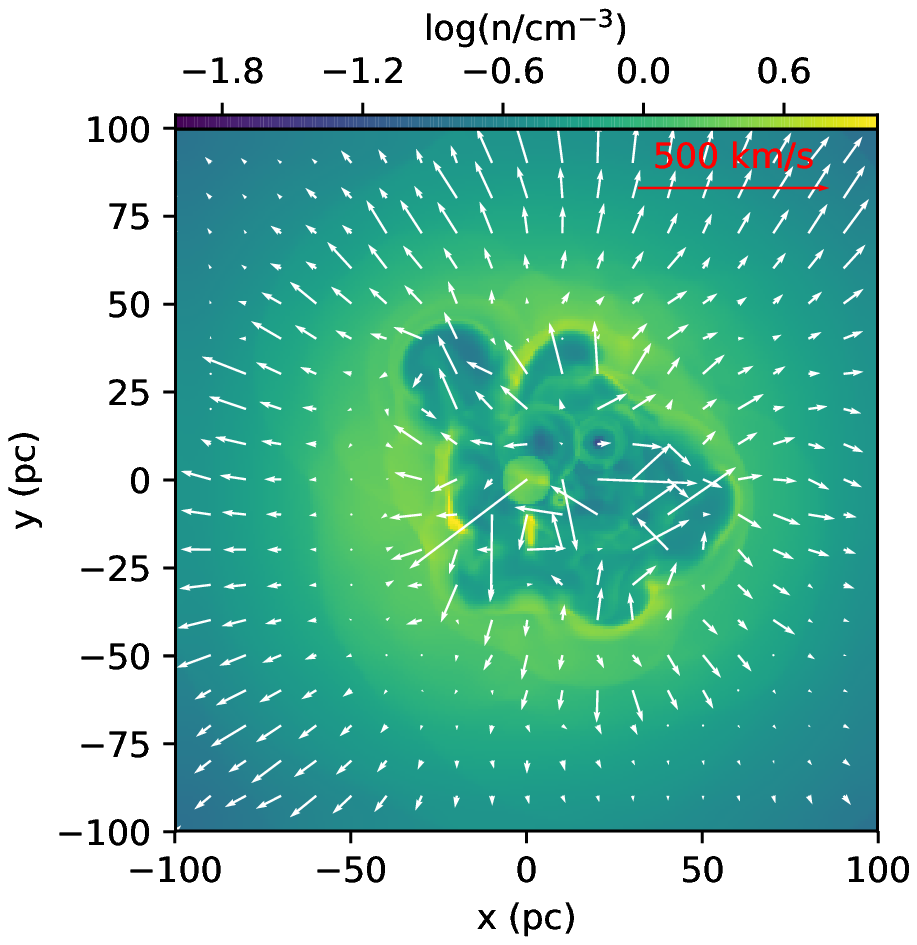}
    \includegraphics[width=0.323\textwidth]{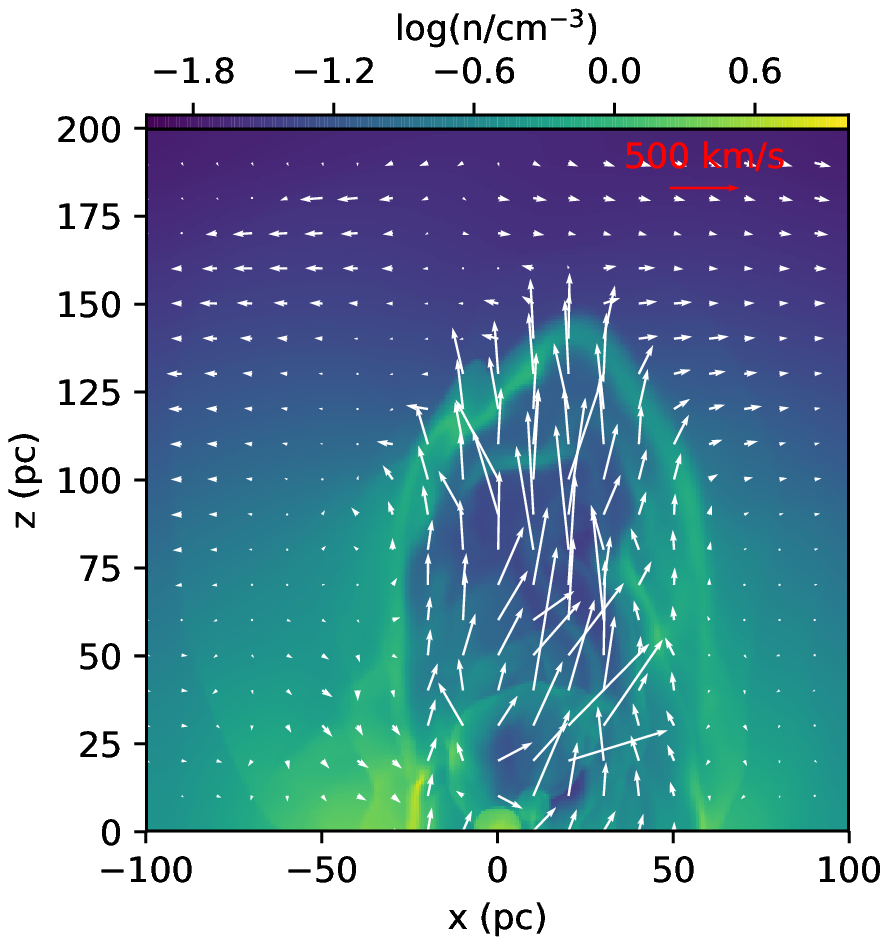}
    \includegraphics[width=0.323\textwidth]{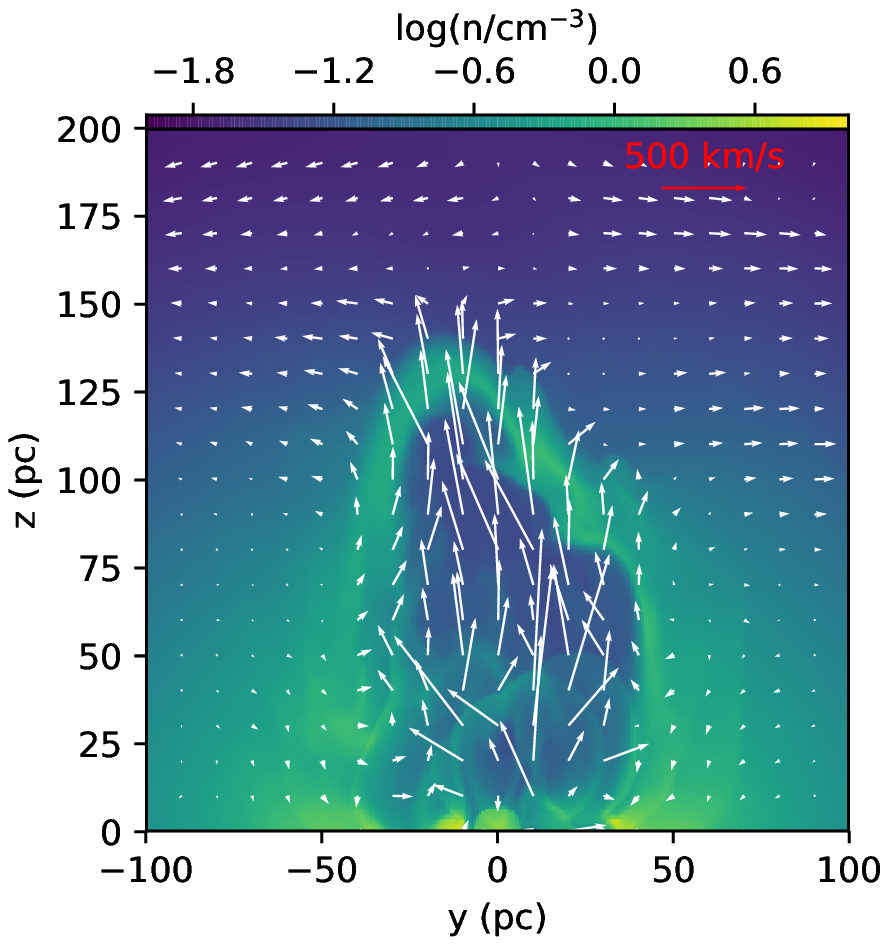}\newline
    \includegraphics[width=0.323\textwidth]{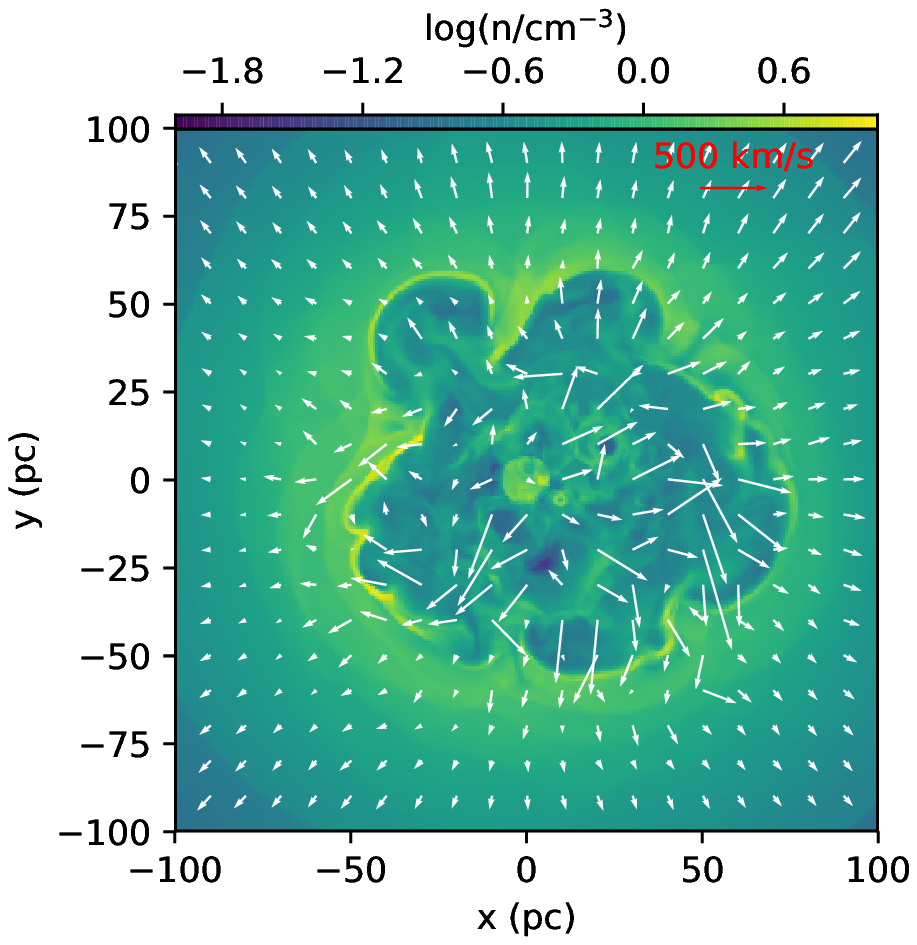}
    \includegraphics[width=0.323\textwidth]{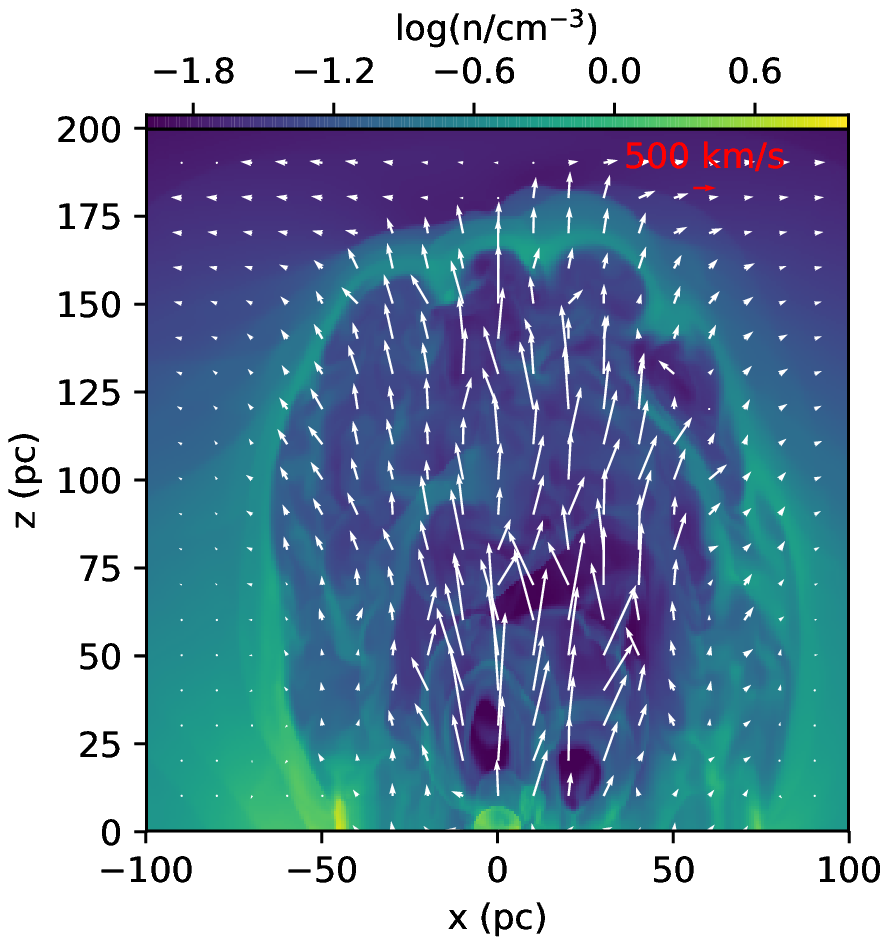}
    \includegraphics[width=0.323\textwidth]{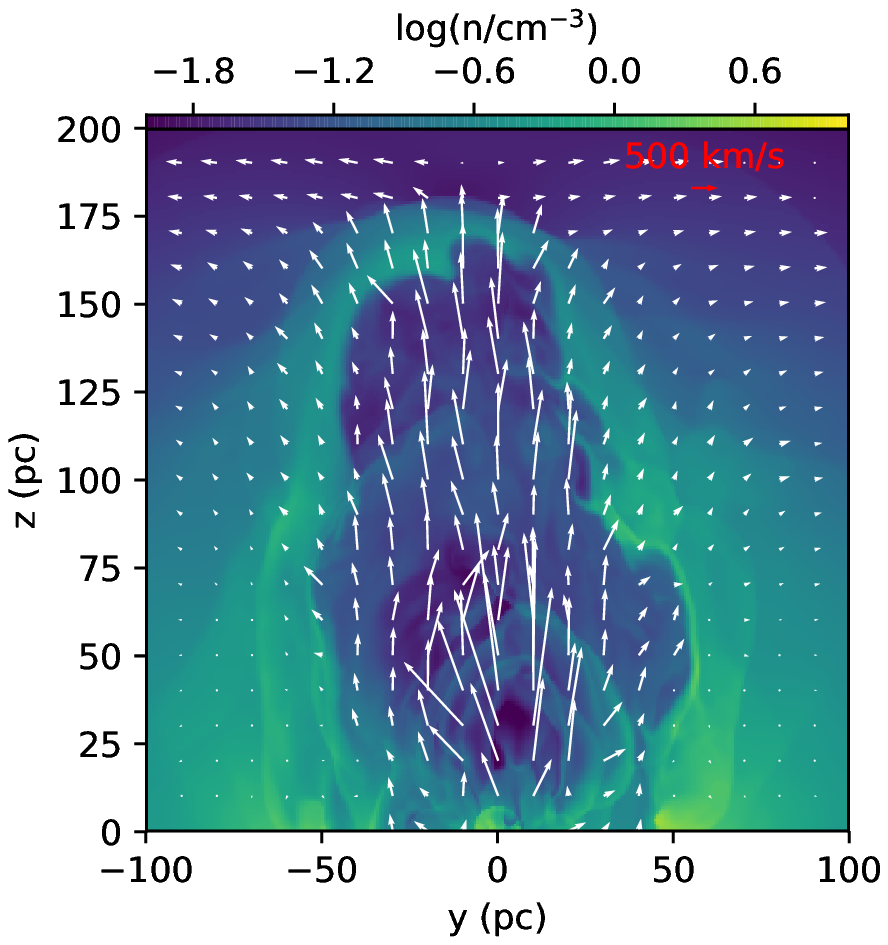}\newline
    \includegraphics[width=0.323\textwidth]{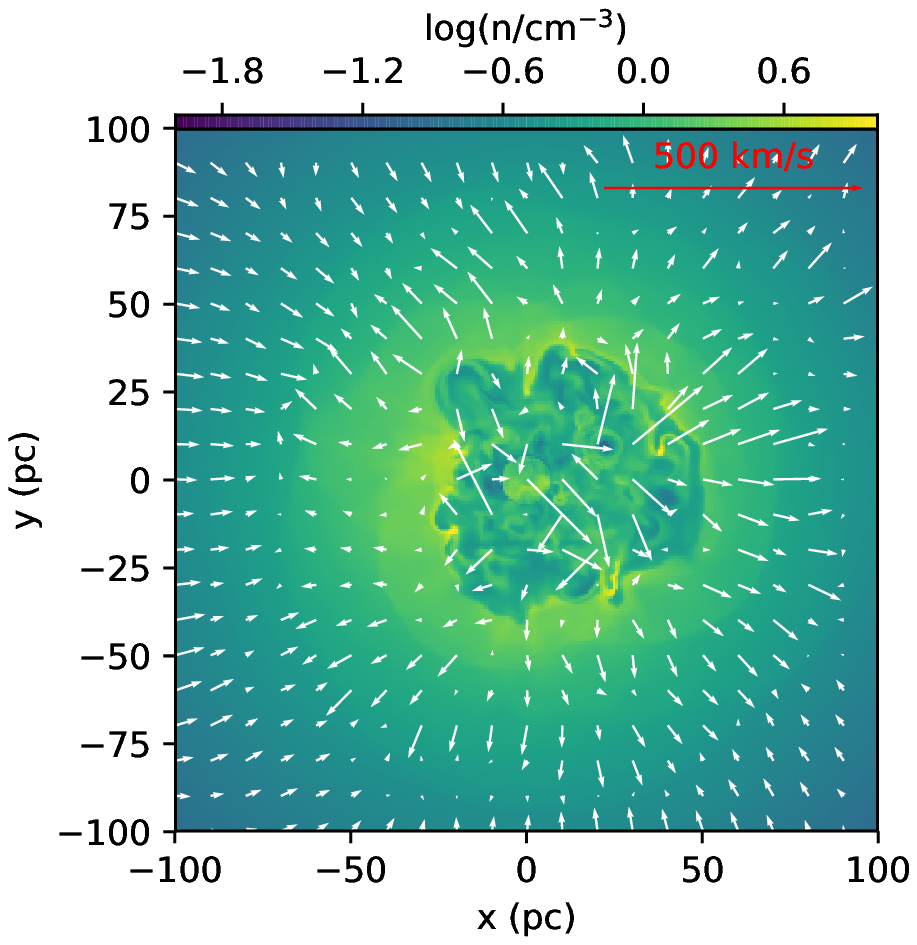}
    \includegraphics[width=0.323\textwidth]{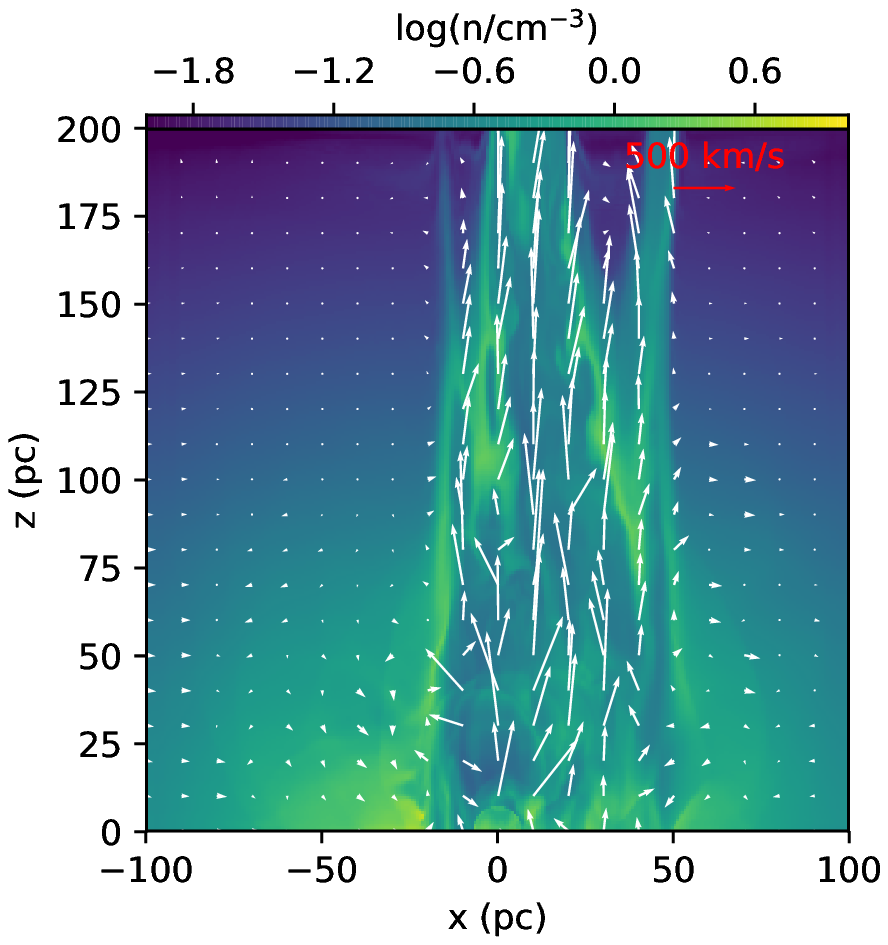}
    \includegraphics[width=0.323\textwidth]{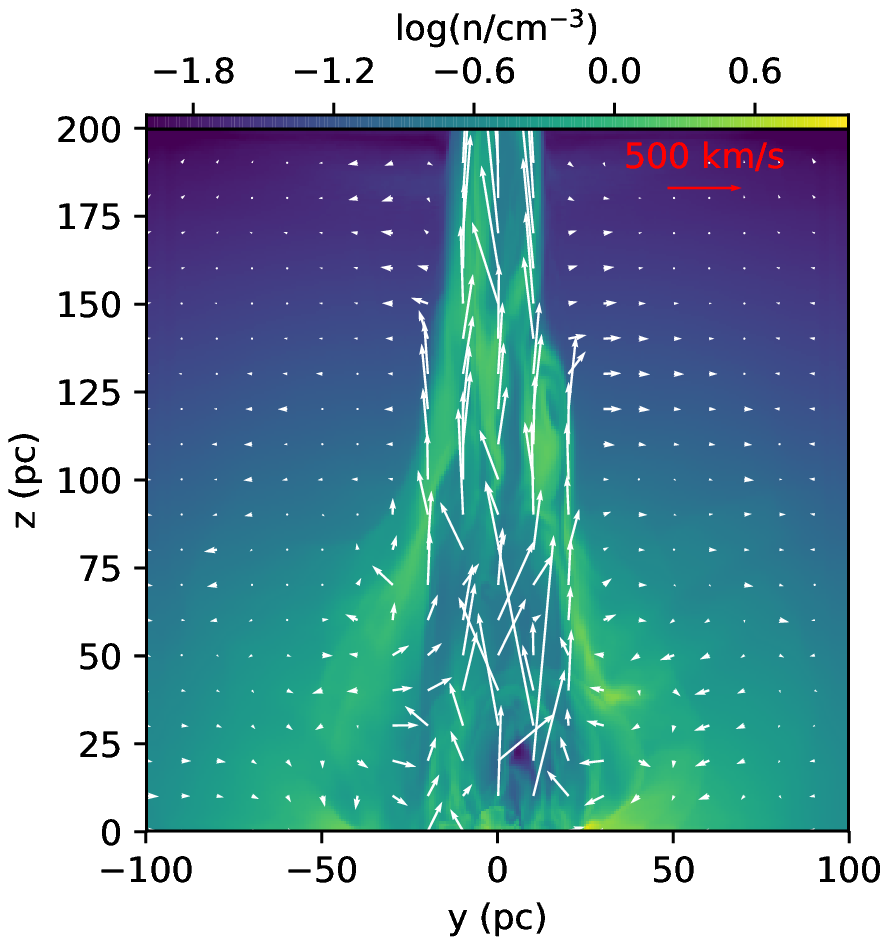}
    \caption{Simulated density-velocity images after 330 kyr. In the upper, middle and lower rows, we show the results of runs \textit{B80I2}, \textit{B50I1} and \textit{B200I1}, respectively. The $x-z$ and $y-z$ panels are slices through the center of the box along each axis, while the $x-y$ panel shows the slice at $z$ = 0. The background is the density distribution in logarithmic scale, and the white arrows indicate the velocity.}
\label{fig:rhoe}
\end{figure*}

\begin{figure*}
    \centering
    \includegraphics[width=0.323\textwidth]{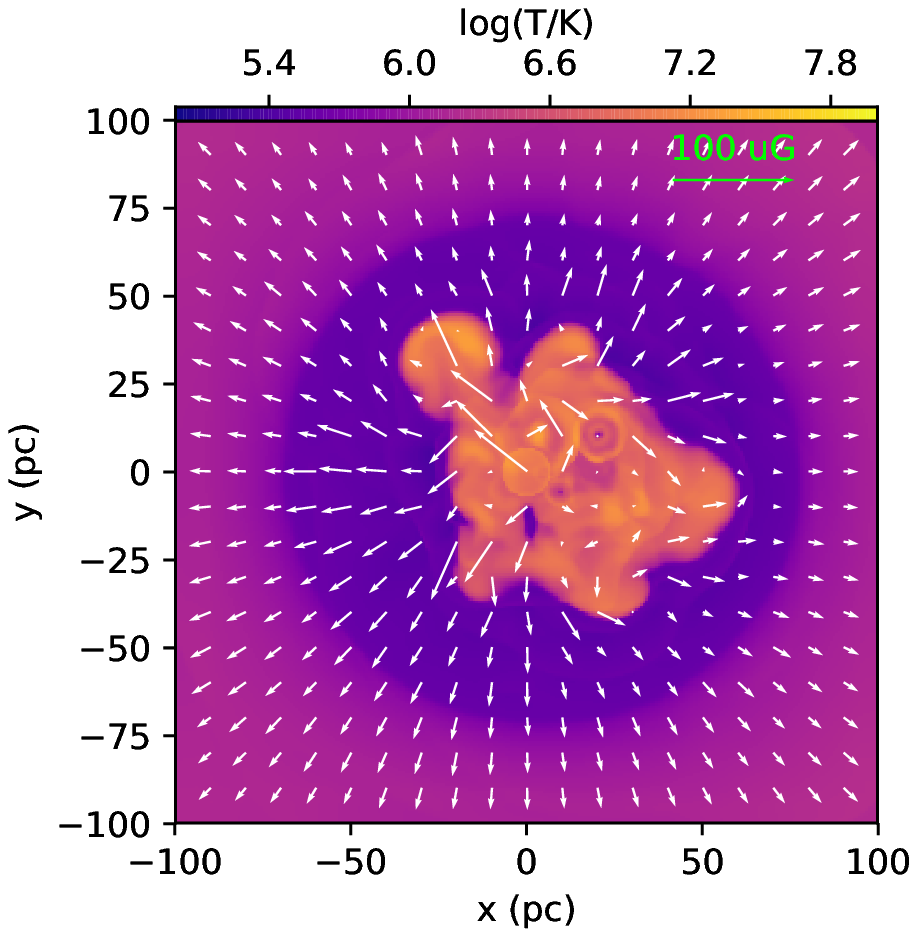}
    \includegraphics[width=0.323\textwidth]{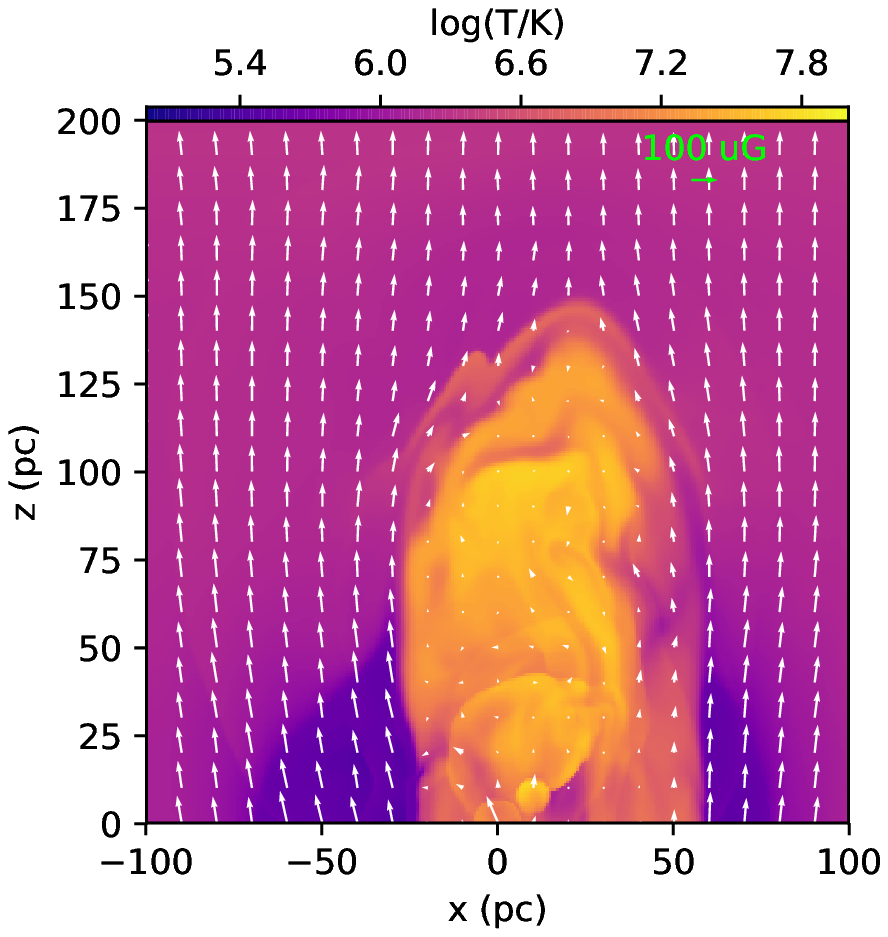}
    \includegraphics[width=0.323\textwidth]{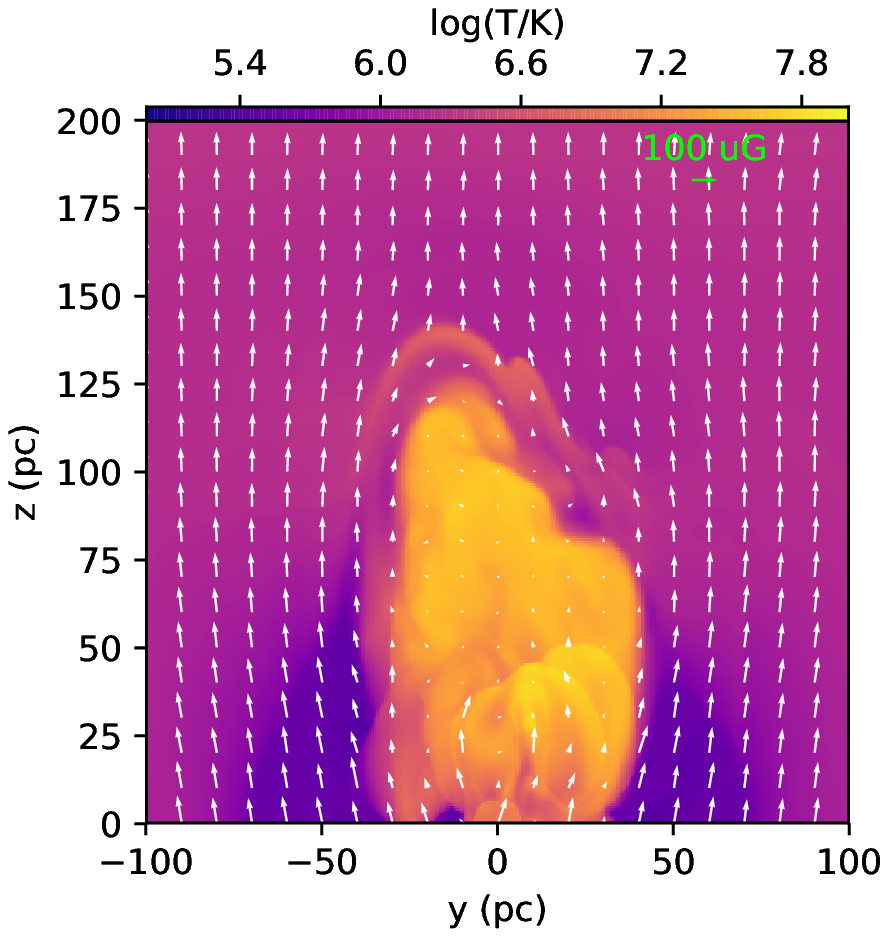}\newline
    \includegraphics[width=0.323\textwidth]{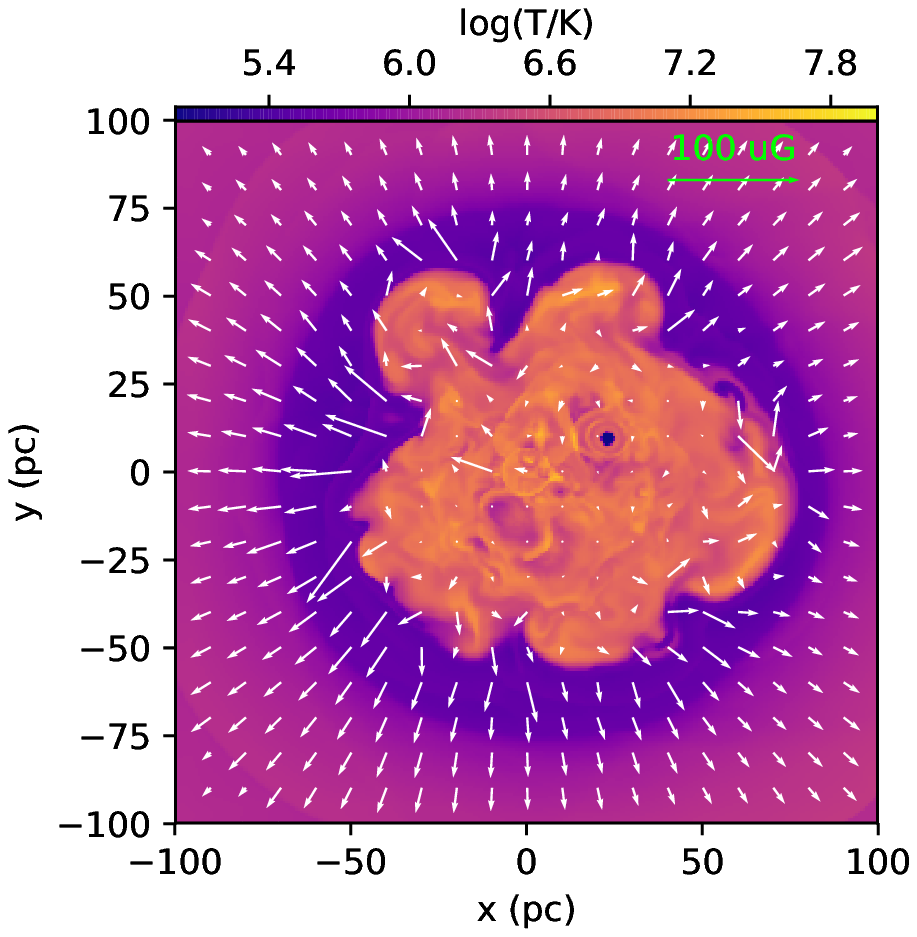}
    \includegraphics[width=0.323\textwidth]{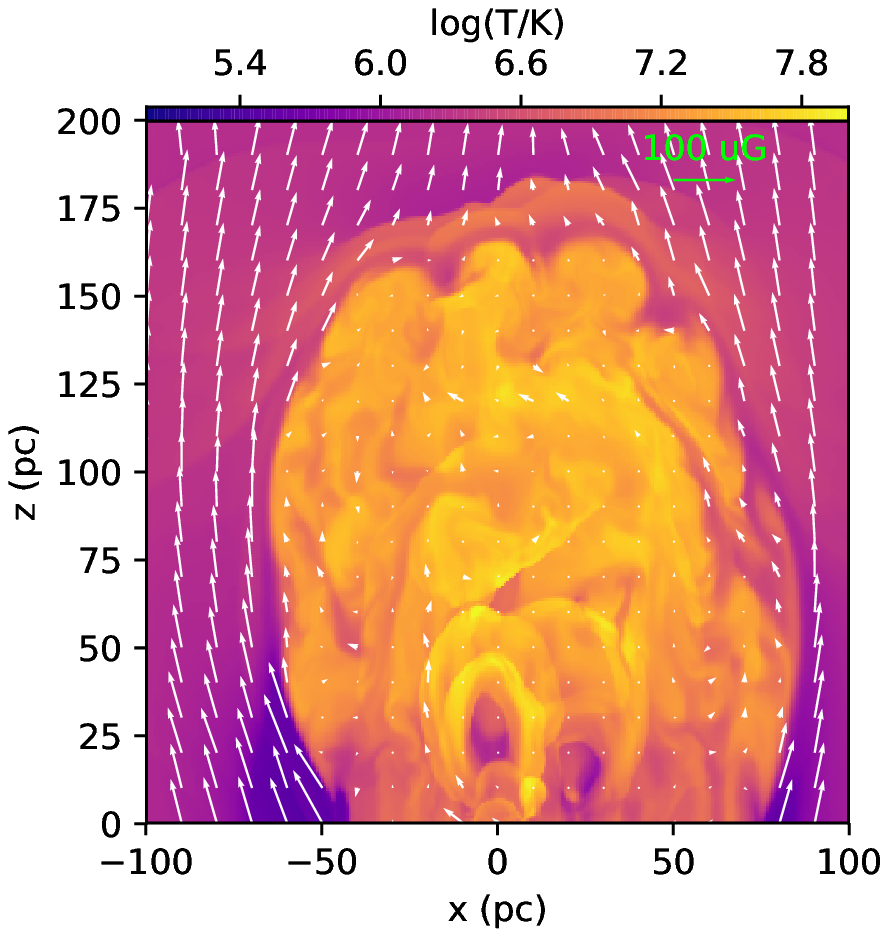}
    \includegraphics[width=0.323\textwidth]{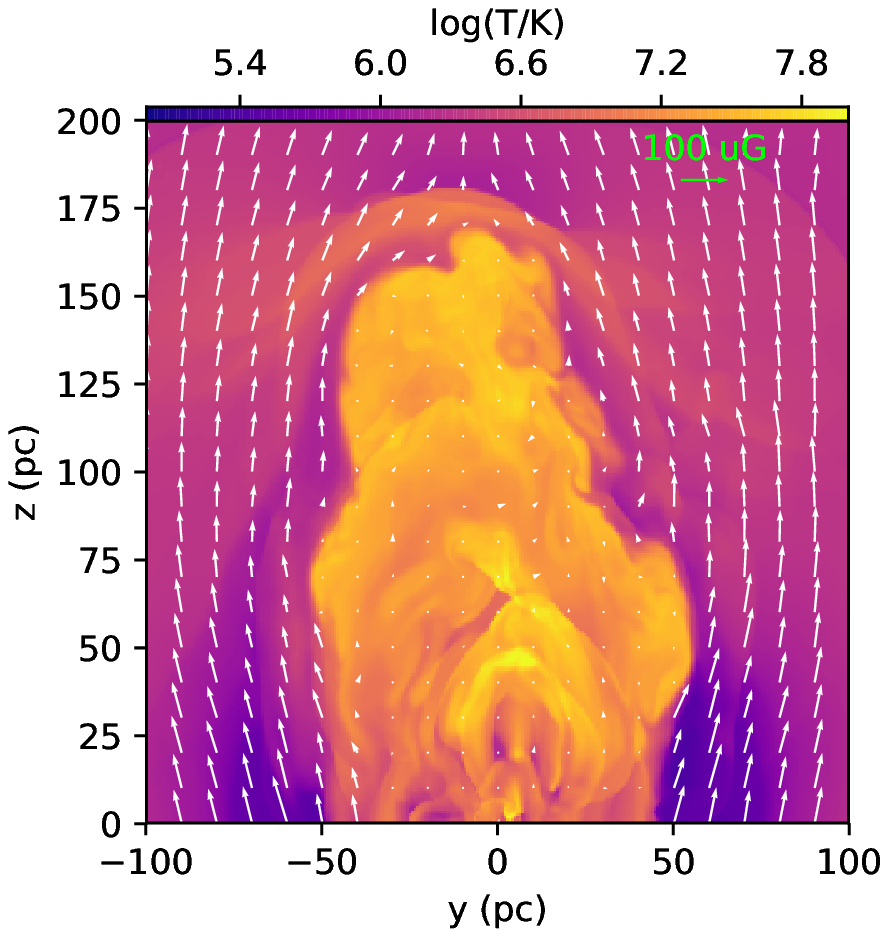}\newline
    \includegraphics[width=0.323\textwidth]{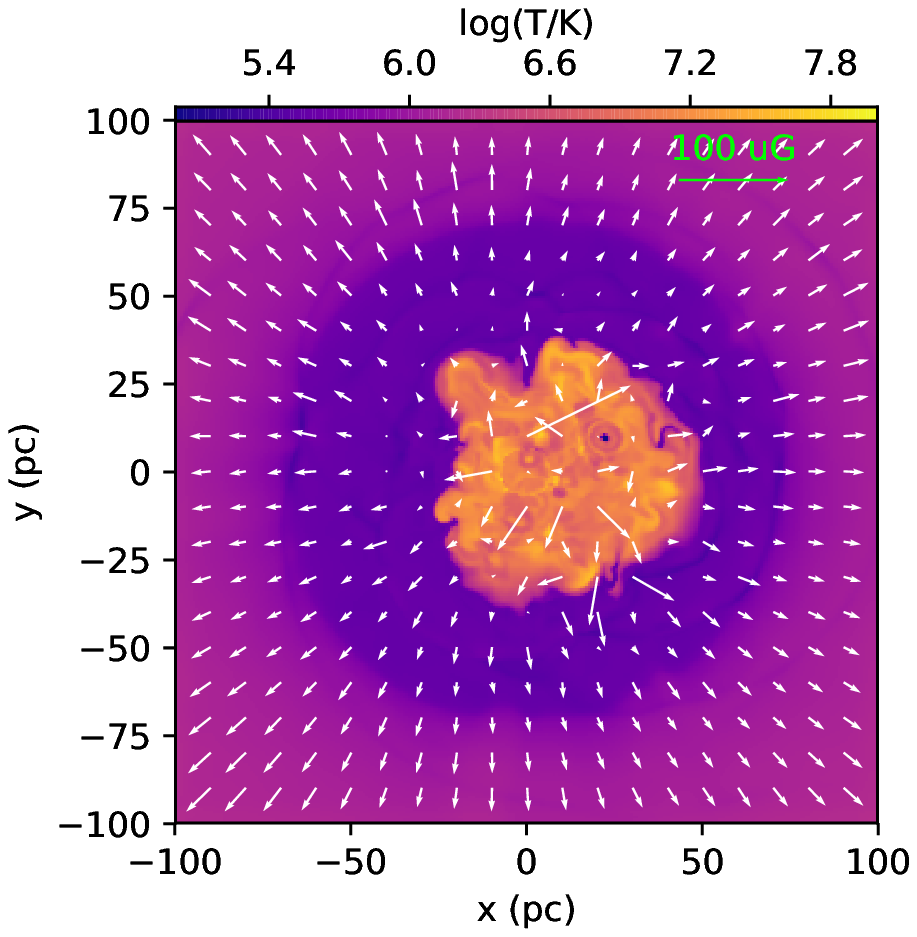}
    \includegraphics[width=0.323\textwidth]{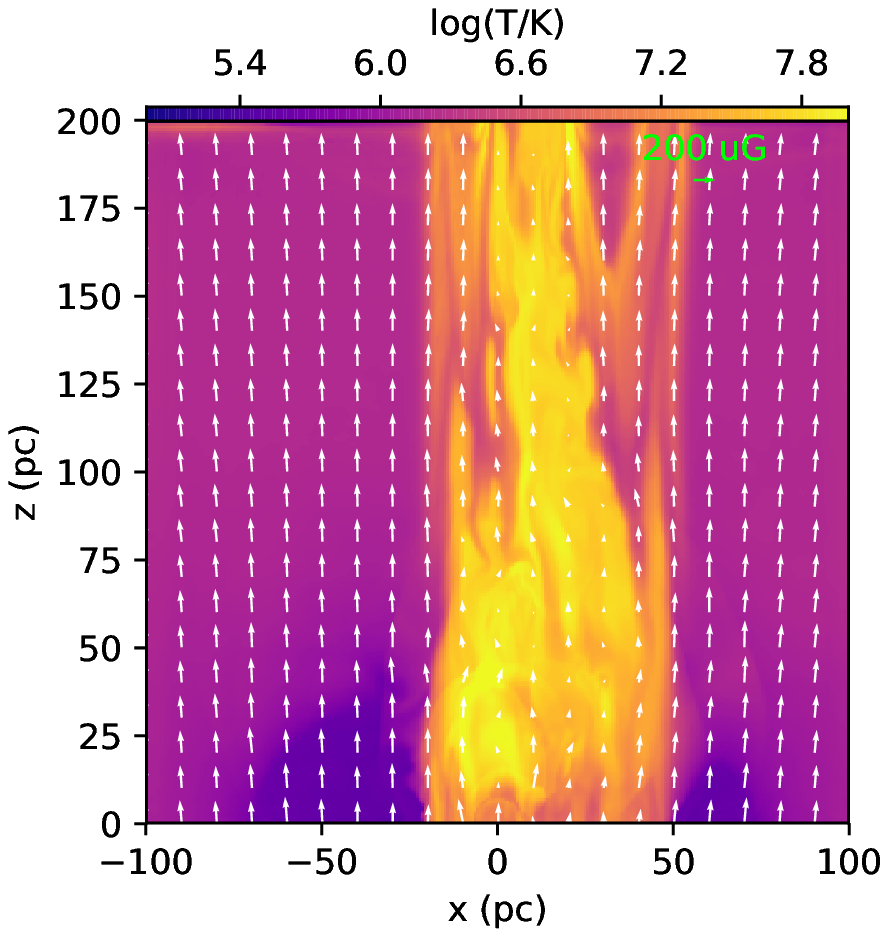}
    \includegraphics[width=0.323\textwidth]{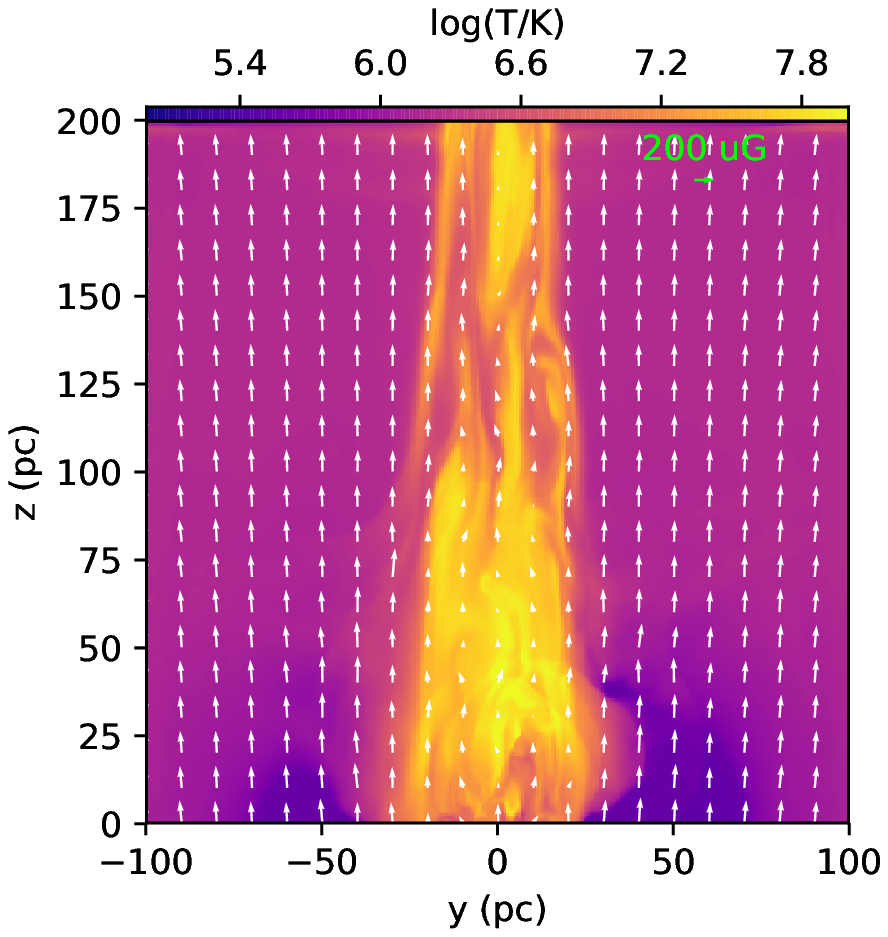}
    \caption{Simulated temperature-magnetic field images after 330 kyr. In the upper, middle and lower rows, we show the results of runs \textit{B80I2}, \textit{B50I1} and \textit{B200I1}, respectively. The $x-z$ and $y-z$ panels are slices through the center of the box along each axis, while the $x-y$ panel shows the slice at $z$ = 0. The background is the temperature distribution in logarithmic scale, and the white arrows indicate the magnetic field.}
\label{fig:Te}
\end{figure*}

Similarly, we show the snapshots of density and temperature maps of simulation runs \textit{B80I2}, \textit{B50I1} and \textit{B200I1} in Figures~\ref{fig:rhoe} and \ref{fig:Te}, all at $t$ = 330 kyr.
These three simulations share some common features with the fiducial simulation. In particular, a vertically-collimated, bubble-like structure is formed in all these simulations.
The bubble is delineated by a dense outer shell with compressed magnetic field and has a low-density, high-temperature interior with vertically-oriented velocities and generally weak magnetic field.
The bubble interior is not smooth, rather, it is filled with chaotic small-scale structures, again due to the sequential SN shocks and mutual interactions between them. Below we shall describe the more unique features in the individual simulations.

Simulation \textit{B80I2} (top row in Figures~\ref{fig:rhoe} and \ref{fig:Te}) adopts a lower explosion frequency than the fiducial case. This leads to a smaller energy injection rate, but is still sufficient to form a bubble.
The bubble evolves more slowly, reaching a height of only 140 pc by the time of 330 kyr.
The width of the bubble is also somewhat smaller than in the fiducial case (thus also narrower than the observed bubble), which remains the case even if we followed the bubble growth to a height of 200 pc.
This occurs because, given a weaker SN energy injection but the same magnetic confinement, reduction in the horizontal expansion is greater than in the vertical expansion.
We note that at a further reduced explosion frequency, much of the SN ejecta would not be able to escape from the strong gravity near the mid-plane, and a bubble would never form.

In \textit{B50I1} (middle row in Figures~\ref{fig:rhoe} and \ref{fig:Te}), which has a weaker magnetic field compared to the fiducial run, the resultant vertical collimation is less effective and thus the bubble appears fatter.
We note that a thinner bubble could still be achieved, should a lower explosion frequency be adopted in combination with the weaker magnetic field, for the reason explained above.
%The ejecta will slow down in $x-y$ plane, but keep a high velocity along $z$-axis. If the velocity difference between $x-y$ plane and $z$-axis become larger in a shorter duration, it will produce a thinner bubble. Frequent energy input will make it difficult to enlarge the velocity difference, on the contrary, a lower explosion frequency will make it easier to produce an asymmetric structure.
However, in this case it would take a much longer time for the bubble to grow to the observed height of 190 pc.

In contrast, \textit{B200I1} results in a significantly thinner structure.
The magnetic field in this run is so strong that it can resist the compression of the SN shocks and consequently
%the magnetic weakening is not obvious as other sets, which leads to a relatively strong magnetic strength inside.
there is little sweeping of magnetic field inside the bubble.
With the strong magnetic collimation, some SN ejecta are able to rapidly propagate along the field lines, forming vertical protrusions (several of these are captured in the $x-z$ and $y-z$ slides).
The overall morphology is obviously inconsistent with the observed bubble.

\subsection{Comparison with Observations}
\label{subsec:compobs}
Here we shall provide a more quantitative comparison with the radio and X-ray observations, with a focus on the fiducial simulation, which has the best morphological agreement with the observed bubble.

\begin{figure*}
    \centering
    \includegraphics[width=0.472\textwidth]{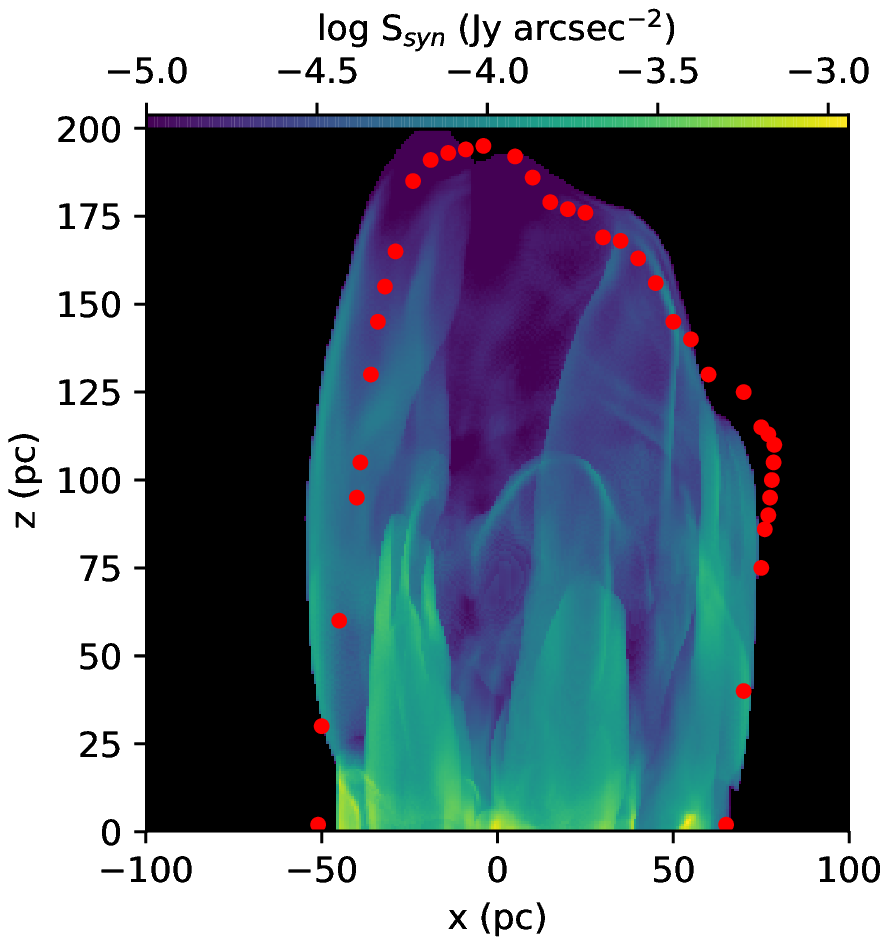}
    \includegraphics[width=0.472\textwidth]{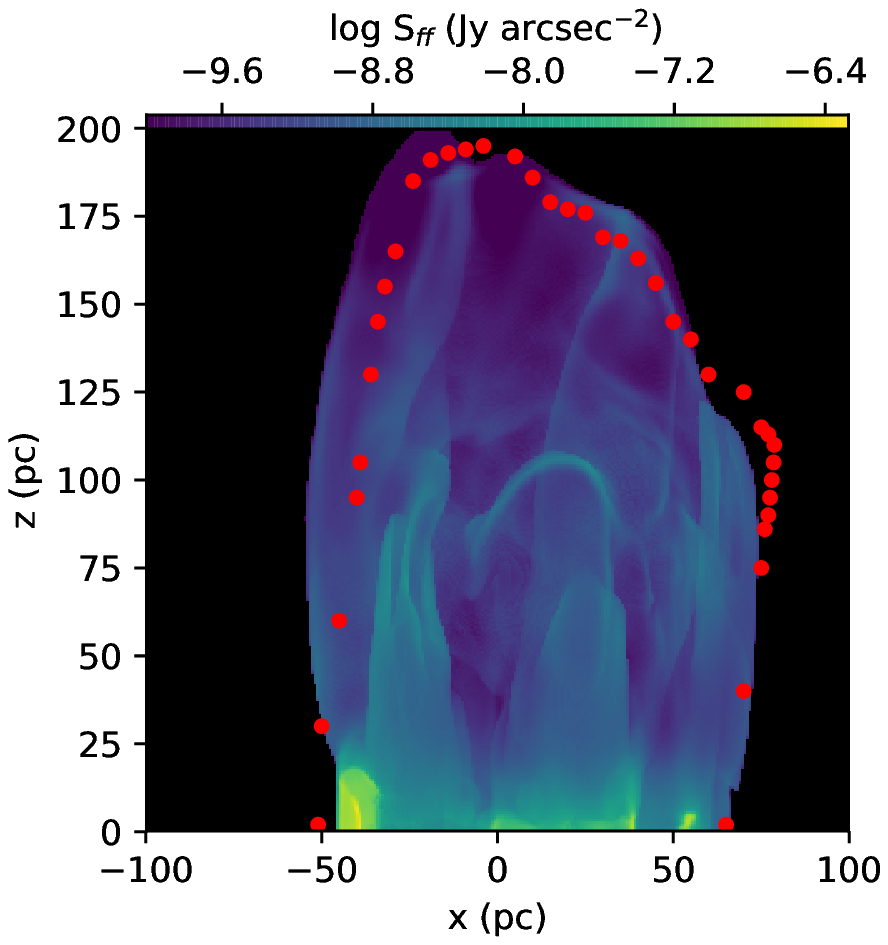}
    \caption{1284 MHz radio intensity distribution in simulation \textit{B80I1} at 330 kyr.
    %The initial magnetic field is 80 $\mu$G and the explosion interval is 1 kyr.
    The red dotted line outlines the rim of the northern radio bubble. \textit{Left:} Synchrotron emission;  \textit{Right:} Free-free emission.  In the left panel, values lower than 10$^{-5}$ Jy arcsec$^{-2}$ are suppressed to enhance visualization of the faint features.
    }
\label{fig:radioflux}
\end{figure*}

\begin{figure*}
    \centering
    \includegraphics[width=0.472\textwidth]{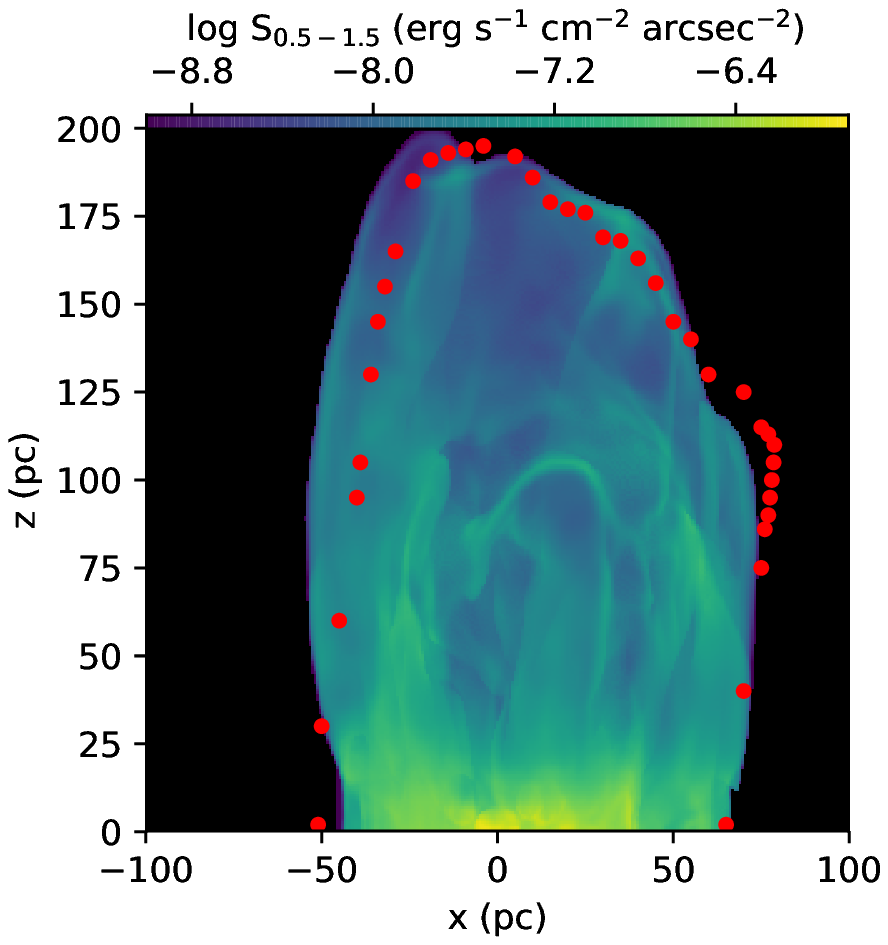}
    \includegraphics[width=0.472\textwidth]{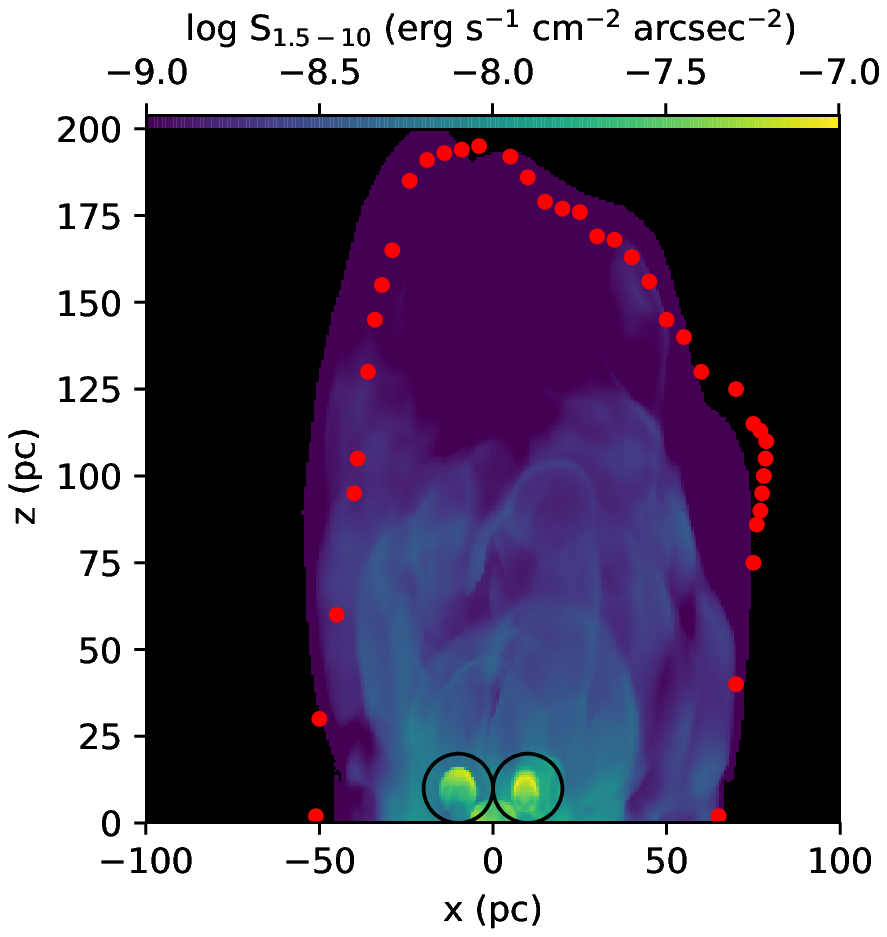}
    \caption{Synthetic $0.5-1.5$ ({\it left}) and $1.5-10$ ({\it right}) keV X-ray intensity distribution in simulation \textit{B80I1} at 330 kyr. The red dotted line outlines the rim of the northern radio bubble, while the black circles highlight two young SNRs.
     Values lower than 10$^{-9}$ erg s$^{-1}$ cm$^{-2}$ arcsec$^{-2}$ are suppressed to enhance visualization of the faint features.
    }
\label{fig:xflux}
\end{figure*}

% \begin{table*}
%   \caption{Simulation Results on the Radio Bubbles for set \textit{B80I1}}
%   \label{table:sum}
%   \centering
%   \begin{tabular}{l l l}
%       \hline\hline
%       Parameters                      & t=0 yr               &t=330 kyr    \\
%       \hline
%       Thermal Energy (erg)                & 1.7$\times$10$^{52}$           &3.4$\times$10$^{52}$ \\
%       Kinetic Energy (erg)                 & 0                              &1.4$\times$10$^{52}$ \\
%       Magnetic Energy (erg)                & 2.4$\times$10$^{52}$           &6.4$\times$10$^{52}$ \\
%       Synchrotron flux density at 1284 MHz (Jy) & 0                          &7830\\
%       Free-free flux density at 1284 MHz (Jy)  & 0.003                     &0.1 \\
%       $1-10$ keV X-ray luminosity (erg s$^{-1}$)      & 0              & $8.3\times 10^{36}$\\
%       \hline
%   \end{tabular}\\
% \end{table*}

The synthetic synchrotron and free-free intensity maps of \textit{B80I1}, after an evolution time $t$ = 330 kyr, are shown in the left and right panels of Figure~\ref{fig:radioflux}.
The overall morphology is quite similar between the synchrotron and free-free emission, which is partially owing to our assumption that the density of relativistic electrons scales with the local gas density.
However, the synchrotron intensity is everywhere orders of magnitude higher than the free-free counterpart in the synthetic maps.
This holds true even considering the uncertainties in the energy density of the relativistic electrons and the magnetic strength.
Consequently, synchrotron dominates the total flux density at 1284 MHz, consistent with the MeerKAT observation \citep{2019Natur.573..235H}.
It is noteworthy that both the hydrogen recombination line, H90$\alpha$, at 8309 MHz and the 8.4 GHz continuum are found to trace the GCL \citep{2019PASJ...71...80N}, which exhibits a loop-like structure spatially coincident with the northern radio bubble. This suggests that the thermal component may have an increasingly larger contribution toward higher frequencies, which can be due to a combined effect of substantial synchrotron cooling at higher frequencies and the presence of ambient cooler gas not taken into account in our simulation.
%We cannot compare the two flux density at different radio band, and more observations will be helpful.
%We note that the stellar wind, an important heating source, is neglected, but the radiative cooling is taken into consideration, which may cause an inaccurate free-free flux density estimation.

The overall extent of the synthetic synchrotron emission highly resembles that of the northern radio bubble (delineated by the red dotted line in Figure~\ref{fig:radioflux}), which, has a width of 120 pc at its base and a height of 190 pc.
%and some diffusive filamentary structures.
% The simulation bubble is a little wider than the radio bubbles, possibly due to the absence of molecular clouds, such as those in the CMZ { [at the base the bubble, the simulated one appears narrower. This is against your argument here][If there is a CMZ, the bubble will reach this scale at z-axis earlier and looks thinner in x-y plane. Then we can use weaker magnetic field to confine the morphology.]}.
Another interesting feature in the simulation is the presence of numerous filaments both at the edge of and inside the bubble,
which closely resemble the NTFs \citep{1984Natur.310..557Y}, although the ones in the simulation appear thicker and fuzzier in general, which may be partly owing to our moderate resolution.
In the simulation, these filaments originate from the sequential SN shocks and their mutual interactions, and are associated with locally amplified magnetic field (Figure~\ref{fig:B}).
%some of the observed NTFs might have been produced by, e.g., colliding stellar winds or magnetic reconnection \citep{1988ApJ...330..718H, 1994ApJ...424L..91S, 1996IAUS..169..247M, 2003ApJ...598..325Y,  BandaBarragan2016, BandaBarragan2018, 2019ApJ...490..L1}, which are not included in our simulation.
Their possible relation with the NTFs will be further addressed in Section~\ref{sec:dis}.

The 1284 MHz synchrotron flux density of the simulated bubble is found to be 5801 Jy, which is to be contrasted with our rough estimate of the observed flux density in the MeerKAT image, 970 Jy, obtained by assuming a mean flux density of 3 mJy beam$^{-1}$ across the projected area of the bubble.
We caution that the MeerKAT mosaic image presented in \citet{2019Natur.573..235H} was not corrected for the primary beam attenuation and that the extended emission from the bubble suffers from potential flux loss in the interferometric image (I. Heywood, private communication), thus our estimate should be treated as a lower limit of the true flux density.
On the other hand, the simulated flux density depends heavily on the assumed energy density of relativistic electrons.
Therefore, the apparently large discrepancy between the observed and simulated radio flux densities should be taken as a point for future improvement rather than a failure of the simulation.

The synthetic 0.5--1.5 keV and 1.5--10 keV X-ray intensity maps are shown in Figure~\ref{fig:xflux}. Compared to its radio morphology, the simulated bubble appears smoother in the X-rays.
The expanding shell of the bubble (Figure~\ref{fig:rho}) leaves no significant sign of limb-brightening in the 1.5-10 keV map, which is roughly consistent with the X-ray observations. This might be due to the fact that the shell is on average cooler than the bubble interior (Figure~\ref{fig:T}). Indeed, in the 0.5-1.5 keV map, which is more sensitive to gas temperatures below $\sim$1 keV, limb-brightening is more evident especially at the northwestern side of the bubble, although this energy band is not directly observable due to the large foreground absorption column density (a few $10^{22}\rm~cm^{-2}$; \citealp{2019Natur.567..347P}).
The 1.5-10 keV map also exhibits
much fewer small-scale structures in the bubble interior, except near the $x-y$ plane where the gas density is high and the most recent SNe freshly deposit a fraction of their kinetic energy.
In particular, remnants of two newly exploded SNe are evident near the center (marked in the right panel of Figure~\ref{fig:xflux}), although they are not clearly seen in the synthetic radio map.
An SNR evolving near Sgr A* will be heavily shaped by the strong gravity, with a large part of the ejecta pulled to the mid-plane, resulting in an appearance
resembling the bipolar X-ray lobes detected in the innermost 15 parsecs of the Galactic center \citep{Ponti2015, 2019Natur.567..347P}.

% { They both have a break-out side, because the gravity is strong there \citep{Yalinewich2017}.}
% There is also no clear sign of limb-brightening.
% { [would there be a limb-brightening effect if we go down to 0.5 keV][I will have a test. However, why do you feel there should be a limb-brightening in X-ray, since hot plasma will form in the center of an SNR]}

%For initial conditions and final results,
The thermal, kinetic and magnetic energy of the bubble is calculated by summing over all ``bubble pixels'' (Section~\ref{subsec:synthetic}),
%and estimate the radio and X-ray emission.
% The results are summarized in Table~\ref{table:sum}.
which is found to be 1.9, 1.2 and 0.1$\times 10^{52}$ erg, respectively.
The initial thermal and magnetic energy within the bubble volume are 0.7 and 1.1 $\times 10^{52}$ erg. A net decrease of the magnetic energy underscores the sweep-up of the magnetic field.
\citet{2019Natur.567..347P} estimated a thermal energy of 4$\times 10^{52}$ erg for the X-ray chimneys (sum of the northern and southern halves), which is well matched by the simulated value of 1.9$\times 10^{52}$ erg for the northern chimney.
%The sum of the differential energy, 1.4$\times 10^{52}$ erg, is much less than the total energy input from the 330 SNe (3.3$\times 10^{53}$ erg), which is understandable because we have adopted an outflow boundary condition and have not taken into account the gravitational energy and the work done during bubble expansion.

\citet{2019Natur.567..347P} also measured density and temperature profiles along selected Galactic longitude, $l = 0\degr$, and Galactic latitude, $b = 0\fdg7$.
For a direct comparison, we construct density and temperature profiles at $l = 0\degr$ and $b = 0\fdg7$ from the simulation, as shown in  Figure~\ref{fig:profile}.
Precisely speaking, Sgr A* is located at $l=0{\fdg}05579$, $b=-0{\fdg}04608$, but here we neglect this small difference and simply take the $x=0$ plane and $z=98$ pc plane for comparison.
We calculate the density-weighted mean density along the line-of-sight as
\begin{equation}
       < n > = \sqrt{\dfrac{\int\ n_{\rm t}^2dV}{\int\ dV}},
\label{eqn:temp}
\end{equation}
where $dV = A dl$, $A$ is the projected area, and the line-of-sight integration ($dl$) is from the farthest side to the nearest side of the bubble.
 The projected area varies across the profiles to approximate the rather irregular spectral extraction regions used in \citet{2019Natur.567..347P}. At $l = 0\degr$, the width is 12.5 pc, and the lengths are 20 pc and 70 pc respectively for $b < 0\fdg26$ and $b > 0\fdg26$. At $b =  0\fdg7$, the width is 12.5 pc, and the length is always 70 pc.
The emissivity-weighted mean temperature is calculated as
\begin{equation}
       < T > = \dfrac{\int\ Tn_{\rm t}^2\Lambda(T, Z)dV}{\int\ n_{\rm t}^2\Lambda(T, Z)dV},
\label{eqn:temp}
\end{equation}
where $\Lambda$ is the tabulated X-ray emissivity as a function of temperature and metallicity extracted from \textit{ATOMDB}.
By examining the distribution of the SN ejecta through the tracer parameter, we have verified that the assumption of a uniform metallicity is a reasonable approximation.

 At $l = 0\degr$, the simulated density profile peaks at the midplane and decreases untill $z \approx$ 40 pc, beyond which it flattens. This general trend is in reasonable agreement with the observed density profile.
%Below $z \sim$ 30 pc, the simulated density profile modestly increases toward the mid-plane, while
Notably, the observed density profile has a significantly higher peak at low $z$. This may be due partly to the smaller line-of-sight depth adopted by \citet{2019Natur.567..347P} for the two inner data points, and partly to contamination from unresolved stellar objects and non-thermal extended features to the apparently diffuse X-ray emission near the mid-plane \citep{2018ApJS..235...26Z}.
The simulated temperature profile appears bumpy around a mean value of $\sim1.0$ keV. The ``bumps'' are most likely due to consecutive SN shocks propagating upward.
Near the top of the expanding shell the temperature quickly drops to $\sim$0.6 keV.
%A relatively high temperature of 1.4 keV is seen at $z \sim$ 15 pc, which can be attributed to the heating by the two young SNe shown in Figure~\ref{fig:xflux}.
The observed temperature profile, on the other hand, appears flatter and has a lower value of 0.7--0.8 keV between 20--150 pc.
We note that the observed temperature was derived using a single-temperature spectral model to the underlying plasma having a range of temperatures \citep{2019Natur.567..347P}.
The Galactic center hot ISM is expected to have a somewhat lower temperature than in the whole bubble interior. Inclusion of the hot ISM in the observed spectrum could have led to a lower observed temperature.
%A relatively low temperature is seen near the base of the bubble, which can be attributed to a combined effect of stronger radiative cooling associated with the high gas density and adiabatic cooling given the outflow boundary condition at the $z=0$ plane.

At $b = 0\fdg7$, the simulated density profile peaks at the eastern and western edges of the bubble shell, which is consistent with Figure~\ref{fig:density}.
However, there is no clear sign of limb-brightening in the observed density profile; an enhanced density is only weakly seen near the eastern edge ($X \approx$ -60 pc) but is absent near the western edge ($x \approx$ 70 pc).   One possibility is that the soft X-ray emission from the denser and cooler western shell has largely dropped out of the observation band ({but could have been seen in the 0.5--1.5 keV band, as shown in the left panel of Figure~\ref{fig:xflux}).
The simulated temperature profile shows a roughly inverse ``U''-shape, with values peaking at $\sim$1.25 keV at $x = 20$ pc.
It is noteworthy that the outermost few points in the simulated profile are actually outside the bubble volume, whose values only reflect the unperturbed ISM.
The observed temperature profile, again derived from a spectral fit using a single-temperature model, appears flat around a mean value of 0.8 keV.

\begin{figure*}
    \centering
    \includegraphics[width=0.472\textwidth]{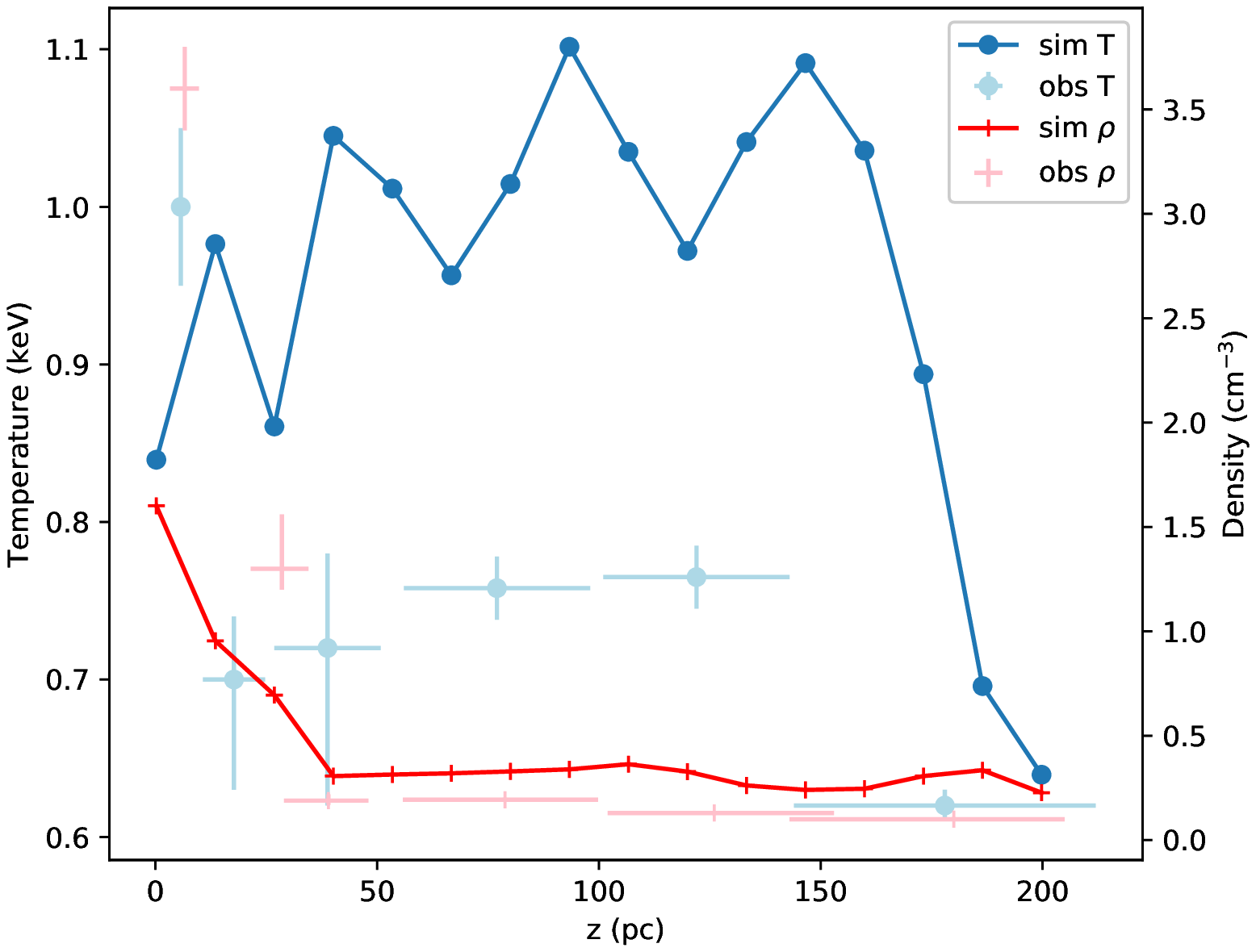}
    \includegraphics[width=0.472\textwidth]{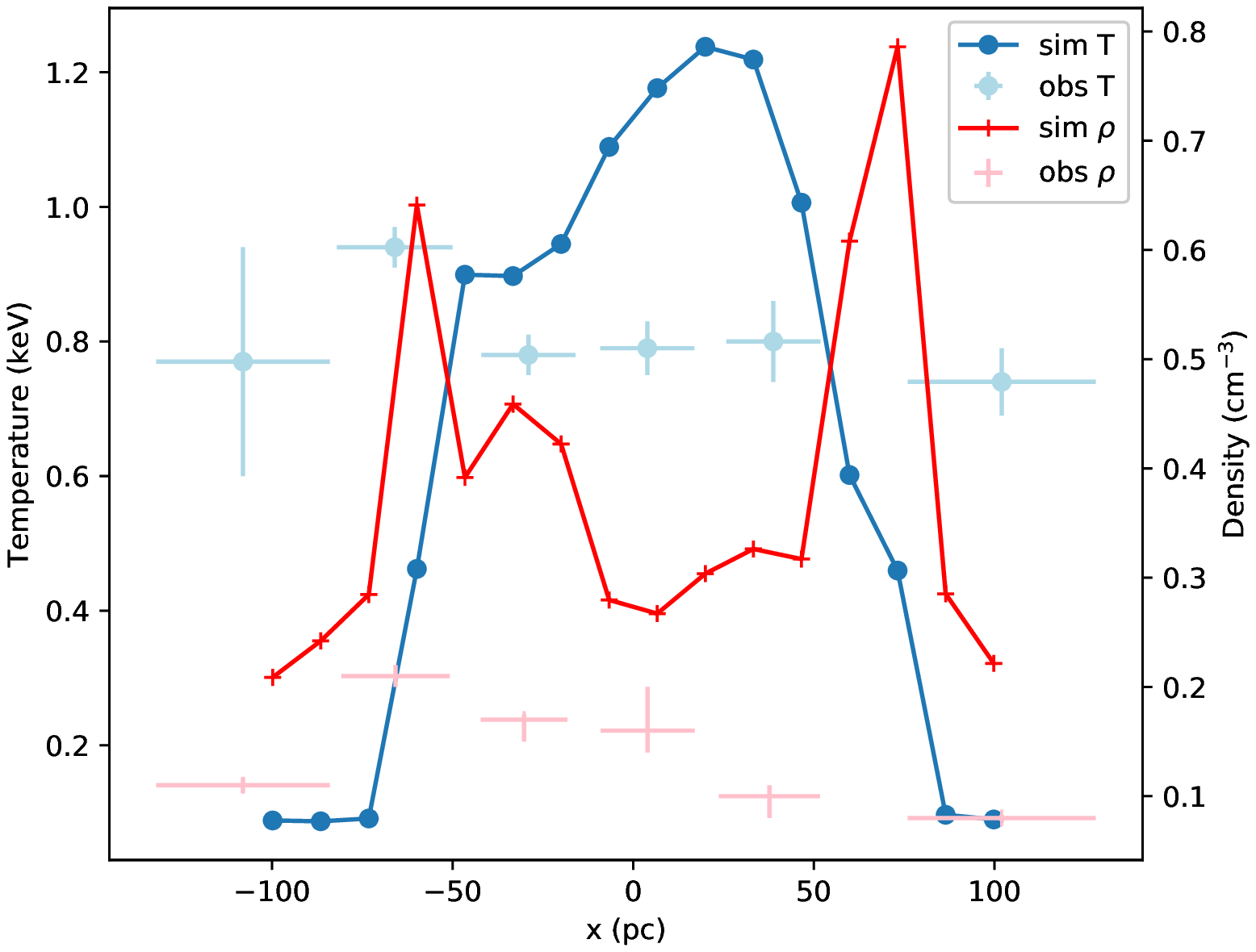}
    \caption{Density and temperature profiles of \textit{B80I1}. The light blue dots and pink pluses indicate the temperature and the total density in the simulation, respectively. The blue dots and red pluses respectively indicate the temperature and the density from the observations. The observed values are manually estimated from \citet{2019Natur.567..347P}. \textit{Left:} The profile at Galactic longitude $l = 0 \degr$. \textit{Right:} The profile at Galactic latitude $b = 0\fdg7$. Note that the observed density/temperature profiles cover a wider range reaching beyond the bubble volume.}
\label{fig:profile}
\end{figure*}

\citet{2019Natur.567..347P} did not provide an explicit total X-ray luminosity of the chimneys. A rough  estimate of this value can be made by adopting a cylinder of 150 pc in both diameter and height, as assumed by \citep{2019Natur.567..347P},
a mean density of 0.1 cm$^{-3}$ and a mean temperature of 1$\times 10^{7}$ K (0.86 keV), which are representative of the X-ray chimneys.
This leads to an estimated 1.5--10 keV luminosity of $\sim$2.8$\times 10^{36}$ erg s$^{-1}$ for the northern chimney, again well matched by the simulated value of 2.0$\times 10^{36}$ erg s$^{-1}$.
%The formation of the X-ray chimney is related with the X-ray emission in SNRs.
%The shock-heated gas was ionized, then produced X-ray emission by bremsstrahlung mechanism.
%Usually the radio shells of SNRs will encompass the X-ray region, but this is not obvious due to the mixture of many SNRs in this case.

For completeness, the synthetic radio and X-ray maps of runs \textit{B80I2}, \textit{B50I1} and \textit{B200I1} are shown in Figure~\ref{fig:fluxe}.
While these maps exhibit some interesting features, it is immediately clear that none of them matches the observed bubble morphology (again approximated by the red dotted line).
%The \textit{B80I1} shows a symmetrical morphology to the radio bubbles, because its time steps are different from other three.
%Different time steps cause different stochastic positions of the SNe explosions.
%Set \textit{B50I1} has more filamentary structures, but its magnetic field is the weakest magnetic field.
%Comparing sets \textit{B80I1}, \textit{B50I1} and \textit{B200I1}, we find that a stronger magnetic field leads to higher velocity along the magnetic field and lower velocity perpendicular to the magnetic field.
%These phenomena may be related to the morphology of the radio bubbles and the formation of NTFs.
%In our views, to form the radio bubbles, we need lower magnetic field, if we use lower supernova rate and SFR.

\begin{figure*}
    \centering
    \includegraphics[width=0.323\textwidth]{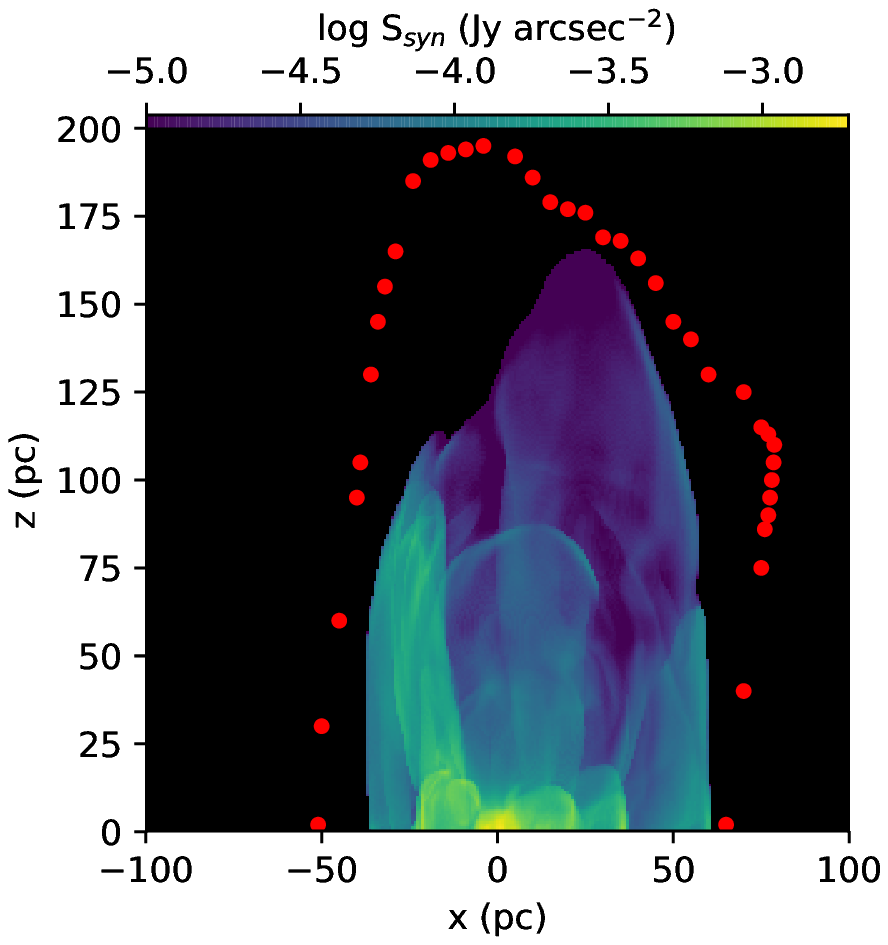}
    \includegraphics[width=0.323\textwidth]{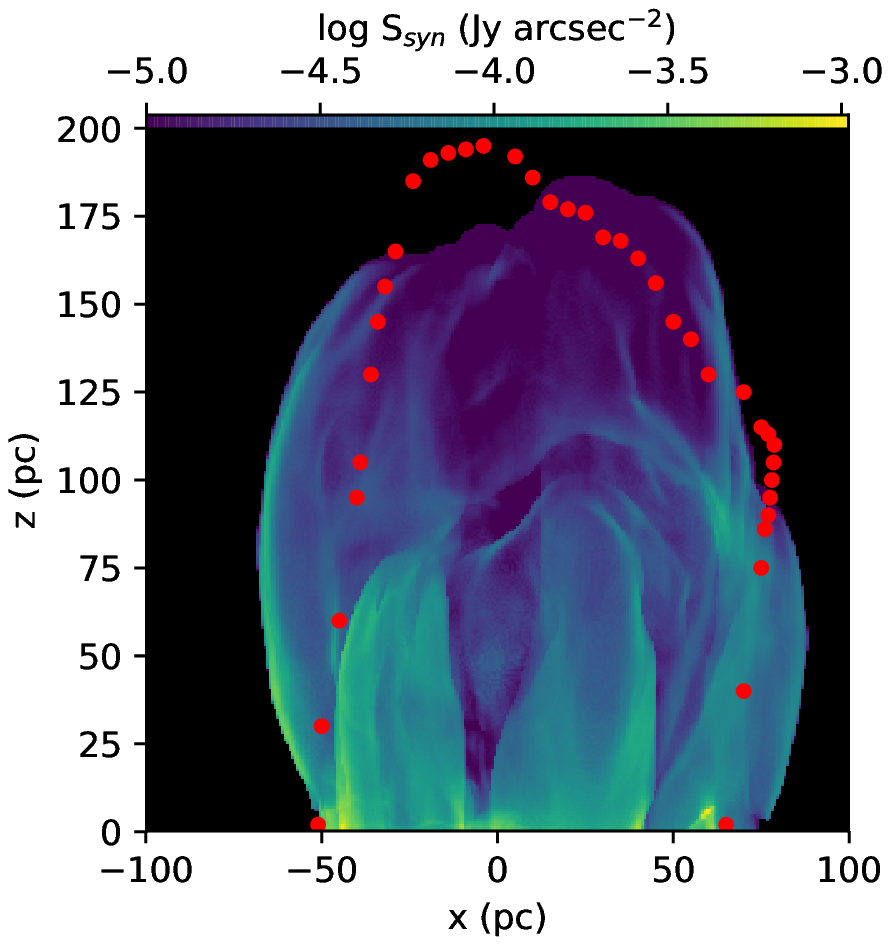}
    \includegraphics[width=0.323\textwidth]{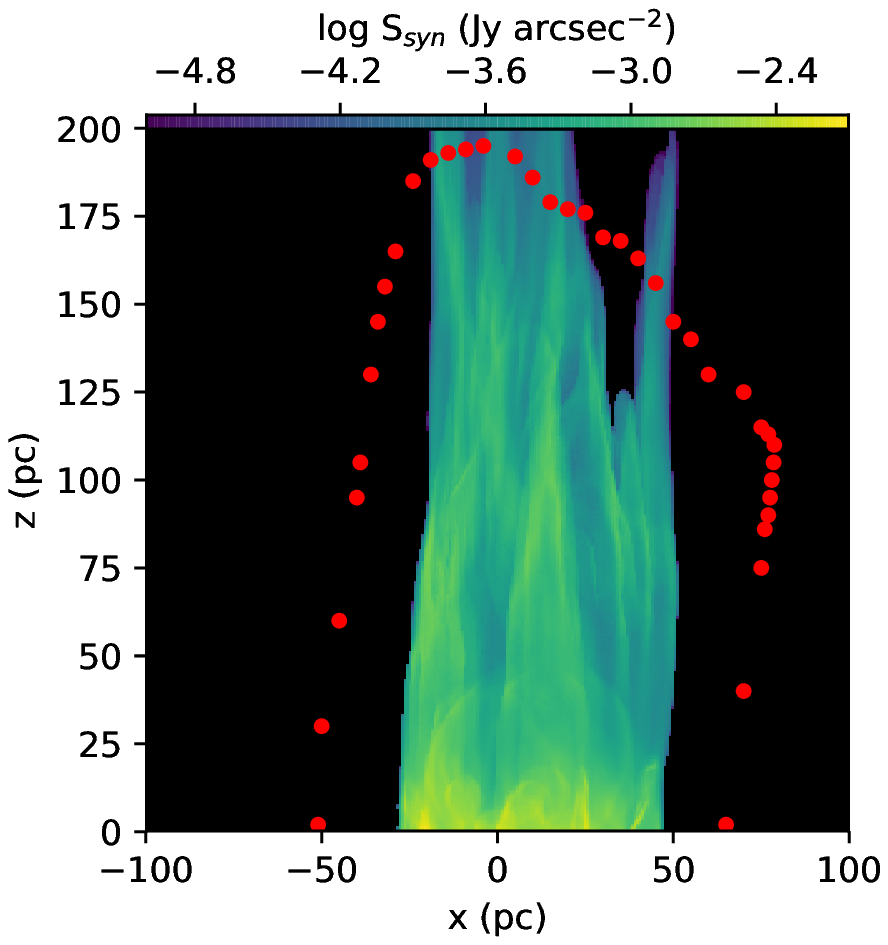}\newline
    \includegraphics[width=0.323\textwidth]{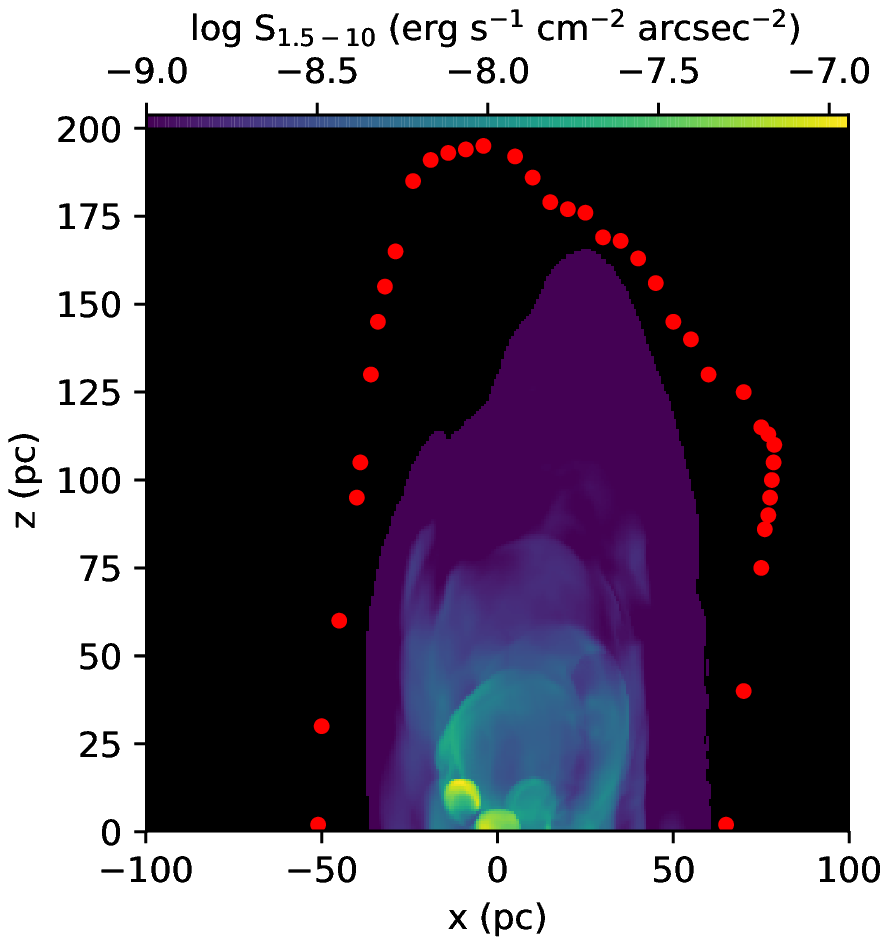}
    \includegraphics[width=0.323\textwidth]{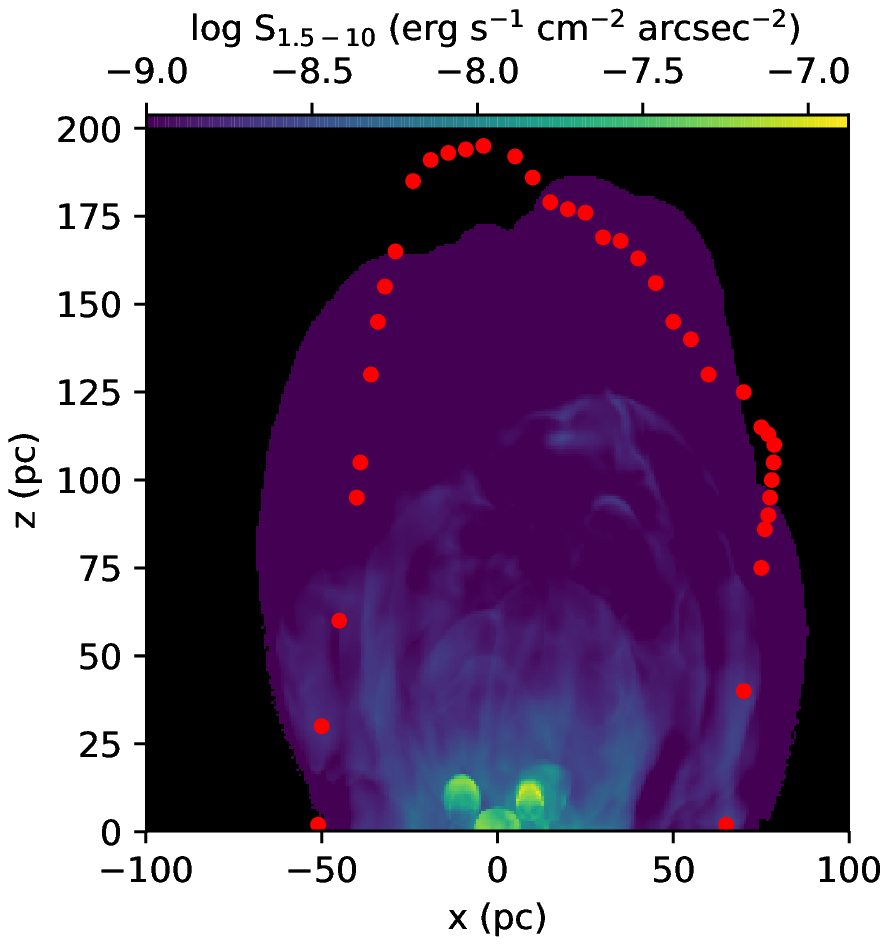}
    \includegraphics[width=0.323\textwidth]{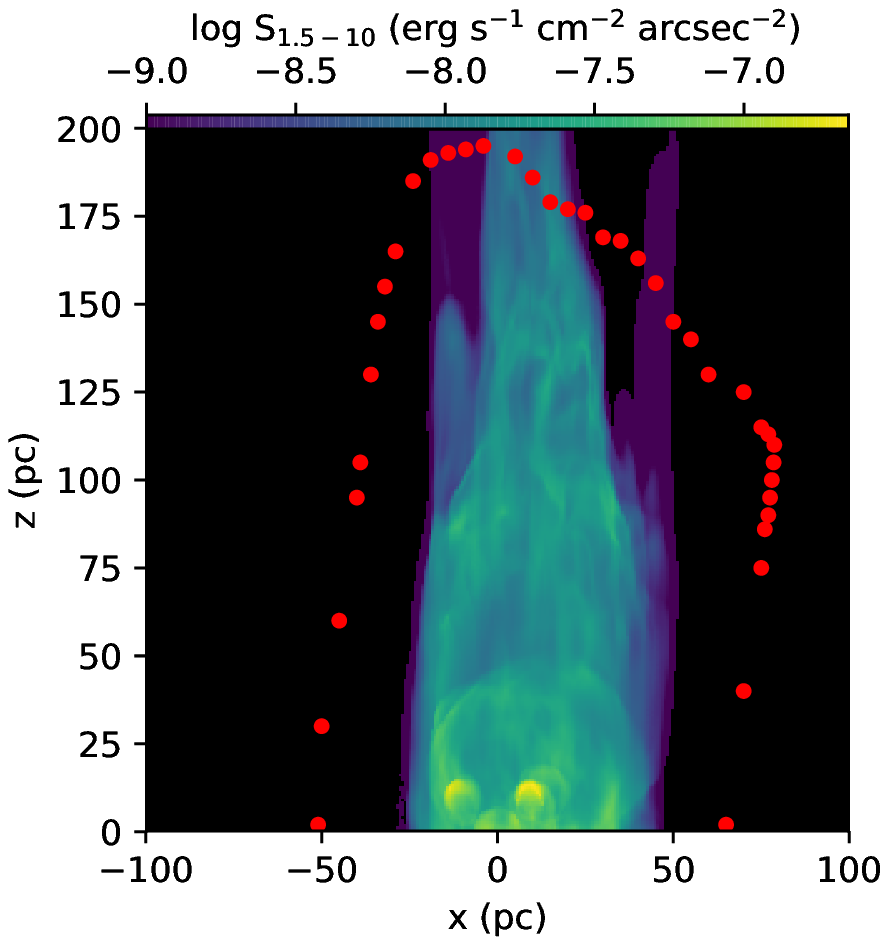}
    \caption{\textit{Upper panels:} Synthetic synchrotron intensity distribution at 1284 MHz. Values lower than 10$^{-5}$ Jy arcsec$^{-2}$ are masked for better visualization. \textit{Lower panels:} Synthetic 1.5--10 keV X-ray intensity distribution. Values lower than 10$^{-9}$ erg s$^{-1}$ cm$^{-2}$ arcsec$^{-2}$ are suppressed to enhance visualization of the faint features}. The red dotted line outlines the morphology of the northern radio bubble. The left, middle and right columns show the results of runs \textit{B80I2}, \textit{B50I1 } and \textit{B200I1}, respectively.
\label{fig:fluxe}
\end{figure*}

\section{Discussion} \label{sec:dis}

\subsection{The Origin and Fate of the Galactic Center Radio Bubbles/X-ray Chimneys}
\label{subsec:disrb}
The simulations presented in the previous section show that an outflow driven by sequential SN explosions and collimated by a vertical magnetic field can provide a reasonable explanation for the observed radio bubbles/X-ray chimneys in the Galactic center.
In particular, the simulations can well reproduce the overall morphology, X-ray luminosity and thermal energy of the northern bubble.

This scenario relies on two key ingredients: SN explosions clustering in the nuclear disk to provide a semi-continuous energy input, and a vertical, moderately strong magnetic field to provide the collimation.
Both ingredients are very likely available in the Galactic center.
Indeed, direct evidence for contemporary SN explosions in the Galactic center was provided by at least a few SNRs clearly visible in radio or X-ray images (e.g., \citealp{Ponti2015}).
Moreover, about two hundred emission-line objects have been detected in the Galactic center, most of which are likely evolved massive stars \citep{2012MNRAS.425..884D}. These stars may belong to the same population that gave rise to the SNe responsible for launching the bubbles.
As for the magnetic field, it is widely thought that it is predominantly poloidal in the Galactic center, at least in regions outside the giant molecular clouds \citep{Ferriere2009}.
In this regard, an SNe-driven, magnetically-collimated outflow
should naturally develop in the Galactic center, provided the correctness of our simulations.

As mentioned in Section~\ref{sec:intro}, a competing driver of a large-scale outflow is the kinetic power from the central SMBH, even though Sgr A* is by no means comparable with a classical AGN.
While our simulations cannot automatically rule out an AGN-driven outflow, they share useful insight on the latter case.
Compared to the distributed SN explosions, energy input from the SMBH is highly concentrated. Thus an AGN-driven outflow on the hundred-parsec scale may either acquire a highly elongated shape in the case of a canonical jet-driven outflow (e.g., \citealp{Zhang2020}), or inflate a fat bubble in the case of a more isotropic wind symbiotic with the hot accretion flow onto a weakly accreting SMBH \citep{2015ApJ...804..101Y}.
Magnetic collimation may also shape the wind-blown bubble, but one expects that the resultant structure is again a highly elongated one. Thus matching the morphology of the radio bubbles with an AGN wind-driven outflow may require some fine-tuning, which awaits a detailed investigation.

%stellar wind
We now turn to consider the fate of the radio bubbles. In the framework of our simulations, the SNe-driven outflow is necessarily an evolving structure. In fact, at the end of our fiducial simulation, the top of the bubble still expands at a speed of $\sim 600\rm~km~s^{-1}$ (Section~\ref{subsec:formation}).
Provided a continuous energy injection from future SNe, which is quite likely given the evolved massive stars near the disk plane \citep{2012MNRAS.425..884D}, the bubbles should continue to grow and gradually evolve into a more ``chimney''-like structure, as long as a moderately strong magnetic field persists to greater heights.
Conversely, if SNe were temporarily shut off, one expects that the bubble/chimney would ultimately disperse and collapse within a time not much greater than the sound-crossing time (a few hundred kyr).
We have run a test simulation to examine such a case. Specifically, we adopt the same setting as the fiducial simulation, except that SN explosions cease after a time of 200 kyr. It is found that the upper edge of the bubble can still climb to a height of $\sim$190 pc with its accumulated momentum. However, the interior of the bubble, especially its lower portion, begins to collapse soon after the shutoff of the SNe, due to the loss of energy injection against the strong central gravity. In addition, the mean gas temperature inside the bubble gradually declines. Such an effect might bring the simulated temperature profile into better agreement with the observed temperature profile (Figure~\ref{fig:profile}), although we have no evidence that the Galactic center is currently experiencing a substantial drop in the SN birth rate.  %Then, after a time of $\sim$ kyr, the entire bubble loses its identity, meanwhile the magnetic field strength restores to the initial configuration through much of the volume previously occupied by the bubble.

It is interesting to ask whether the radio bubbles/X-ray chimneys have a causal relation with the Fermi bubbles \citep{2010ApJ...724.1044S} and eROSITA bubbles \citep{Predehl2020} found on much larger scales.
We note that the age of the radio bubbles inferred from our simulations is only a few hundred kyr, much shorter than the dynamical timescale of a few Myr originally suggested by \citet{2019Natur.573..235H}.
However, \citet{2019Natur.573..235H}'s estimate was based on the assumption of a constant expansion velocity of the bubbles, which is implausible, hence a shorter timescale is expected.
The estimated age of the Fermi bubbles, on the other hand, ranges from 1 Myr \citep{2013MNRAS.436.2734Y} to 1 Gyr \citep{2011PhRvL.106j1102C}.
Thus, in the context of our supernova-based model for the origin of the radio bubbles/chimneys, the radio bubbles would be a dynamically younger and independent structure simply evolving in the interior of the Fermi/eROSITA bubbles, which themselves were formed by older activities in the Galactic center.

Alternatively, as suggested by \citet{2019Natur.567..347P}, the X-ray chimney may be a channel that transports energy from the Galactic center to the high-latitude region currently occupied by the Fermi bubbles.
In this case, the channel should have existed for tens of Myr, so that star formation in the Galactic center can be sufficient to supply the total energy content of the Fermi bubbles, $\sim 10^{56}$ erg \citep{2013Natur.493...66C}.
However, such a picture contradicts with the capped morphology of the radio bubbles (the southern bubble is not obviously capped in X-rays; \citealp{2021A&A...646A..66P}), which, according to our simulations, is naturally explained as the expanding shell of a newly born outflow.
This picture may be reconciled if star formation in the Galactic center has been episodic on a timescale of $\sim$10 Myrs \citep{2015MNRAS.453..739K}.
In this case, the ``chimney'' is (re)established by consecutive generations of mini-starbursts and collapses inbetween.
Of course, over such a long interval, the activity of Sgr A* can also play an important role in contributing to the inflation of the chimneys, especially in view of the fact it was likely much more active in the recent past \citep{2010ApJ...714..732P, Ponti2013, 2018ApJ...856..180C}.
In a hybrid scenario, Sgr A*, with supernovae and even stellar winds, can simultaneously sustain the ``chimney'' and transport energy to larger scales, implying X-ray emission beyond the edge of the radio bubbles, which is also suggested by \citet{2021A&A...646A..66P}.

\subsection{Origin of the Non-thermal Filaments}
\label{subsec:disNTFs}
The origin of the NTFs has been extensively debated since their discovery nearly four decades ago.
Proposed models for the NTFs include expanding magnetic loops \citep{1988ApJ...330..718H}, induced electric fields \citep{1988ApJ...333..735B, 1989ApJ...343..703M}, thermal instability in relativistic gas \citep{1993A&A...270..416R}, cosmic strings \citep{1986PhRvD..34..944C}, magnetic reconnection \citep{1992A&A...264..493L, 1994ApJ...424L..91S, 1996IAUS..169..247M,  BandaBarragan2016, BandaBarragan2018}, analogs of cometary plasma tails \citep{1999ApJ...521..587S}, a turbulent magnetic field \citep{2006ApJ...637L.101B}, stellar winds or SNe of the young star cluster \citep{2003ApJ...598..325Y, 2019ApJ...490..L1}, pulsar wind nebulae  \citep{2019MNRAS.489L..28B}, and the tidal destruction of gas clouds \citep{2021MNRAS.501.1868C}.
Of course, a multi-SNe hypothesis has also been suggested \citep{2020PASJ...72L...4S}.

In our simulations, filamentary features resembling the observed NTFs trigger and form primarily at the interface of colliding shocks of individual SNe (Figure~\ref{fig:B}).
Magnetic fields are compressed and amplified in these filaments, where particle acceleration (e.g., due to diffusive shock acceleration) is expected to take place.
Also the Radio Arc finds its possible counterpart in the simulations, which arises from the piling of consecutive SN shocks at the sides of the bubble (Figure~\ref{fig:rho}).
%The interaction region is usually threadlike in this case, so it it easy to produce many magnetized filaments.
%Figure~\ref{fig:T} also show some filaments with strong magnetic field in the slices, which implies the formation of NTFs.
Comparing Figure~\ref{fig:radioflux} and Figure~\ref{fig:fluxe}, it occurs that an SN-driven outflow evolving in a weaker magnetic field produces more filaments.
This is because a strong magnetic field can more easily confine an SN shock and reduce its chance of encountering other shocks. We note that in the simulation many filaments are indeed one-dimensional structures, i.e, they have a distinct long-axis roughly oriented vertically, but some others arise from a projection effect, i.e., a two-dimensional surface viewed edge-on. Such a surface is also the result of colliding shock fronts.
We stress that the moderate resolution of our simulation would smear the appearance of the shock fronts, so we anticipate that additional apparent filaments would show up with higher resolution.  The viability of this formation mechanism for the NTFs could be assessed by direct comparison of the cross-sectional profiles of the filaments appearing in the simulations with those of observed NTFs, but a higher  resolution simulation is needed for such a comparison.

%suggested it has a similar origin with the NTFs, but the surrounding environment is different from NTFs.
% For example, there is high-density molecular clouds or higher magnetic field.
%Many NTFs formed in this region then produce a complex, the radio arc.
%In the density distribution shows the multi-shell structures on the two sides, akin to the radio arc.

%It has been suggested that the Radio Arc derives its size and brightness from its rooting on a giant molecular cloud. In our simulation, it appears that a similar arc-like structure can form without such an environmental effect.

We note that there are NTFs found outside the radio bubbles \citep{2019Natur.573..235H}. These might have been formed in a past generation of clustering SN explosions, and they exist for a longer time than the associated outflow.
Of course, we cannot rule out the aforementioned alternative models for all NTFs.
In reality, the NTFs can have a mixed origin, i.e., different processes, including
SN shocks, stellar winds and pulsar winds can produce seeds of NTFs which are further shaped by the compressed magnetic field or other mechanisms.

\subsection{Strength of the Galactic Center Magnetic Field}
\label{subsec:disul}
The magnetic field is a crucial component of the Galactic center environment.
At present, the average field strength is still quite uncertain. The assumption of energy equipartition between the magnetic field and relativistic particles leads to estimates up to $\sim$ 1 mG in the brightest NTFs and as low as 10~${\mu}$G in the more diffuse background.
\citet{2010Natur.463...65C} derived a lower limit of $\sim 50~\mu$G based on the diffuse $\gamma$-ray flux and suggested a typical value of $\sim 100~\mu$G in the central 400 pc region.

In our simulation \textit{B50I1}, which adopts a field strength of 50 $\mu$G, an outflow can be developed, although the resultant bubble appears fatter due to the reduced magnetic confinement compared to the fiducial simulation (Section~\ref{subsec:compset}).
This lends some support to the above lower limit.

On the other hand, simulation \textit{B200I1}, which assumes a field strength of $200~\mu$G, is obviously inconsistent with the observation (Figure~\ref{fig:fluxe}).
This conclusion holds even if the other parameter, the SN birth rate, were adjusted within a reasonable range.
%The supernova frequency, 1 kyr$^{-1}$, is so large that it can be taken as an upper limit of the supernova frequency here.
%Therefore, we can only use a lower supernova frequency.
%However, set \textit{B200I1} shows a thinner irregular morphology,
Qualitatively, at a lower SN birth rate, the shock and ejecta of individual SNe would be less resistant to the magnetic pressure, thus they are less likely to evolve into a mutual network. The resultant outflow hardly takes a  bubble shape, rather it would consist of many barrel-like structures, through which individual SN ejecta propagate.
Only a much higher SN birth rate can counteract the magnetic pressure, but this would be inconsistent with the currently accepted star formation rate in the Galactic center ($\sim 0.1\rm~M_\odot~yr^{-1}$).
Therefore, our simulations provide a meaningful constraint on the average magnetic field on 100 pc scales in the Galactic center, $50\rm~{\mu}G \lesssim B_0 \lesssim 200~{\mu}G$.

%Except for the multi-SNe model, jets and stellar winds will also be largely shaped by the strong magnetic field and cannot produce such a radio bubble,
%if the energy input rate is lower than SNe.
%In other words, the observed radio bubbles cannot be formed, unless there is a mechanism with an energy input rate larger than one SN per millennium.

%\citet{Thomas2020} suggested a magnetic strength of 200 $\mu$G in the NTFs and the magnetic strength in the NTFs is usually larger than the surrounding ISM, which  also implies an upper limit of 200 $\mu$G.

Our fiducial run \textit{B80I1} demonstrates localized magnetic field amplification across the bubble, reaching a maximum field strength of 175 $\mu$G.
It is expected that the global magnetic field  would gradually restore to the initial configuration
after the termination of clustering SN explosion and the dispersion/collapse of the outflow.
%which means our model can indeed produce some regions with strong magnetic field and the NTFs can be produced by the multi-SNe.
%The shock wave cannot compress the magnetic field efficiently, if the magnetic field is strong.
%Consequently, the DSA theory predict the amplified magnetic field should be four times larger than the initial value, however, it is only two times at most in our simulation.
%We here focus on the mean diffusive magnetic field, and the local magnetic amplification caused by instabilities, turbulence or other mechanisms is not included.
%In some localized filaments, the magnetic strength can even reach 1 mG \citep{Ferriere2009}.
%However, the resolution of our simulation is not adequate to reproduce these local features.

\subsection{Caveats} \label{subsec:caveat}
Despite the satisfactory reproduction of the major observed properties of the radio bubbles/X-ray chimneys, some notable discrepancies exist between our simulation results and the observations, which warrant the following remarks.

The observed edge-brightened radio bubbles have a low-surface-brightness interior, while in our simulation the edge-interior contrast is less significant.
A possible cause is that we have ignored synchrotron cooling.
Using a magnetic field of 20 $\mu$G, \citet{2019Natur.573..235H} derived a synchrotron cooling time of 1--2 Myr by assuming that the electron energy density distribution has a power-law index of 2.
Based on the same method, we estimate a cooling time of 250 kyr for 80 $\mu$G, which is comparable to the evolution time of the bubble in our simulation. Hence the relativistic electrons produced at the early stage and now filling the bubble interior should be subject to radiative cooling, an effect that is not taken into account but otherwise would enhance the edge-interior contrast.

An alternative and more likely cause is the absence of a cool gas shell in our simulation. The presence of cool gas (with a temperature of $\sim 10^4$ K) in the outer part of the GCL has been known for some time \citep{2010ApJ...708..474L, 2019ApJ...875...32N}.
This cool gas is not found in our simulations, owing to the very moderate radiative cooling even in the dense shell of post-shock gas.
This is also the reason why the free-free emission predicted by our simulation is negligible compared to the synchrotron (Section~\ref{subsec:compobs}).
Hence the detected cool gas probably has an external origin that is missing in the framework of our simulation.
Indeed a substantial amount of both cool and cold gas exist in the NSD/CMZ \citep{ 2007A&A...467..611F}, and part of this gas may be swept into the bubble shell and/or entrained into the bubble interior.
For example, \citet{2021A&A...646A..66P} argued that a gas cloud associated with the bright 25 $\mu$m source AFGL5376 has been accelerated and is now defining part of the wall of the bubble.
An additional source of cool gas is the stellar wind of the massive stars distributed in the nuclear disk.

In principle, the Galactic center outflow may also be driven by stellar winds \citep{1992ApJ...397L..39C}.
Stellar winds as an additional energy and momentum source have not been included in our simulation.
We can give a rough estimate of the collective energy input from the massive stars in the Galactic center.
The stellar winds should be dominated by the Wolf–Rayet stars, which have a typical mass loss rate of $10^{-5}\rm~M_{\odot}~yr^{-1}$ and a wind velocity of $2000\rm~km~s^{-1}$.
Thus the $\sim$200 evolved massive stars found by \citet{2012MNRAS.425..884D} in the nuclear disk have a total kinetic power of $2.5\times10^{39}\rm~erg~s^{-1}$ and would release a kinetic energy of $2.6\times10^{52}\rm~erg$ in 330 kyr.
The massive stars in the central parsec provide an additional kinetic energy of $3\times10^{51}\rm~erg$ in 330 kyr, assuming a collective mass loss rate of $10^{-3}\rm~M_{\odot}~yr^{-1}$ and a wind velocity of $1000\rm~km~s^{-1}$ \citep{1997A&A...325..700N, 2004ApJ...613..322Q}.
Therefore, the energy input from the massive stars is about one order of magnitude smaller than that of the SNe in our simulation.
Nevertheless, massive stars may start launching strong winds a few Myr before their core collapse, significantly shaping the ambient gas into which the bubbles expand.
%but the dissipation of stellar winds is much large, due to the strong viscosity, gravitation, radiative cooling, which will limit their influence.
A self-consistent implementation of the stellar winds requires a reliable stellar evolution model and a much higher resolution, thus awaits future work.

\section{Summary} \label{sec:sum}
The recently discovered radio bubbles and X-ray chimneys in the Galactic center both point to a dynamically young outflow.
%The AGN model and starburst model can both rough explain their origin, but we pay more attention to investigate multi-SNe explosions after the starburst in this work.
In this work we have used three-dimensional MHD simulations, carefully tailored to the physical conditions of the Galactic center, to explore the scenario in which a SN-driven, magnetically-collimated outflow produces the observed bubbles/chimneys.
The main results and implications of our study include:
\begin{enumerate}

  \item A SN-driven, magnetically-collimated outflow is naturally formed in almost all simulations performed. The morphology, X-ray luminosity and thermal energy of the radio bubbles/X-ray chimneys can be well reproduced for a reasonable choice of two parameters, namely, the SN birth rate and the strength of the vertical magnetic field. Meanwhile, we have examined the effect of changing these two parameters on the formation of the bubble.

  %\item The periodic starburst can also transport energy to the Fermi and eROSITA bubbles, and even dominates their formation.

  \item Dense filamentary features are seen both at the edge and in the interior of the simulated bubble, which are the sites of colliding shocks of individual SNe. This offers a plausible explanation for at least a fraction of the observed NTFs and the Radio Arc.

  \item In the framework of our simulations, the magnetic field in the Galactic center is likely to have a strength between 50--200 $\mu$G, consistent with previous estimates based on independent arguments.

\end{enumerate}
%Although this work can help us to simultaneously figure out the origin of the radio bubbles, the X-ray chimney, the NTFs and the radio arc, but the low resolution makes it inaccurate to explain some details, such as the formation of the NTFs and the radio arc.
%In addition, the stellar winds and SMBH are both essential energy sources, but we do not include them in our simulation, which largely limits our conclusions, such as the estimation of X-ray luminosity.

% The simulation lasts for 330 kyr, but we are also curious about the subsequent evolution, especially its relation with the Fermi bubbles.
% There were possibly many  starbursts
% However, to study a larger scale, we have to use a lower resolution, which is impossible to reflect the nature of the SNe, or we must spend unacceptable long time to perform the simulation.
% To study the large scale, we have to give up the small scale.
% Even the adaptive mesh refinement (AMR) is also useless in this case, because we must cover three magnitudes in one simulation box.
% Another solution is a step-by-step simulation.
% To run it many times at different scales, while use the results at smaller scale as the initial conditions at larger scale.
% This is feasible, but we have to choose more changeable parameters manually.

In conclusion, we are able to provide a viable formation mechanism for the radio bubbles/X-ray chimneys. This  invites future work to explore the possible physical connection between Galactic outflows on various scales.

\acknowledgements
This work is supported by the National Key Research and Development Program of China (grant 2017YFA0402703) and National Natural Science Foundation of China (grant 11873028).
We acknowledge the computing resources of Nanjing University, Purple Mountain observatory and National Astronomical Observatories of China.
We thank Miao Li and Feng Yuan for their helpful discussions,
and G. Ponti and I. Heywood for their communications on the estimation of the X-ray luminosity and radio flux density, respectively.

\bibliographystyle{aasjournal}
\bibliography{mydb}

\begin{thebibliography}{}
\expandafter\ifx\csname natexlab\endcsname\relax\def\natexlab#1{#1}\fi
\providecommand{\url}[1]{\href{#1}{#1}}

\bibitem[{{Baganoff} {et~al.}(2003){Baganoff}, {Maeda}, {Morris}, {Bautz},
  {Brandt}, {Cui}, {Doty}, {Feigelson}, {Garmire}, {Pravdo}, {Ricker}, \&
  {Townsley}}]{2003ApJ...591..891B}
{Baganoff}, F.~K., {Maeda}, Y., {Morris}, M., {et~al.} 2003, \apj, 591, 891

\bibitem[{Banda-Barrag{\'a}n {et~al.}(2018)Banda-Barrag{\'a}n, Federrath,
  Crocker, \& Bicknell}]{BandaBarragan2018}
Banda-Barrag{\'a}n, W.~E., Federrath, C., Crocker, R.~M., \& Bicknell, G.~V.
  2018, \mnras, 473, 3454

\bibitem[{Banda-Barrag{\'a}n {et~al.}(2016)Banda-Barrag{\'a}n, Parkin,
  Federrath, Crocker, \& Bicknell}]{BandaBarragan2016}
Banda-Barrag{\'a}n, W.~E., Parkin, E.~R., Federrath, C., Crocker, R.~M., \&
  Bicknell, G.~V. 2016, \mnras, 455, 1309

\bibitem[{{Barkov} \& {Lyutikov}(2019)}]{2019MNRAS.489L..28B}
{Barkov}, M.~V., \& {Lyutikov}, M. 2019, \mnras, 489, L28

\bibitem[{Barnes {et~al.}(2017)Barnes, Longmore, Battersby, Bally, Kruijssen,
  Henshaw, \& Walker}]{Barnes2017}
Barnes, A.~T., Longmore, S.~N., Battersby, C., {et~al.} 2017, \mnras, 469, 2263

\bibitem[{{Benford}(1988)}]{1988ApJ...333..735B}
{Benford}, G. 1988, \apj, 333, 735

\bibitem[{{Bland-Hawthorn} \& {Cohen}(2003)}]{2003ApJ...582..246B}
{Bland-Hawthorn}, J., \& {Cohen}, M. 2003, \apj, 582, 246

\bibitem[{Blasi(2013)}]{Blasi2013}
Blasi, P. 2013, \aapr, 21, 70

\bibitem[{{Boldyrev} \& {Yusef-Zadeh}(2006)}]{2006ApJ...637L.101B}
{Boldyrev}, S., \& {Yusef-Zadeh}, F. 2006, \apjl, 637, L101

\bibitem[{{Camilo} {et~al.}(2018){Camilo}, {Scholz}, {Serylak}, {Buchner},
  {Merryfield}, {Kaspi}, {Archibald}, {Bailes}, {Jameson}, {van Straten},
  {Sarkissian}, {Reynolds}, {Johnston}, {Hobbs}, {Abbott}, {Adam}, {Adams},
  {Alberts}, {Andreas}, {Asad}, {Baker}, {Baloyi}, {Bauermeister}, {Baxana},
  {Bennett}, {Bernardi}, {Booisen}, {Booth}, {Botha}, {Boyana}, {Brederode},
  {Burger}, {Cheetham}, {Conradie}, {Conradie}, {Davidson}, {De Bruin}, {de
  Swardt}, {de Villiers}, {de Villiers}, {de Villiers}, {de Villiers}, {De
  Waal}, {Dikgale}, {du Toit}, {du Toit}, {Esterhuyse}, {Fanaroff}, {Fataar},
  {Foley}, {Foster}, {Fourie}, {Gamatham}, {Gatsi}, {Geschke}, {Goedhart},
  {Grobler}, {Gumede}, {Hlakola}, {Hokwana}, {Hoorn}, {Horn}, {Horrell},
  {Hugo}, {Isaacson}, {Jacobs}, {Jansen van Rensburg}, {Jonas}, {Jordaan},
  {Joubert}, {Joubert}, {J{\'o}zsa}, {Julie}, {Julius}, {Kapp}, {Karastergiou},
  {Karels}, {Kariseb}, {Karuppusamy}, {Kasper}, {Knox-Davies}, {Koch},
  {Kotz{\'e}}, {Krebs}, {Kriek}, {Kriel}, {Kusel}, {Lamoor}, {Lehmensiek},
  {Liebenberg}, {Liebenberg}, {Lord}, {Lunsky}, {Mabombo}, {Macdonald},
  {Macfarlane}, {Madisa}, {Mafhungo}, {Magnus}, {Magozore}, {Mahgoub}, {Main},
  {Makhathini}, {Malan}, {Malgas}, {Manley}, {Manzini}, {Marais}, {Marais},
  {Marais}, {Maree}, {Martens}, {Matshawule}, {Matthysen}, {Mauch}, {McNally},
  {Merry}, {Millenaar}, {Mjikelo}, {Mkhabela}, {Mnyand u}, {Moeng}, {Mokone},
  {Monama}, {Montshiwa}, {Moss}, {Mphego}, {New}, {Ngcebetsha}, {Ngoasheng},
  {Niehaus}, {Ntuli}, {Nzama}, {Obies}, {Obrocka}, {Ockards}, {Olyn}, {Oozeer},
  {Otto}, {Padayachee}, {Passmoor}, {Patel}, {Paula}, {Peens-Hough},
  {Pholoholo}, {Prozesky}, {Rakoma}, {Ramaila}, {Rammala}, {Ramudzuli},
  {Rasivhaga}, {Ratcliffe}, {Reader}, {Renil}, {Richter}, {Robyntjies},
  {Rosekrans}, {Rust}, {Salie}, {Sambu}, {Schollar}, {Schwardt}, {Seranyane},
  {Sethosa}, {Sharpe}, {Siebrits}, {Sirothia}, {Slabber}, {Smirnov}, {Smith},
  {Sofeya}, {Songqumase}, {Spann}, {Stappers}, {Steyn}, {Steyn}, {Strong},
  {Struthers}, {Stuart}, {Sunnylall}, {Swart}, {Taljaard}, {Tasse}, {Taylor},
  {Theron}, {Thondikulam}, {Thorat}, {Tiplady}, {Toruvanda}, {van Aardt}, {van
  Balla}, {van den Heever}, {van der Byl}, {van der Merwe}, {van der Merwe},
  {van Niekerk}, {van Rooyen}, {van Staden}, {van Tonder}, {van Wyk}, {Wait},
  {Walker}, {Wallace}, {Welz}, {Williams}, {Xaia}, {Young}, \&
  {Zitha}}]{2018ApJ...856..180C}
{Camilo}, F., {Scholz}, P., {Serylak}, M., {et~al.} 2018, \apj, 856, 180

\bibitem[{{Carretti} {et~al.}(2013){Carretti}, {Crocker}, {Staveley-Smith},
  {Haverkorn}, {Purcell}, {Gaensler}, {Bernardi}, {Kesteven}, \&
  {Poppi}}]{2013Natur.493...66C}
{Carretti}, E., {Crocker}, R.~M., {Staveley-Smith}, L., {et~al.} 2013, \nat,
  493, 66

\bibitem[{Cheng {et~al.}(2015)Cheng, Chernyshov, Dogiel, \& Ko}]{Cheng2015}
Cheng, K.~S., Chernyshov, D.~O., Dogiel, V.~A., \& Ko, C.~M. 2015, \apj, 804,
  135

\bibitem[{Cheng {et~al.}(2011)Cheng, Chernyshov, Dogiel, Ko, \& Ip}]{Cheng2011}
Cheng, K.~S., Chernyshov, D.~O., Dogiel, V.~A., Ko, C.~M., \& Ip, W.~H. 2011,
  \apjl, 731, L17

\bibitem[{{Chevalier}(1992)}]{1992ApJ...397L..39C}
{Chevalier}, R.~A. 1992, \apjl, 397, L39

\bibitem[{{Chudnovsky} {et~al.}(1986){Chudnovsky}, {Field}, {Spergel}, \&
  {Vilenkin}}]{1986PhRvD..34..944C}
{Chudnovsky}, E.~M., {Field}, G.~B., {Spergel}, D.~N., \& {Vilenkin}, A. 1986,
  \prd, 34, 944

\bibitem[{{Coughlin} {et~al.}(2021){Coughlin}, {Nixon}, \&
  {Ginsburg}}]{2021MNRAS.501.1868C}
{Coughlin}, E.~R., {Nixon}, C.~J., \& {Ginsburg}, A. 2021, \mnras, 501, 1868

\bibitem[{{Crocker} \& {Aharonian}(2011)}]{2011PhRvL.106j1102C}
{Crocker}, R.~M., \& {Aharonian}, F. 2011, \prl, 106, 101102

\bibitem[{Crocker {et~al.}(2015)Crocker, Bicknell, Taylor, \&
  Carretti}]{Crocker2015}
Crocker, R.~M., Bicknell, G.~V., Taylor, A.~M., \& Carretti, E. 2015, \apj,
  808, 107

\bibitem[{{Crocker} {et~al.}(2010){Crocker}, {Jones}, {Melia}, {Ott}, \&
  {Protheroe}}]{2010Natur.463...65C}
{Crocker}, R.~M., {Jones}, D.~I., {Melia}, F., {Ott}, J., \& {Protheroe}, R.~J.
  2010, \nat, 463, 65

\bibitem[{{di Teodoro} {et~al.}(2020){di Teodoro}, {McClure-Griffiths},
  {Lockman}, \& {Armillotta}}]{2020Natur.584..364D}
{di Teodoro}, E.~M., {McClure-Griffiths}, N.~M., {Lockman}, F.~J., \&
  {Armillotta}, L. 2020, \nat, 584, 364

\bibitem[{{Di Teodoro} {et~al.}(2018){Di Teodoro}, {McClure-Griffiths},
  {Lockman}, {Denbo}, {Endsley}, {Ford}, \& {Harrington}}]{2018ApJ...855...33D}
{Di Teodoro}, E.~M., {McClure-Griffiths}, N.~M., {Lockman}, F.~J., {et~al.}
  2018, \apj, 855, 33

\bibitem[{{Dong} {et~al.}(2012){Dong}, {Wang}, \&
  {Morris}}]{2012MNRAS.425..884D}
{Dong}, H., {Wang}, Q.~D., \& {Morris}, M.~R. 2012, \mnras, 425, 884

\bibitem[{{Fabian}(2012)}]{2012ARA&A..50..455F}
{Fabian}, A.~C. 2012, \araa, 50, 455

\bibitem[{{Ferland} {et~al.}(2017){Ferland}, {Chatzikos}, {Guzm{\'a}n},
  {Lykins}, {van Hoof}, {Williams}, {Abel}, {Badnell}, {Keenan}, {Porter}, \&
  {Stancil}}]{2017RMxAA..53..385F}
{Ferland}, G.~J., {Chatzikos}, M., {Guzm{\'a}n}, F., {et~al.} 2017, \rmxaa, 53,
  385

\bibitem[{{Ferri{\`e}re}(2009)}]{Ferriere2009}
{Ferri{\`e}re}, K. 2009, \aap, 505, 1183

\bibitem[{{Ferri{\`e}re} {et~al.}(2007){Ferri{\`e}re}, {Gillard}, \&
  {Jean}}]{2007A&A...467..611F}
{Ferri{\`e}re}, K., {Gillard}, W., \& {Jean}, P. 2007, \aap, 467, 611

\bibitem[{{Genzel} {et~al.}(2010){Genzel}, {Eisenhauer}, \&
  {Gillessen}}]{2010RvMP...82.3121G}
{Genzel}, R., {Eisenhauer}, F., \& {Gillessen}, S. 2010, Reviews of Modern
  Physics, 82, 3121

\bibitem[{Guo \& Mathews(2012)}]{Guo2012}
Guo, F., \& Mathews, W.~G. 2012, \apj, 756, 181

\bibitem[{{Heckman} \& {Best}(2014)}]{2014ARA&A..52..589H}
{Heckman}, T.~M., \& {Best}, P.~N. 2014, \araa, 52, 589

\bibitem[{{Heckman} \& {Thompson}(2017)}]{2017hsn..book.2431H}
{Heckman}, T.~M., \& {Thompson}, T.~A. 2017, {Handbook of Supernovae}, ed.
  A.~W. {Alsabti} \& P.~{Murdin} (Springer International Publishing), 2431

\bibitem[{{Heyvaerts} {et~al.}(1988){Heyvaerts}, {Norman}, \&
  {Pudritz}}]{1988ApJ...330..718H}
{Heyvaerts}, J., {Norman}, C., \& {Pudritz}, R.~E. 1988, \apj, 330, 718

\bibitem[{{Heywood} {et~al.}(2019){Heywood}, {Camilo}, {Cotton}, {Yusef-Zadeh},
  {Abbott}, {Adam}, {Aldera}, {Bauermeister}, {Booth}, {Botha}, {Botha},
  {Brederode}, {Brits}, {Buchner}, {Burger}, {Chalmers}, {Cheetham}, {de
  Villiers}, {Dikgale-Mahlakoana}, {du Toit}, {Esterhuyse}, {Fanaroff},
  {Foley}, {Fourie}, {Gamatham}, {Goedhart}, {Gounden}, {Hlakola}, {Hoek},
  {Hokwana}, {Horn}, {Horrell}, {Hugo}, {Isaacson}, {Jonas}, {Jordaan},
  {Joubert}, {J{\'o}zsa}, {Julie}, {Kapp}, {Kenyon}, {Kotz{\'e}}, {Kriel},
  {Kusel}, {Lehmensiek}, {Liebenberg}, {Loots}, {Lord}, {Lunsky}, {Macfarlane},
  {Magnus}, {Magozore}, {Mahgoub}, {Main}, {Malan}, {Malgas}, {Manley},
  {Maree}, {Merry}, {Millenaar}, {Mnyandu}, {Moeng}, {Monama}, {Mphego}, {New},
  {Ngcebetsha}, {Oozeer}, {Otto}, {Passmoor}, {Patel}, {Peens-Hough},
  {Perkins}, {Ratcliffe}, {Renil}, {Rust}, {Salie}, {Schwardt}, {Serylak},
  {Siebrits}, {Sirothia}, {Smirnov}, {Sofeya}, {Swart}, {Tasse}, {Taylor},
  {Theron}, {Thorat}, {Tiplady}, {Tshongweni}, {van Balla}, {van der Byl}, {van
  der Merwe}, {van Dyk}, {Van Rooyen}, {Van Tonder}, {Van Wyk}, {Wallace},
  {Welz}, \& {Williams}}]{2019Natur.573..235H}
{Heywood}, I., {Camilo}, F., {Cotton}, W.~D., {et~al.} 2019, \nat, 573, 235

\bibitem[{{Insertis} \& {Rees}(1991)}]{1991MNRAS.252...82I}
{Insertis}, F.~M., \& {Rees}, M.~J. 1991, \mnras, 252, 82

\bibitem[{{Ko} {et~al.}(2020){Ko}, {Breitschwerdt}, {Chernyshov}, {Cheng},
  {Dai}, \& {Dogiel}}]{2020ApJ...904...46K}
{Ko}, C.~M., {Breitschwerdt}, D., {Chernyshov}, D.~O., {et~al.} 2020, \apj,
  904, 46

\bibitem[{Kroupa(2001)}]{Kroupa2001}
Kroupa, P. 2001, \mnras, 322, 231

\bibitem[{{Kruijssen} {et~al.}(2015){Kruijssen}, {Dale}, \&
  {Longmore}}]{2015MNRAS.447.1059K}
{Kruijssen}, J.~M.~D., {Dale}, J.~E., \& {Longmore}, S.~N. 2015, \mnras, 447,
  1059

\bibitem[{{Krumholz} \& {Kruijssen}(2015)}]{2015MNRAS.453..739K}
{Krumholz}, M.~R., \& {Kruijssen}, J.~M.~D. 2015, \mnras, 453, 739

\bibitem[{{Lacki}(2014)}]{2014MNRAS.444L..39L}
{Lacki}, B.~C. 2014, \mnras, 444, L39

\bibitem[{{Launhardt} {et~al.}(2002){Launhardt}, {Zylka}, \&
  {Mezger}}]{2002A&A...384..112L}
{Launhardt}, R., {Zylka}, R., \& {Mezger}, P.~G. 2002, \aap, 384, 112

\bibitem[{{Law}(2010)}]{2010ApJ...708..474L}
{Law}, C.~J. 2010, \apj, 708, 474

\bibitem[{{Lesch} \& {Reich}(1992)}]{1992A&A...264..493L}
{Lesch}, H., \& {Reich}, W. 1992, \aap, 264, 493

\bibitem[{Li \& Bryan(2020)}]{Li2020}
Li, M., \& Bryan, G.~L. 2020, \apjl, 890, L30

\bibitem[{{Li} {et~al.}(2017){Li}, {Bryan}, \&
  {Ostriker}}]{2017ApJ...841..101L}
{Li}, M., {Bryan}, G.~L., \& {Ostriker}, J.~P. 2017, \apj, 841, 101

\bibitem[{{Longair}(2011)}]{Longair2011}
{Longair}, M.~S. 2011, {High Energy Astrophysics} (Cambridge University Press)

\bibitem[{{Mannucci} {et~al.}(2005){Mannucci}, {Della Valle}, {Panagia},
  {Cappellaro}, {Cresci}, {Maiolino}, {Petrosian}, \&
  {Turatto}}]{2005A&A...433..807M}
{Mannucci}, F., {Della Valle}, M., {Panagia}, N., {et~al.} 2005, \aap, 433, 807

\bibitem[{{Mignone} {et~al.}(2007){Mignone}, {Bodo}, {Massaglia}, {Matsakos},
  {Tesileanu}, {Zanni}, \& {Ferrari}}]{Mignone2007}
{Mignone}, A., {Bodo}, G., {Massaglia}, S., {et~al.} 2007, \apjs, 170, 228

\bibitem[{{Mignone} {et~al.}(2012){Mignone}, {Zanni}, {Tzeferacos}, {van
  Straalen}, {Colella}, \& {Bodo}}]{Mignone2012}
{Mignone}, A., {Zanni}, C., {Tzeferacos}, P., {et~al.} 2012, \apjs, 198, 7

\bibitem[{{Morris}(1996)}]{1996IAUS..169..247M}
{Morris}, M. 1996, in IAU Symposium, Vol. 169, Unsolved Problems of the Milky
  Way, ed. L.~{Blitz} \& P.~J. {Teuben}, 247

\bibitem[{{Morris} \& {Yusef-Zadeh}(1989)}]{1989ApJ...343..703M}
{Morris}, M., \& {Yusef-Zadeh}, F. 1989, \apj, 343, 703

\bibitem[{Mou {et~al.}(2014)Mou, Yuan, Bu, Sun, \& Su}]{Mou2014}
Mou, G., Yuan, F., Bu, D., Sun, M., \& Su, M. 2014, \apj, 790, 109

\bibitem[{Mou {et~al.}(2015)Mou, Yuan, Gan, \& Sun}]{Mou2015}
Mou, G., Yuan, F., Gan, Z., \& Sun, M. 2015, \apj, 811, 37

\bibitem[{{Nagoshi} {et~al.}(2019){Nagoshi}, {Kubose}, {Fujisawa}, {Sorai},
  {Yonekura}, {Sugiyama}, {Niinuma}, {Motogi}, \& {Aoki}}]{2019PASJ...71...80N}
{Nagoshi}, H., {Kubose}, Y., {Fujisawa}, K., {et~al.} 2019, \pasj, 71, 80

\bibitem[{{Najarro} {et~al.}(1997){Najarro}, {Krabbe}, {Genzel}, {Lutz},
  {Kudritzki}, \& {Hillier}}]{1997A&A...325..700N}
{Najarro}, F., {Krabbe}, A., {Genzel}, R., {et~al.} 1997, \aap, 325, 700

\bibitem[{{Nakashima} {et~al.}(2019){Nakashima}, {Koyama}, {Wang}, \&
  {Enokiya}}]{2019ApJ...875...32N}
{Nakashima}, S., {Koyama}, K., {Wang}, Q.~D., \& {Enokiya}, R. 2019, \apj, 875,
  32

\bibitem[{Nogueras-Lara {et~al.}(2020)Nogueras-Lara, Sch{\"o}del,
  Gallego-Calvente, Gallego-Cano, Shahzamanian, Dong, Neumayer, Hilker,
  Najarro, Nishiyama, Feldmeier-Krause, Girard, \& Cassisi}]{NoguerasLara2019}
Nogueras-Lara, F., Sch{\"o}del, R., Gallego-Calvente, A.~T., {et~al.} 2020,
  Nature Astronomy, 4, 377

\bibitem[{{Orlando} {et~al.}(2007){Orlando}, {Bocchino}, {Reale}, {Peres}, \&
  {Petruk}}]{Orlando2007}
{Orlando}, S., {Bocchino}, F., {Reale}, F., {Peres}, G., \& {Petruk}, O. 2007,
  \aap, 470, 927

\bibitem[{{Ponti} {et~al.}(2021){Ponti}, {Morris}, {Churazov}, {Heywood}, \&
  {Fender}}]{2021A&A...646A..66P}
{Ponti}, G., {Morris}, M.~R., {Churazov}, E., {Heywood}, I., \& {Fender}, R.~P.
  2021, \aap, 646, A66

\bibitem[{Ponti {et~al.}(2013)Ponti, Morris, Terrier, \& Goldwurm}]{Ponti2013}
Ponti, G., Morris, M.~R., Terrier, R., \& Goldwurm, A. 2013, in Cosmic Rays in
  Star-Forming Environments, ed. D.~F. {Torres} \& O.~{Reimer}, Vol.~34, 331

\bibitem[{{Ponti} {et~al.}(2010){Ponti}, {Terrier}, {Goldwurm}, {Belanger}, \&
  {Trap}}]{2010ApJ...714..732P}
{Ponti}, G., {Terrier}, R., {Goldwurm}, A., {Belanger}, G., \& {Trap}, G. 2010,
  \apj, 714, 732

\bibitem[{{Ponti} {et~al.}(2015){Ponti}, {Morris}, {Terrier}, {Haberl},
  {Sturm}, {Clavel}, {Soldi}, {Goldwurm}, {Predehl}, {Nandra}, {B{\'e}langer},
  {Warwick}, \& {Tatischeff}}]{Ponti2015}
{Ponti}, G., {Morris}, M.~R., {Terrier}, R., {et~al.} 2015, \mnras, 453, 172

\bibitem[{{Ponti} {et~al.}(2019){Ponti}, {Hofmann}, {Churazov}, {Morris},
  {Haberl}, {Nandra}, {Terrier}, {Clavel}, \& {Goldwurm}}]{2019Natur.567..347P}
{Ponti}, G., {Hofmann}, F., {Churazov}, E., {et~al.} 2019, \nat, 567, 347

\bibitem[{{Poznanski}(2013)}]{Poznanski2013}
{Poznanski}, D. 2013, \mnras, 436, 3224

\bibitem[{{Predehl} {et~al.}(2020){Predehl}, {Sunyaev}, {Becker}, {Brunner},
  {Burenin}, {Bykov}, {Cherepashchuk}, {Chugai}, {Churazov}, {Doroshenko},
  {Eismont}, {Freyberg}, {Gilfanov}, {Haberl}, {Khabibullin}, {Krivonos},
  {Maitra}, {Medvedev}, {Merloni}, {Nandra}, {Nazarov}, {Pavlinsky}, {Ponti},
  {Sanders}, {Sasaki}, {Sazonov}, {Strong}, \& {Wilms}}]{Predehl2020}
{Predehl}, P., {Sunyaev}, R.~A., {Becker}, W., {et~al.} 2020, \nat, 588, 227

\bibitem[{{Quataert}(2004)}]{2004ApJ...613..322Q}
{Quataert}, E. 2004, \apj, 613, 322

\bibitem[{{Rosso} \& {Pelletier}(1993)}]{1993A&A...270..416R}
{Rosso}, F., \& {Pelletier}, G. 1993, \aap, 270, 416

\bibitem[{{Rozyczka} \& {Tenorio-Tagle}(1995)}]{1995MNRAS.274.1157R}
{Rozyczka}, M., \& {Tenorio-Tagle}, G. 1995, \mnras, 274, 1157

\bibitem[{Sarkar {et~al.}(2015)Sarkar, Nath, \& Sharma}]{Sarkar2015}
Sarkar, K.~C., Nath, B.~B., \& Sharma, P. 2015, \mnras, 453, 3827

\bibitem[{{Sarkar} {et~al.}(2017){Sarkar}, {Nath}, \&
  {Sharma}}]{2017MNRAS.467.3544S}
{Sarkar}, K.~C., {Nath}, B.~B., \& {Sharma}, P. 2017, \mnras, 467, 3544

\bibitem[{{Serabyn} \& {Morris}(1994)}]{1994ApJ...424L..91S}
{Serabyn}, E., \& {Morris}, M. 1994, \apjl, 424, L91

\bibitem[{{Shore} \& {LaRosa}(1999)}]{1999ApJ...521..587S}
{Shore}, S.~N., \& {LaRosa}, T.~N. 1999, \apj, 521, 587

\bibitem[{Simpson {et~al.}(2001)Simpson, Lane, Immer, \&
  Youngquist}]{simpson2001simple}
Simpson, J.~C., Lane, J.~E., Immer, C.~D., \& Youngquist, R.~C. 2001, Simple
  analytic expressions for the magnetic field of a circular current loop (NASA
  technical documents)

\bibitem[{{Smith} {et~al.}(2001){Smith}, {Brickhouse}, {Liedahl}, \&
  {Raymond}}]{2001ApJ...556L..91S}
{Smith}, R.~K., {Brickhouse}, N.~S., {Liedahl}, D.~A., \& {Raymond}, J.~C.
  2001, \apjl, 556, L91

\bibitem[{{Sofue}(2020)}]{2020PASJ...72L...4S}
{Sofue}, Y. 2020, \pasj, 72, L4

\bibitem[{{Sofue} \& {Handa}(1984)}]{1984Natur.310..568S}
{Sofue}, Y., \& {Handa}, T. 1984, \nat, 310, 568

\bibitem[{{Sormani} {et~al.}(2020){Sormani}, {Tress}, {Glover}, {Klessen},
  {Battersby}, {Clark}, {Hatchfield}, \& {Smith}}]{2020MNRAS.497.5024S}
{Sormani}, M.~C., {Tress}, R.~G., {Glover}, S. C.~O., {et~al.} 2020, \mnras,
  497, 5024

\bibitem[{{Stolte} {et~al.}(2008){Stolte}, {Ghez}, {Morris}, {Lu}, {Brand ner},
  \& {Matthews}}]{2008ApJ...675.1278S}
{Stolte}, A., {Ghez}, A.~M., {Morris}, M., {et~al.} 2008, \apj, 675, 1278

\bibitem[{{Stone} \& {Norman}(1992)}]{1992ApJ...389..297S}
{Stone}, J.~M., \& {Norman}, M.~L. 1992, \apj, 389, 297

\bibitem[{{Su} {et~al.}(2010){Su}, {Slatyer}, \&
  {Finkbeiner}}]{2010ApJ...724.1044S}
{Su}, M., {Slatyer}, T.~R., \& {Finkbeiner}, D.~P. 2010, \apj, 724, 1044

\bibitem[{Thomas {et~al.}(2020)Thomas, Pfrommer, \& En{\ss}lin}]{Thomas2020}
Thomas, T., Pfrommer, C., \& En{\ss}lin, T. 2020, \apjl, 890, L18

\bibitem[{{Truelove} \& {McKee}(1999)}]{Truelove1999}
{Truelove}, J.~K., \& {McKee}, C.~F. 1999, \apjs, 120, 299

\bibitem[{Wu \& Zhang(2019)}]{Wu2019}
Wu, D., \& Zhang, M.-F. 2019, RAA, 19, 124

\bibitem[{{Yang} {et~al.}(2013){Yang}, {Ruszkowski}, \&
  {Zweibel}}]{2013MNRAS.436.2734Y}
{Yang}, H. Y.~K., {Ruszkowski}, M., \& {Zweibel}, E. 2013, \mnras, 436, 2734

\bibitem[{{Yuan} {et~al.}(2015){Yuan}, {Gan}, {Narayan}, {Sadowski}, {Bu}, \&
  {Bai}}]{2015ApJ...804..101Y}
{Yuan}, F., {Gan}, Z., {Narayan}, R., {et~al.} 2015, \apj, 804, 101

\bibitem[{{Yusef-Zadeh}(2003)}]{2003ApJ...598..325Y}
{Yusef-Zadeh}, F. 2003, \apj, 598, 325

\bibitem[{{Yusef-Zadeh} {et~al.}(1984){Yusef-Zadeh}, {Morris}, \&
  {Chance}}]{1984Natur.310..557Y}
{Yusef-Zadeh}, F., {Morris}, M., \& {Chance}, D. 1984, \nat, 310, 557

\bibitem[{Yusef-Zadeh \& Wardle(2019)}]{2019ApJ...490..L1}
Yusef-Zadeh, F., \& Wardle, M. 2019, \mnras, 490, L1

\bibitem[{{Zhang}(2018)}]{2018Galax...6..114Z}
{Zhang}, D. 2018, Galaxies, 6, 114

\bibitem[{{Zhang} {et~al.}(2017){Zhang}, {Tian}, {Leahy}, {Zhu}, {Cui}, \&
  {Shan}}]{Zhang2017}
{Zhang}, M.~F., {Tian}, W.~W., {Leahy}, D.~A., {et~al.} 2017, \apj, 849, 147

\bibitem[{Zhang \& Guo(2020)}]{Zhang2020}
Zhang, R., \& Guo, F. 2020, \apj, 894, 117

\bibitem[{{Zhou} {et~al.}(2021){Zhou}, {Leung}, {Li}, {Nomoto}, {Vink}, \&
  {Chen}}]{2021ApJ...908...31Z}
{Zhou}, P., {Leung}, S.-C., {Li}, Z., {et~al.} 2021, \apj, 908, 31

\bibitem[{{Zhu} {et~al.}(2018){Zhu}, {Li}, \& {Morris}}]{2018ApJS..235...26Z}
{Zhu}, Z., {Li}, Z., \& {Morris}, M.~R. 2018, \apjs, 235, 26

\bibitem[{Zubovas {et~al.}(2011)Zubovas, King, \& Nayakshin}]{Zubovas2011}
Zubovas, K., King, A.~R., \& Nayakshin, S. 2011, \mnras, 415, L21

\bibitem[{{Zubovas} \& {Nayakshin}(2012)}]{2012MNRAS.424..666Z}
{Zubovas}, K., \& {Nayakshin}, S. 2012, \mnras, 424, 666

\end{thebibliography}
\end{document}